\documentclass[lscape,numberedappendix,appendixfloats]{emulateapj}

\usepackage{graphicx,ifthen,url,float,lscape}

\usepackage{psfig}
\usepackage{amsmath}
\usepackage{amssymb}
\usepackage{xspace}
\newcommand {\um}{$\mu$m~}
\newcommand {\ums}{$\mu$m}

\newcommand {\Mpy}{M$_\odot$\,yr$^{-1}$}
\def\ts     {\thinspace}
\def\uJy    {$\mu$Jy}
\def\kms    {\ifmmode{{\rm \ts km\ts s}^{-1}}\else{\ts km\ts s$^{-1}$}\fi}
\def\msol   {\ifmmode{{\rm M}_{\odot}}\else{M$_{\odot}$}\fi}
\def\lsol   {\ifmmode{{\rm L}_{\odot}}\else{L$_{\odot}$}\fi}
\def\zsol   {\ifmmode{{\rm Z}_{\odot}}\else{Z$_{\odot}$}\fi}

\def\ltsima{$\; \buildrel < \over \sim \;$}
\def\simlt{\lower.5ex\hbox{\ltsima}}
\def\gtsima{$\; \buildrel > \over \sim \;$}
\def\simgt{\lower.5ex\hbox{\gtsima}}

\newcommand{\sfr}{{\rm\,M$_\odot$\,yr$^{-1}$}}
\newcommand{\lsun}{{\rm\,L$_\odot$}}

\newcommand{\ratio}{$S_{500\mu m}$/$S_{24\mu m}$}
\def\ppm    {$\pm$}
\shortauthors{Shu et al.}

\newcommand{\herschel}{{\it Herschel~}}
\RequirePackage[normalem]{ulem} %DIF PREAMBLE
\RequirePackage{color}\definecolor{RED}{rgb}{1,0,0}\definecolor{BLUE}{rgb}{0,0,1} %DIF PREAMBLE
\providecommand{\DIFaddend}{} %DIF PREAMBLE
\begin{document}


%\title{\textnormal ASTRODEEP: \\
\title{Identification of \lowercase{z}$\simgt2$ \herschel 500\um sources using color-deconfusion}
%on the prevalence of high-redshift dusty starbursts}


\author{X.W. Shu\altaffilmark{1,2}, 
D. ~Elbaz\altaffilmark{1},
N. ~Bourne\altaffilmark{5},
C. ~Schreiber\altaffilmark{1},
T. ~Wang\altaffilmark{1,10},
%N. ~Bourne\altaffilmark{5},
%R. Leiton\altaffilmark{6},
%A. ~Fontana\altaffilmark{3},
%H. C. ~Ferguson\altaffilmark{4},
J. ~S. ~Dunlop\altaffilmark{5},
A. ~Fontana\altaffilmark{3},
%C. ~Schreiber\altaffilmark{1},
%T. ~Wang\altaffilmark{1,11},
R. Leiton\altaffilmark{6}, 
M. Pannella\altaffilmark{1,7},
K. Okumura\altaffilmark{1},
%N. ~Bourne\altaffilmark{5},
M. J. ~Micha{\l}owski\altaffilmark{5},
P. ~Santini\altaffilmark{3},
E. ~Merlin\altaffilmark{3},
F. ~Buitrago\altaffilmark{5},
V. ~A. ~Bruce\altaffilmark{5},
R. ~Amorin\altaffilmark{3},
M. ~Castellano\altaffilmark{3},
S. ~Derriere\altaffilmark{8},
A. ~Comastri\altaffilmark{9},
N. ~Cappelluti\altaffilmark{9}, 
J. X. ~Wang\altaffilmark{11}, 
H. C. ~Ferguson\altaffilmark{4},
%S. ~M. ~Faber\altaffilmark{10}
}
\altaffiltext{1}{Laboratoire AIM-Paris-Saclay, CEA/DSM/Irfu - CNRS - Universit\'e Paris Diderot, CEA-Saclay, pt courrier 131, F-91191 Gif-sur-Yvette, France, xinwen.shu@cea.fr}
\altaffiltext{2}{Department of Physics, Anhui Normal University, Wuhu, Anhui, 241000, China}
\altaffiltext{3}{INAF - Osservatorio Astronomico di Roma, Via Frascati 33, I - 00040 Monte Porzio Catone (RM), Italy}
\altaffiltext{4}{Space Telescope Science Institute, 3700 San Martin Drive, Baltimore, MD 21218, USA}
\altaffiltext{5}{SUPA\thanks{Scottish Universities Physics Alliance}, Institute for Astronomy, University of Edinburgh, Royal Observatory, Edinburgh, EH9 3HJ, U.K.}
\altaffiltext{6}{Astronomy Department, Universidad de Concepci\'{o}n, Casilla 160-C,
Concepción, Chile}
\altaffiltext{7}{Ludwig-Maximilians-Universität, Department of
Physics, Scheinerstr. 1, 81679 M\"{u}nchen, Germany}
\altaffiltext{8}{Observatoire astronomique de Strasbourg, Université de Strasbourg, CNRS, UMR 7550, 11
rue de l’Université, F-67000 Strasbourg, France}
\altaffiltext{9}{INAF - Osservatorio Astronomico di Bologna, Via Ranzani 1, I - 40127, Bologna, Italy}
%\altaffiltext{10}{UCO/Lick Observatory, University of California, 1156 High Street, Santa Cruz, CA 95064, USA}
\altaffiltext{10}{School of Astronomy and Astrophysics, Nanjing University, Nanjing, 210093, China}
\altaffiltext{11}{Department of Astronomy, University of Science and Technology of China, Hefei, Anhui 230026, China}
%\author{D. ~Elbaz\altaffilmark{1}}
%\affil{Laboratoire AIM-Paris-Saclay, CEA/DSM/Irfu - CNRS - Universit\'e Paris Diderot, CEA-Saclay, pt courrier 131, F-91191 Gif-sur-Yvette, France}
\setcounter{footnote}{15}
%\affil{National Optical Astronomy Observatories, Tucson, AZ 85719}
%\email{aastex-help@aas.org}

%\and

%\author{R. J. Hanisch\altaffilmark{5}}
%\affil{Space Telescope Science Institute, Baltimore, MD 21218}


%\altaffiltext{1}{Visiting Astronomer, Cerro Tololo Inter-American Observatory.
%CTIO is operated by AURA, Inc.\ under contract to the National Science
%\altaffiltext{2}{Society of Fellows, Harvard University.}
%\altaffiltext{3}{present address: Center for Astrophysics,
%    60 Garden Street, Cambridge, MA 02138}
%\altaffiltext{4}{Visiting Programmer, Space Telescope Science Institute}
%\altaffiltext{5}{Patron, Alonso's Bar and Grill}


\begin{abstract}
We present a new method to search for candidate $z\simgt2$ \herschel 500\um sources  
in the GOODS-North field, using a \ratio~``color deconfusion" technique. 
%Our method which is based on the \ratio color deconfusion,  
%Our method is to search for candidates in the map-based ratio image which 
%can reduce effectively the source confusion from low-redshift galaxies, making  
%a high-$z$ {\sc Spire}/500\um source more evident to be identified. 
%Since there is an increased far-infrared emission towards longer wavelengths, with increases of 
%redshift The \ratio flux density ratio is a strong function of redshift 
%We constructed a map of 24\um sources beam-smeared to match the \ratio~ratio map by dividing the beam-
%Since redshift domintes shift of far-infrared emission peak of galaxies, 
%The observed \ratio~flux ratios can be used to select potential high-$z$ sources 
%against the low-redshift ones, since  
%Our approach is to construct a \ratio~ratio map to 
{ Potential high-$z$ sources are selected} 
against low-redshift ones 
from their large 500\um to 24\um flux density ratios. 
By effectively reducing the contribution
from low-redshift populations to the observed 500\um emission, 
%
%We construct a \ratio~ratio map to identify distant galaxies 
%against the low-redshift ones from large 500\um to 24\um flux density ratios. 
we are able to identify counterparts to high-$z$ 500\um sources whose 
24\um fluxes are { relatively faint}. 
The recovery of known $z\simgt4$ starbursts confirms the efficiency of this 
approach in selecting high-$z$ \herschel sources. 
{ The resulting sample consists of {  34 dusty star-forming galaxies at $z\simgt2$. 
The inferred infrared luminosities are in the range  $1.5\times10^{12}-1.8\times10^{13}$\lsun,} 
corresponding to { dust-obscured} star formation rates (SFRs) of $\sim260-3100$\sfr~for  
a Salpeter initial mass function. }
{ Comparison with previous SCUBA 850$\mu\rm m$-selected galaxy samples shows 
that our method is more efficient at selecting high-$z$ dusty galaxies,}
with a median redshift {  of $z=3.07\pm0.83$ and 10 }of the sources at $z\simgt4$.
%We find that the redshift distribution of the \ratio-selected sources is, by 
%selection, higher than previous SCUBA 850$\mu\rm m$-selected SMGs, 
%with a median redshift of $z=3.19\pm0.83$ and 11 of the sources at $z\simgt4$.
%This suggests that the fraction of dusty starburst galaxies at high redshifts may be 
%higher than previously thought, though the cosmic variance should be taken into account given  
%the small sky area we probed.  
%We estimate the number of sources which are possibly missed among the 
%$z\simgt$2 {\sc Spire} population using a nominal identification method 
%is $\sim30$\%, but larger survey fields and 
%high-resolution imaging and spectroscopic observations are needed 
%to confirm the results. %With the photometric redshift information
%The derived characteristic dust temperatures ($T_d$) are found to correlate with IR luminosities, albeit the scatter is large.
{ We find that }at a fixed luminosity, the dust temperature is $\sim5$K cooler than that expected 
from the $T_d-L_{\rm IR}$ relation at $z\simlt$1, though different temperature selection 
effects should be taken into account.  
%$T_D=33.3\pm4.0$K,
The radio-detected { subsample} (excluding three strong AGN) follows the far-infrared/radio 
correlation at lower redshifts, and no evolution with redshift is observed out to $z\sim5$, 
suggesting that the far-infrared emission is star formation dominated. 
{ The contribution of the high-$z$ \herschel 500\um sources to the cosmic SFR density is 
comparable to that of (sub)millimeter galaxy populations} 
at $z\sim2.5$ and at least 40\% of the extinction-corrected UV samples at $z\sim4$. 
Further investigation into the nature of these high-$z$ 
dusty galaxies will {be crucial for our understanding} of the 
star formation histories and the buildup of stellar mass at the earliest 
cosmic epochs.  
\end{abstract}



\section{Introduction}
%Submillimeter galaxies (SMGs) are ultraluminous, dust obscured star-forming systems with 
%extreme star formation rates (SFRs) in the range of $\sim$100-1000 M yr$^{-1}$ 
% \citep{blain02}. 
Ultraluminous infrared galaxies (ULIRGs; with rest-frame 8-1000\um luminosities in excess of $10^{12}$\lsun) 
are among the brightest far-infrared (far-IR) emitters in the Universe (Sanders \& Mirabel 1996; Casey, Narayanan \& 
Cooray 2014 and references therein). 
The far-IR luminosities of ULIRGs are dominated by reprocessed thermal 
dust emission due to star-formation activity, corresponding to a star formation rate (SFR) of $>$170\sfr. 
Locally, these luminous, dusty star-forming galaxies are rare, but they become more abundant 
with increasing redshift and dominate the IR luminosity density of galaxies around $z\sim2$
%thousands times more abundant 
%in the early Universe than at present 
\citep{lagache05, chapman05, lefloch09, magnelli09, magnelli13}. 
The formation process of ULIRGs is currently in dispute.  
%While merges are believed to a dominant process in local ULIRGs, 
%in the local universe, such situation is less clear in high-$z$ ULIRGs ($z\sim2$), 
%they are not necessary for the high luminosities of 
%they are playing a minor role in ULIRGs at high redshift ($z\simgt2$), 
%at least for those on the ``main-sequence"
%Studies have suggested that the driver of star formation in ULIRGs at high redshift ($z\simgt2$) 
%appears different from the local ULIRGs, and mergers being a dominant process for the latter 
While mergers are believed to be a dominant process in local ULIRGs, 
studies have suggested that the driver of star formation in ULIRGs at high redshift ($z\simgt2$) 
appears different (e.g., Dekel et al. 2009), with a significant fraction being pure disks (e.g.,  
Kartaltepe et al. 2012; Targett et al. 2013). 
%, with a significant faction being disks (Kartaltepe et al. 2012) 
%possibly being driven by the mergers   
%(e.g., Engel et al. 2010, see also Dekel et al. 2009 for secular gas accretion model). 
Reproducing the number counts and extreme far-IR luminosities for this population, especially those at 
the highest redshifts, can place tight constraints on galaxy formation models \citep{baugh05, swinbank08, coppin09}.
Since a significant fraction of cosmic star formation is
likely hidden by dust (e.g., Lagache et al. 2005; Chapman et al. 2005;
Casey et al. 2012a, 2012b, 2014; Barger et al. 2012; Lutz et al. 2014;
Madau \& Dickinson 2014), understanding high-$z$ far-IR luminous galaxies is
crucial in order to construct a complete picture of galaxy evolution.
 
%(Lagache, Puget \& Dole 2005; Chapman et al. 2005; LeFloc'h et al. 2009; Magnelli et al. 2013).  
%Since their discovery more than a decade ago, the study of SMGs has been revolutionizing 
%our knowledge of the early formation history of stars and galaxies
%(e.g., Lagache, Puget, \& Dole 2005).
 %Dusty SMGs are a thousand times more abundant at $z\sim2$ than 
%they are at present day (Magnelli et al. 2013), and contribute significantly to 
%the rapid build-up of stellar mass in the early Universe. They are thought to be 
%linked to both QSO activity and the formation
%of today's massive elliptical galaxies (e.g., Hopkins et al. , Hickox et al. 2012), 
%putting powerful constraints on models of galaxy formation and evolution and the 
%environments of violent star-bursts. 

%ULIRGs at high redshift were first studied in the submillimeter (submm) and millimeter (mm) bands, 
%where the Rayleigh-Jeans portion of the dust spectral energy distribution (SED) 
Observations at submillimeter (submm) and millimeter (mm) 
wavelengths have identified a population of ULIRGs at high redshift (e.g., Borys et al. 2003; 
Pope et al. 2005; Perera et al. 2008; Micha{\l}owski et al. 2010a,b; 
Yun et al. 2012; Wardlow et al. 2011; Roseboom et al. 2013; Casey et al. 2013), 
which benefits from a strong negative $K$-correction in the (sub)mm. 
% 
%produced by the Rayleigh-Jeans portion of the dust spectral energy distribution (SED). 
%For sources at a fixed luminosity, this effect makes the observed submm flux  
%almost independent of redshift over a wide range of $1<z<10$ (Blain et al. 2002). 
%By identifying radio counterparts to precisely localize the (sub)millimeter sources for successful optical 
%However, determining the nature of
%One of the challenges in studying SMGs has been the difficulty 
%to obtain an identification of the counterparts at other wavelengths, and hence redshift for SMGs. 
%While the precise localization of a (sub)millimeter source can be obtained by identifying radio counterparts (e.g., Ivison et al. 2007; 
%Roseboom et al. 2010) to precisely localize the submm sources for successful optical spectroscopy. 
%, the bulk of SMG population is between $z\sim2$-3 (e.g., Chapman et al. 2005).
%This method, however, is severely biased galaxies at higher redshifts ($z>3$) because of the 
Follow-up studies of submm-selected galaxies (SMGs) have shown that the bulk of SMGs 
is between $z\sim2-3$ (e.g., Chapman et al. 2005), 
with less than $\sim30$\% being at $z>3$ (Chapin et al. 2009; Biggs et al. 2011; Wardlow et al. 2011; Micha{\l}owski 
et al. 2012). 
While radio identification has been the common technique to precisely localize         
SMGs at other wavelengths for follow-up observations, 
a significant number of them remain undetected even in the current 
deepest radio observations (e.g., Swinbank et al. 2014; Barger et al. 2014), 
indicating that they are possibly at higher redshifts. 
Interferometric submm imaging of SMGs has been used to unambiguously 
identify counterparts in the optical/near-IR in a relatively unbiased way, 
revealing a significant population of SMGs at $z\simgt$3
 (e.g., Younger et al. 2007, 2009; Smol\v{c}i\'{c} et al. 2012a, 2012b, 2015; Miettinen et al. 2015), 
challenging the previously derived redshift distribution of SMGs (see also Vieira et al. 2013).
On the other hand, the selection of SMGs is known to be severely biased against galaxies 
with warmer dust temperatures (e.g., Chapman et al. 2003; Casey et al. 2009; 
Magdis et al. 2010; Chapman et al. 2010; Magnelli et al. 2010). 
Therefore, the properties of the { whole high-$z$ ULIRG population,} 
%IR-luminous ($L_{\rm IR}>10^{12}$\lsol)  
%star-forming galaxies, 
such as the redshift distribution, luminosity and number density, 
are still not well characterized.
The {\it Herschel Space Observatory} (hereafter {\it Herschel}) carried out observations 
in the far-IR bands up to 500\um and 
mapped much larger sky areas down to the confusion limit than previous 
(sub)mm surveys. 
%By identifying galaxies at 70--500 $\mu$m, Herschel sampled 
By sampling the peak of heated dust emission for the first time,  
{\it Herschel} enables a direct assessment of 
the far-IR spectral energy distribution (SED) of SMGs 
(e.g., Magnelli et al. 2010, 2012, 2014; Huang et al. 2014; Swinbank et al. 2014). 
%{ \herschel
%{  with that of the $Spitzer$-MIPS 24$\mu$m 
%observed with $Spitzer$-MIPS 
%with the (sub)millimeter photometry at longer wavelengths 
%enables us to robustly characterize the far-IR SED and 
%measure the total IR luminosities of galaxies. 
It has provided unbiased measurements of obscured SFR, dust temperature and dust mass  
for a large number of star-forming galaxies up to $z\sim2$ (e.g., Elbaz et al. 2010, 2011; 
Oliver et al. 2012; and references therein). 

%
%and is probing extreme starbursts at 
%$z>4$ by their red {\sc SPIRE} colors \citep{Riechers13, Dowell14}. 
%especially for those at $z>2$, 
%allowing for direct measurements of the bolometric output of star-forming galaxies, hence of 
%their star formation rate (SFR) (see Elbaz et al. 2010, 2011 and references therein). 
%The properties and evolutionary history of IR-luminous star-forming galaxies 
%can be studied in a great detail with \herschel data.
 
%thus need to be revised 
%tested against a complete sample of high-$z$ dusty starbursts 
%selected with \herschel in order to untangle the dependence of SMG analysis on spectroscopic and selection biases. 
%In addition to the populations of FIR luminous galaxies newly discovered, \herschel  
%supplies up to six photometric data points in the FIR range for $0<z<\sim4$ galaxies detected by 
%The combination of \herschel photometry with those at 24 $\mu$m and (sub)millimeter (if available)  
%enables us to robustly measure characterize their far-infrared SEDs, 
%redshift distribution, star formation rates and dust temperatures (Magnelli et al. 2010; 
%Roseboom et al. 2012; Elbaz et al. 2010, 2011 and references therein; Huang et al. 2014).   
%In addition, studies have confirmed 
It has { also} been shown that \herschel is efficient at detecting galaxies at very high redshift 
when a well-characterized selection is used (e.g., Pope \& Chary 2010). 
For instance, Riechers et al. (2013) pre-selected sources with red SPIRE colors 
(i.e., $S_{500\mu m}/S_{350\mu m}$ vs. $S_{350\mu m}/S_{250\mu m}$), 
yielding a sample of potential $z>4$ sources, including one at $z=6.34$, 
the most distant known ULIRG to date. 
Nevertheless, the requirement of $S_{500\mu m}>S_{350\mu m}>S_{250\mu m}$ means that 
such selection is biased against some $z>4$ dusty star-forming galaxies with 
warm dust temperatures (Daddi et al. 2009b; Capak et al. 2011; Smol{\v c}i{\'c} et 
al. 2015), and limited to the brightest 500\um sources ($\simgt$30 mJy, Dowell et al. 2014). 
%Therefore, due to the lack of a more conservative and efficient way to
Therefore, a more conservative and efficient way to 
select high-$z$ ULIRGs is required to better understand the properties of
this important population of dust obscured galaxies in the early Universe. 

%A efficient selection of high-redshift dust obscured galaxies is still
%lacking, 

%Without supplementary data, \herschel is not sufficient to constrain the physical properties 
%of these high redshift candidates. 

A challenge in studying the properties of high-$z$ \herschel
sources has been the low resolution of the data, %However, the photometric measurements of \herschel suffer from the same confusion 
%issue as in SMGs. 
%Due to large beam sizes, 
especially in the SPIRE passbands (full width at half maximum, FWHM, $\sim$18\arcsec, 25\arcsec 
and 36\arcsec at 250, 350 and 500\ums, respectively). 
Source confusion is a severe issue in the \herschel data, 
which makes it difficult to identify the correct counterparts 
and measure fluxes for individual \herschel sources. 
% and can cause the ambiguity in assigning 
%the correct counterparts at other wavelengths.   
The most common technique to deconvolve the \herschel sources is 
the use of {\it  Spitzer}-MIPS 24$\mu$m data of better resolution ($\sim$5.7\arcsec) as the position priors 
to fit their fluxes (e.g., Magnelli et al. 2009, 2013; Roseboom et al. 2010; Elbaz et al. 2011; 
Leiton et al. 2015). 
%One limitation of this approach is 
This de-blending approach assumes that all \herschel sources are detectable in the deep 24$\mu$m images, 
and thus introduces a bias against 24\ums-faint galaxies which fall below the detection limit. 
Moreover, the extraction of 24\um catalogs usually requires IRAC source positions as priors, 
which can introduce additional identification biases. 
% that are not necessarily associated, which would cause additional identification biases.   
%In addition, 24$\mu$m sources that could not be separated at the resolution of the \herschel bands 
%are removed, keeping only the brightest source at 24$\mu$m for each blended group. 
%This raises the possibility that a high-$z$ galaxy which is 
%Based on the deep MIPS and PACS imaging of the GOODS fields, 
Magdis et al. (2011) have demonstrated that in the Great Observatories Origins Deep Survey (GOODS) 
fields, about 2\% of the PACS sources are missed in the MIPS-24\um catalog, most of which 
being at $z\sim1.3$ and $z\sim0.4$ with strong silicate absorption features. 
%with most of them being at $z\sim1.3$
% and $z\sim0.4$ where the strong redshifted silicate absorption features 
%%at 9.7 and 18  $\mu$m suppress the MIPS-24$\mu$m fluxes, 
%Note that there is a trend for the fraction of SPIRE galaxies detected in the PACS bands 
%to decrease with increasing redshift (e.g., Dannerbauer et al. 2010). 
%The sample of 24$\mu$m dropouts could consist of sources that lie at extreme
%There are potential 24$\mu$m dropout, \herschel sources being at the very high 
%redshift, which are not seen by PACS, but significantly detected in SPIRE imaging (e.g., 
%Daddi et al. 2009a, b). 
% 
%The SPIRE bright and 24\um faint \herschel sources are thought to be at redshifts $z\simgt4$, 
{ There has been no systematic study to address this issue for SPIRE sources, 
though simulations suggest that statistically very few sources would be missed  
when requiring a 24\um prior (Leiton et al. 2015).  }
%some of which are likely at $z$\simgt2. 
% on such a high-redshift population. 
%and thus lack of direct evidence for such a high-redshift population. 

%but there has been a lack of direct evidence for such a high-redshift populationi 
%Currently, there are no systematic studies on this subsample of high-$z$ galaxies. 
  \begin{figure}
%\centering{
%\includegraphics[scale=0.38, angle=-90]{/Users/xshu/paper/goodsn/submit/1028/figure/sedvsredshift/sedvsz/sedvsz_rl.ps}
%\includegraphics[scale=0.5, angle=0]{/Users/mac/paper/goodsn/submit/revise/0921/simulate/sed/ms12/sex/pdplot_500to24.ps}
%\includegraphics[scale=0.38, angle=-90]{/Users/mac/paper/goodsn/submit/revise/0921/figure/sedvsredshift/sedvsz/sedvsz_rl.ps}
% \vspace{0.5cm}
\includegraphics[scale=0.38, angle=-90]{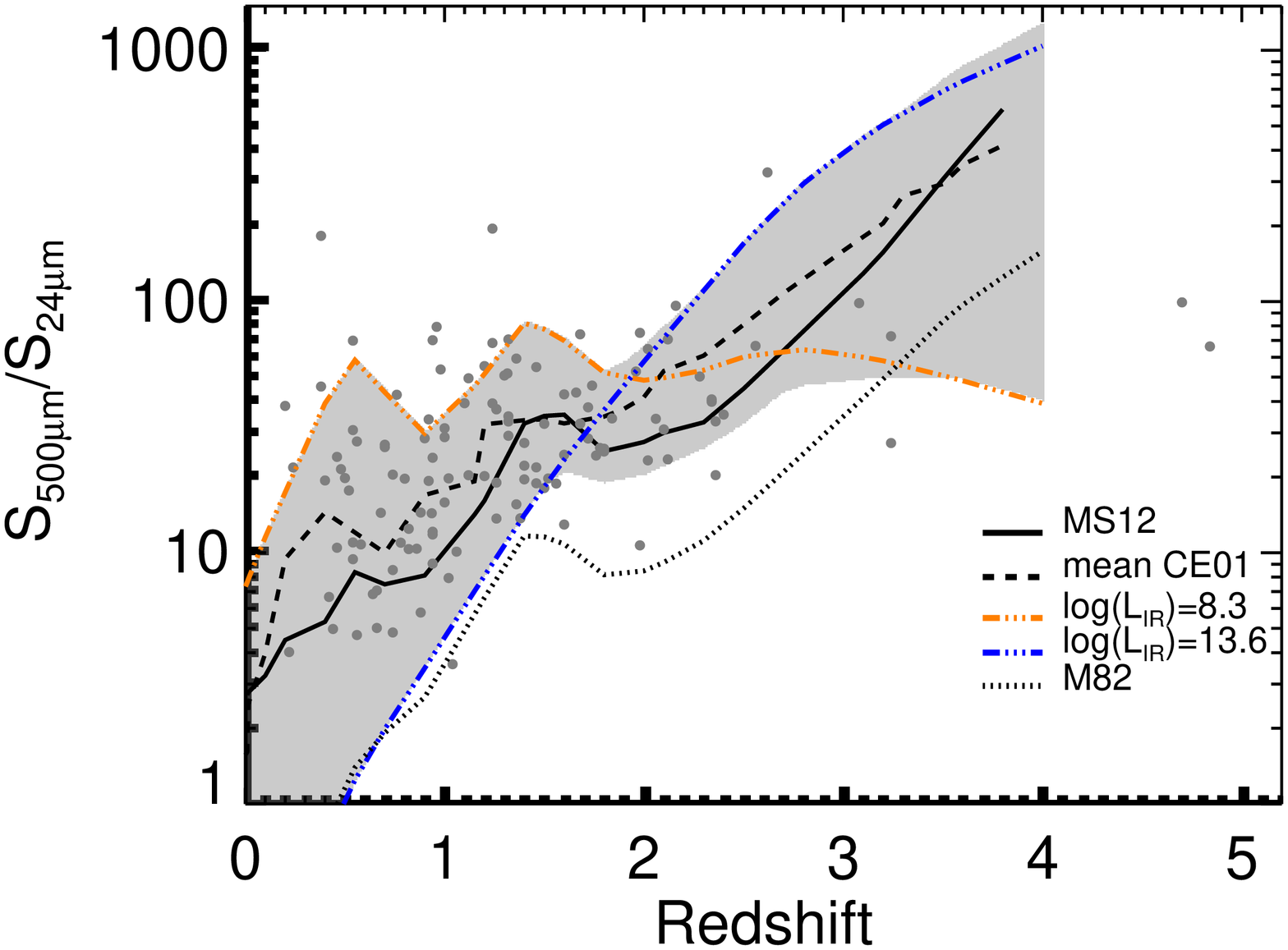}
\caption{
%$Left:$ Image cutouts centered on a simulated source at $z=2$, 3 and 4 (from top to
%bottom). From left to right, 24\um, 500\um and the ratio map.
%The \ratio~flux ratio as a function of redshift. 
%Solid black curve are model SED of main-sequence galaxies from Magdis et al. (2012) 
%and dashed line is the mean trend of Chary \& Elbaz (2001) templates. 
%The grey curve shows the $S_{350\mu\rm m}/S_{24\mu\rm m}$ ratio as a function of redshift 
%drawn from the model SED of main-sequence galaxies for comparison.   
%is Apr220 (dashed), 
%main-sequence galaxy from Magdis et al. 2012 (thick solid), M82 (dot-dashed), Mrk231 (dotted) 
%and mean trend of CE01 templates (thin solid). 
%The shaded region indicates the possible range of \ratio~ratio for galaxies at
%$z\simgt2$, which are poorly studied. 
%The grey dots are the measurements for ``clean" sources 
%from the GOODS-\herschel catalog used in Elbaz et al. (2011). 
%It appears that the observed color distribution follows the trend of model SED of 
%main-sequence galaxies, though the scatter is large. 
%The shaded region indicates the range of \ratio~ratios for galaxies at
%$z\simgt2$.  
%Joint 99\% confidence contours of the \fekalfa emission line intensity vs. velocity width FWHM, obtained
%from Gaussian fits to the line observed with the \chandra HEG, as described in the text.
%The vertical dotted lines correspond to the FWHM of H$\beta$ line (Wang et al. 2007).}
 \ratio~flux ratio as a function of redshift for the GOODS-Herschel (Elbaz et al. 2011) galaxies with ``clean" 500\um 
measurements (grey dots).
Solid black line: main-sequence SED from Magdis et al. (2012). 
Dashed black line: mean trend of the Chary \& Elbaz (2001) SEDs {  that 
are parameterized by the total IR luminosity}. 
%\DIFdelbegin \DIFdel{Dotted black line: $S_{350\mu\rm m}/S_{24\mu\rm m}$ ratio as a function of redshift 
%for the main-sequence SED of Magdis et al. (2012) for comparison. }\DIFdelend \DIFaddbegin \DIFadd 
%[I am not sure of what you used to plot the thin grey line] 
{  Shaded region: range of \ratio~ratios for CE01 templates, with 
orange dot-dashed line representing the SED of the minimum IR luminosity (log($L_{\rm IR}$/\lsun)=8.3) 
and blue dot-dashed line for the maximum IR luminosity (log($L_{\rm IR}$/\lsun)=13.6). 
Black dotted line shows the SED from the local starburst galaxy M82 that is out of the range of 
CE01 libraries at $z\simlt3.2$.}  
%$z\simgt$2 galaxies, as suggested by the  SED evolution trends.}\DIFaddend
}
\label{fig:sedvsz_rl}
\vspace{0.1cm}
\end{figure}

 \begin{figure*}
 \centering
 \includegraphics[scale=0.5, angle=0]{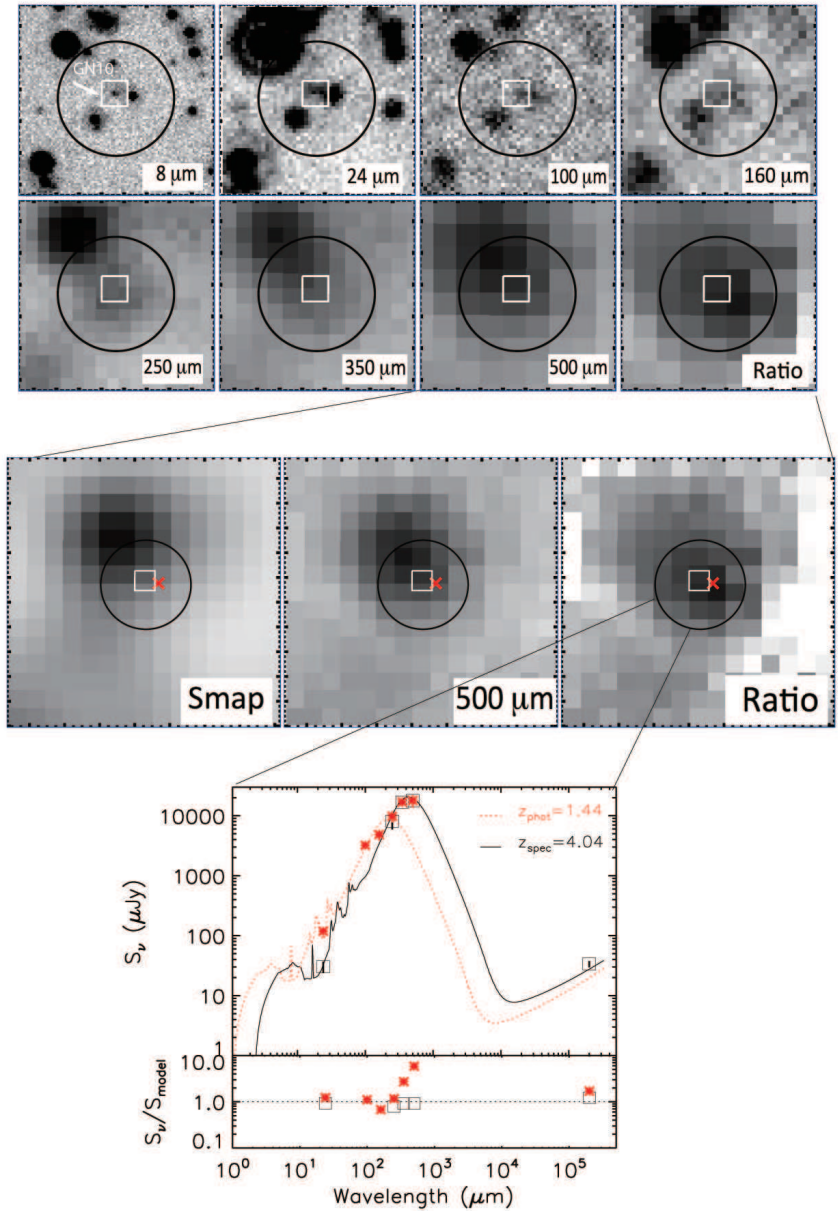}
 \caption{
{\it Upper panel:} Multiwavelength image cutouts of a high-$z$ source, GN10 (Daddi et al. 2009a), in our sample.
 Each cutout is 30\arcsec on a side. The white square 
%with a size of 9\arcsec\ 
denotes the position of the 8\um (24\um) source and
 the black circle indicates the beam FWHM (36\arcsec) at 500\ums.
 %The 500\um flux distribution is shown with the red dashed contour, while the red solid
 %represents the flux contour in the \ratio ratio map.
 %It is evident that the significant 500\um emission of GN10 has been
 %unambiguously uncovered in the ratio map. 
{In the ratio map it is evident that there is an excess emission centred on GN10, which is 
detected with greater significance in the ratio map than in the 500\um map. 
{\it Middle panel:} zoomed-out
view of the 500\um and ratio image, with a size of 1.8x1.8 arcmin$^2$. 
On the left, we also show the ``beam-smoothed" 24\um image (Section 2.2) for comparison.  
Red cross marks the position for a $z_{\rm phot}=1.44$ galaxy close to GN10 
($\sim$5 arcsec separation), for which the SPIRE 350\um and 500\um 
flux is associated in the official \herschel catalog (Elbaz et al. 2011). 
% overlaid with the contours of sources detected in the 8\um. 
{\it Lower panel:} 
24\um and the \herschel photometry (red stars) for the $z_{\rm phot}=1.44$ galaxy. 
Red dotted curve shows the best-fitting SED at $z_{\rm phot}=1.4$. 
Black curve is the best-fitting SED to the far-IR photometry at the redshift of GN10 ($z_{\rm spec}=4.04$).    
The SED fittings suggest that GN10 (instead of the nearby source) dominates the far-IR emission at 350\um and 500\ums. 
Details of the counterpart identifications to the \herschel 500\um sources detected in the ratio map will be 
presented in Section 2.3.  
%Note that the 500\um flux of the source in the \herschel catalog is wrongly associated with a nearby low-$z$ galaxy 
%on the lower right ($\sim$5 arcsec separation). The lower panel is a zoomed-in 
%view of the 500\um and ratio image, overlaid with the contours of sources detected in the 8\um.  
}\DIFaddend
 %Joint 99\% confidence contours of the \fekalfa emission line intensity vs. velocity width FWHM, obtained
 %from Gaussian fits to the line observed with the \chandra HEG, as described in the text.
 %The vertical dotted lines correspond to the FWHM of H$\beta$ line (Wang et al. 2007).}
 }
 \label{fig:contour}
 \end{figure*}

Further problems arise when priors are too close for the \herschel\ flux to be deconvolved: 
when two 24\um priors are closer than the FWHM/3 of any given \herschel band, 
the flux is usually attributed to one or the other.
Previous studies have systematically favored the brightest of the two 24\um counterparts (e.g., 
Magnelli et al. 2009, 2013). 
While such a choice is generally justified for sources at low redshift, 
it can mis-identify the counterpart of a very distant source which is faint at 24\um but much brighter at \herschel wavelengths, especially in the SPIRE 500\um band.
%In addition, the nominal deconvolution method removes 24$\mu$m sources that could not be
%separated at the resolution of the \herschel bands (within 1.5 pixels), keeping only the brightest source
%at 24$\mu$m as counterpart for each blended group. This raise the possibility, especially 
%at the {\sc SPIRE} 500\um where the resolution is worst, that a high-$z$ \herschel source 
%would be misidentified and thus simply missed because of its relatively faint 24$\mu$m flux.
% if it is fainter at the 24$\mu$m.
%Leiton et al. have shown that 
%Assuming the main-sequence SED given by Magdis et al. (2010), 
%a  
For the same \herschel 500\um flux, a $z=3.5$ galaxy can be a factor of $\sim$40 fainter 
in the 24\um band than a galaxy at $z=0.8$ (e.g. Leiton et al. 2015).     
Indeed, the recent study of Yan et al. (2014) has shown that 
the brightest 24\um source does not necessarily contribute 
predominately to the far-IR emission seen in the {SPIRE} bands.   
It is therefore important to investigate in detail the mis-associated and/or missed 
population of luminous, high-$z$ \herschel sources which 
%It has also been suggested that extremely dusty galaxies at $z\simgt3$ 
could play a significant role in the star formation history in the early Universe (Casey et al. 2012b, 2014; 
Barger et al. 2012; Dowell et al. 2014). %, but an efficient selection method is still lacking.  
%
%This raises the possibility that high-$z$ galaxies in blended groups might be missed with high \herschel-to-MIPS flux ratio will be 
%missed 
%In addition, it is expected that more 24$\mu$m sources are  
%In addition,  
%(e.g., Daddi et al. 2009) 
%but no systematic studies on this subsample of SMGs have been conducted so far.

% \begin{figure*}[]
% \centering
% \includegraphics[scale=0.55, angle=0]{fig2b_imgshow.eps}
% \caption{
%{  Examples of ``clean" sources detected in the ratio map. 
%Image cutouts from left to right: the 24\ums, the ``beam-smoothed" 
%24\um (Section 2.2), the 500\ums, the ratio and the radio 1.4 GHz. 
%The upper panel shows the two sources in the GOODS-North field, 
%while 
%The lower panel shows the two sources in the COSMOS field where 
%the SCUBA2-450\um image is available which is displayed in the last column.
%} 
%}\DIFaddend
 %Joint 99\% confidence contours of the \fekalfa emission line intensity vs. velocity width FWHM, obtained
 %from Gaussian fits to the line observed with the \chandra HEG, as described in the text.
 %The vertical dotted lines correspond to the FWHM of H$\beta$ line (Wang et al. 2007).}
% \label{fig:contour}
% \end{figure*}

%implying that source extraction based on 24 $\mu$m positions 
%Motivated by this limitation, w
In this paper, we have undertaken a systematic search for 
high-$z$ \herschel 500\um sources in the GOODS-{\it North} field, which has 
the deepest SPIRE observations currently available, using a map-based 
``color deconfusion" technique. 
We show that our method, by constructing a \ratio~ratio map, can effectively
extract high-$z$ sources contributing to the 500\um emission.
%, which is
%conservative but sufficient if we are mainly interested in studying the high-$z$ 
%ULIRG population. 
% as well as the counterpart identifications.  
%\herschel 
%sources, 
%Our method, by constructing a \ratio~map, which effectively reduces
%the bulk of the low-redshift population contributing to the 500\um
%emission, 
%allows for a conservative and 
%raising the signal from 
%candidate distant sources seen in the ratio~map considerably. 
%The reason we focused on the 500\um sources is because 
%compared to shorter wavelengths, 500\um is able to probe the peak of FIR emission 
%for most distant galaxies (i.e., $z\simgt3.5$), which are expected to 
%have much higher \ratio~ratio than the low-$z$ ones (Figure 1).
% are expected to have compared to 
%shorter wavelengths, 500\um is more efficient to select distant galaxies 
%against low-$z$ ones. 
%similar \ratio~ and $S_{350\um}/S_{24\um}$ ratio, 
The resulting sample consists of 36 galaxies identified at $z\sim2-4.6$. 
%dust obscured galaxies as observed with the deepest 
%far-IR imaging with \herschel in the GOODS-North field using a map-based 
%color deconfusion technique.
%Totally 34 \herschel/{\sc Spire} sources at $z\simgt2$ are identified with this technique, 
%which is detailed in Section 2. 
In Section 2 we present the data and our method; in Sections 3 \&4 we present and discuss the derived properties of 
the galaxies, and we summarize the results in Section 5. 
Throughout this paper, we adopt $H_{\rm 0}=70$ km $s^{-1}$ Mpc$^{-1}$, $\Omega_{\rm M}=0.3$, 
$\Omega_{\Lambda}=0.7$, and we use a Salpeter initial-mass function (IMF).  

% In addi- tion, the Herschel bands effectively fill the gap between MIPS and (sub-)mm surveys (with SCUBA-2, LABOCA, or AzTEC). By combining all these datasets, we can now use the several IR photometric data points to estimate a photometric redshift based on dust emission alone, and compare it with the esti- mations based on UV-to-NIR data. 
 %combination of Herschel observations with the MIPS 
%data (and also IRAC) to diminish the effects of source confusion in IR surveys. 
%These data can be used to constrain the fits to the dust emission templates. 
%Even more novel and rele- vant is the combination of Herschel observations with the MIPS

%However, the photometric measurements of \herschel suffer from exactly the same issue as in SMGs, 
%which can still be severely limited by source confusion above the {\it Herschel}~ 250\,$\mu$m, 
%causing ambiguity in assigning the correct counterparts. 
%An individual detection with very high FIR flux densities may result from 
%the blending of multiple objects, each being less luminous.
%While high-resolution interferometric imaging at the wavelength of the original detection 

\section{Data and method}
\subsection{Data}
The data used in this paper come from the \herschel observations of GOODS-North field as 
a part of the GOODS-\herschel program. 
The observations cover a total area of 10\arcmin$\times$17\arcmin~down to depths of 
1.1, 2.7, 5.7, 7.2 and 9 mJy (3$\sigma$) at 100, 160, 250, 350 and 500$\mu$m, respectively. 
The SPIRE images of GOODS-North are the deepest observations undertaken by \herschel to date, 
while the PACS data are slightly shallower than 
the central region of GOODS-South obtained in GOODS-Herschel that was combined with 
the PEP survey in Magnelli et al. (2013). 
A detailed description of the observations is given in Elbaz et al. (2011). 
We take advantage of these deep \herschel observations, along with the deep 
Spitzer/MIPS 24$\mu$m observation of the GOODS-North field (PI: M. Dickinson), 
to identify the best candidates of distant dusty star-forming galaxies seen by {\it Herschel}. 
Other ancillary data are used including $Spitzer$/IRAC at 3.6, 4.5, 5.8 
and 8\ums, VLA at 1.4 GHz (5$\sigma\sim20\mu$Jy, Morrison et al. 2010),
SCUBA at 850$\mu$m ($3.5\sigma$ catalog, Pope et al. 2006; Wall et al. 
2008), SCUBA-2 at 850\um (4$\sigma$ catalog, Barger et al. 2014), and 
AzTEC at 1.1mm (3.8$\sigma$ catalog, Perera et al. 2008; Chapin et
al. 2009). 
%We also utilized ultradeep (11.5 $\mu$Jy at
%$5\sigma$) 1.4 GHz data presented in 
%Barger et al. (2014) to identify counterparts to the \herschel sources
%(Section 2.1). 

Photometric redshifts used in this paper are drawn from a $K_S$-band selected multi-wavelength catalog in the 
GOODS-North field, spanning 20 photometric bands from $GALEX$ NUV to IRAC 8\ums. 
We refer the reader to Pannella et al. (2015) for a more detailed description of 
the multi-wavelength data, catalog production and photometric redshift estimation. 
This catalog contains 14828 galaxies, $\sim81$\% of which are brighter than the 5$\sigma$ limiting magnitude of 
$K_s=24.5$, and 3775 galaxies having spectroscopic redshifts.   
By comparing to the spectroscopic subsample, Pannella et al. found that 
the relative accuracy [$\Delta z=(z_{\rm phot}-z_{\rm spec})/(1+z_{\rm spec})$] 
of photometric redshifts reaches $\sim$3\%, with less than 3\% catastrophic outliers (e.g., objects with 
$\Delta z>0.2$). 
{  Based on the multi-wavelength photometric catalog, stellar masses were determined 
with $FAST$ for each of $K_{\rm s}$-detected sources (see Pannella et al. 2015 for more details).} 
%were determined with FAST (Kriek et al.
%2009) on the U to 4.5 \um PSF-matched aperture photometry
%from Pannella et al. (2015), using Bruzual \& Charlot (2003) delayed
%exponentially declining star-formation histories}
%Pannella et al. found that 

%A comparison between spectroscopic and photometric redshifts suggests 

% are was built by using the $K_s$-band image as the detection image, 
%and run Sextractor (Bertin \& Arnouts 1996) 

\subsection{Construction of the \ratio~ratio map}
In this section, we present our method to identify candidate high-$z$ dusty 
galaxies. 
% based on the correlation of 500\ums/24\um colour with redshift. 
The observed far-IR/24\um colours of individual galaxies are known to be correlated 
with redshift (Leiton et al. 2015).  
The trend is strongest for the 500\ums/24\um colour. 
In Figure 1, we show the evolution of the 500\ums/24\um ratio as a function of redshift 
for the main-sequence SED from Magdis et al. (2012) (solid black line) and the mean trend 
of Chary \& Elbaz (2001; CE01) templates (dashed black line).
%The evolution of the 350\ums/24\um ratio for main sequence SED is also plotted for comparison (solid grey line). 
%, which are widely used to characterize 
%the FIR properties of star-forming galaxies. 
While the SEDs for galaxies at $z>4$ are poorly constrained, 
distant galaxies are characterized by large 500\ums/24\um ratios 
(e.g., \ratio$>30$ at $z>2$). 
%The are globally in agreement with the observed 
These SEDs are globally in agreement with the observed 500\ums/24\um colour 
distribution of the ``clean" sources in the GOODS-North field (grey dots)\footnote{Clean galaxies are defined as sources with $\leq1$ neighbour of 
$S_{\rm neighbour}>0.5S_{\rm central}$ within a distance of 20\arcsec (e.g., Elbaz et al. 2011), for which 
blending effects are small.}. 
{  We note that there exists a population of low-luminosity galaxies below $z\sim2$ with 
relatively high \ratio~ratios, causing the large scatter of the distribution. 
They might be cold galaxies with abnormal low dust temperature of 10-20 K ($\sim$25 per degree$^2$, e.g., Rowan-Robinson et al. 2010). 
However, given the very small area covered by GOODS-North, we estimated that the number of such galaxies is low ($\sim2$) in the field. 
Some silicate-break galaxies at $z\sim0.5$ and $z\sim1.5$ may contribute the elevated \ratio~ratios 
due to the shift of the silicate absorption features into the MIPS/24\um band at these redshifts, 
but their fraction is relatively small (less than 10\%, Magdis et al. 2011). 
Alternatively, the large scatter could be partly attributed to the catastrophic photometric redshift outliers, or 
chance associations of a high-$z$ 500\um galaxy with a lower redshift 24\um prior. 
These require further investigations.
} 
 
%The shaded region 
%It can be seen that the 500\ums/24\um ratio for bot SEDs at $z>2$,   
%The color evolution as a function of redshift is strongest for the SPIRE/500\um. 
%, /24\um colour, especially for distant 
%galaxies (i.e., $z\simgt3.5$) where 500\um probes the peak of FIR emission.  
% show well-defined trends as 
%a function of redshift (Leiton et al. 2014). 
%The reason  on the 500\um sources is because 
%compared to shorter wavelengths, 500\um is able to probe the peak of FIR emission 
%for most distant galaxies (i.e., $z\simgt3.5$), which are expected to 
%have much higher \ratio~ratio than the low-$z$ ones (Figure 1).
%For galaxies with a main sequence-like SED, $S_{500\mu m}/S_{24\mu m}$ varies by a factor 
%of $>200$ between $z=0$ and $z=4$ whereas it only changes by a factor of 10 when 
%the dust temperature rises from $\sim$30K to $50$K at $z\sim0$. 
%Therefore, w
Therefore, with the assumption that redshift dominates over temperature evolution in
producing the observed shift of the IR SEDs, we are able to select distant galaxies 
against the low-redshift ones based on 
%the 500\um/24\um colour.  
their large 500\,$\mu$m to 24\,$\mu$m flux ratios. 
%In this paper, we are mainly focusing on \herschel/500\um sources, 
%The reason for mainly focusing on the \ratio~color is because 
%compared to shorter wavelengths, 500\um probes the peak of far-IR emission 
%for higher redshift galaxies. 
We choose this color over the shorter SPIRE wavelengths because the 500\um band probes the peak of far-IR emission up to the highest redshifts.
%idensity ratios, as shown in Figure~\ref{fig:sedvsz_rl} for the trend with
%various SED templates. 
%, we selected objects which
%which have large 500\,$\mu$m to 24\,$\mu$m flux ratios against the low-redshift ones (see Figure 2).
%As can be seen in Figure 1, where the $S_{500\um m}/S_{24\um m}$ ratio is plotted as a function of redshift 
%for various templates (see the caption for details), $S_{500\um m}/S_{24\um m}$ varies by a factor  
%Under the hypothesis that redshift dominates over temperature evolution in
%producing the observed shift of the IR SEDs of distant galaxies, we selected objects which
%have large 500\,$\mu$m to 24\,$\mu$m flux ratios (see Figure 2).
%~\ref{fig:firsed}).
%
%While the \herschel/500\,$\mu$m probes peak of FIR emission 
%for most distant galaxies (i.e., $z\simgt3.5$), and thus crucial 
%for our study on band provides a unique probe of 
%the dust emission peak for most distant galaxies 
%and hence a more accurate determination of IR luminosity, 
%the source blending, typically $\sim$5 sources in the 500\um beam, down to 20$\mu$ Jy below each 500\um bean 
However, due to the source blending issues, it is difficult to unambiguously assign a 
measured 500\,$\mu$m flux density to a specific object seen in higher resolution images. 
%Based on the projected source density of 24\um priors in the GOODS-North field, 
%In the GOODS-North field, From the projected source density,
A single 500\um source is usually made of multiple 24\um counterparts that are not necessarily associated.  
%typically 5 sources down to 20$\mu$Jy are within each 500\um beam in the GOODS-North field, 
%though galaxies below $z=2$ are  
%While galaxies below $z=2$ are believed to be fainter in the 500\,$\mu$m, s
Simulations have shown that 500\um flux densities can be systematically overestimated 
due to source blending (Leiton et al. 2015). 
%key regime of dust emission in distant galaxies 
%that remains close enough to the peak emission to optimize the determination of L$_{\rm IR}$.
%while shorter bands are too close to the peak or on the Wien side for distant galaxies. 
%The PACS data (6.7" FWHM at 100\,$\mu$m) and $Spitzer$ (5.7" FWHM at 24\,$\mu$m ) provide 
%the presently best information to identify individual sources responsible for significant 
%contributions to each one of the 500\,$\mu$m beams. 
%The sources were selected by showing 
%large 500/24\,$\mu$m ratios based on the $(S_{500}-S_{24})/S_{500}$ colour map on 
%which a blind source detection was produced. 
%The sources were selected 
%by showinglarge $S_{500\,\mu m}$/$S_{24\,\mu m}$ ratios 
%($(S_{500\mu m}-S_{24\mu m})/S_{500\mu m}>0.5$) 
%based on the $(S_{500\mu m}-S_{24\mu m})/S_{500\mu m}$ colour map 
%(hereafter ``ratio map") on which a blind source detection was produced 
%\footnote{A similar map-based method to search for DSFGs at $z>4$ in the HerMES field is presented in Dowell et al. 2013.}. 
Therefore, we do not simply search for targets using pre-existing 500\um photometric catalogs where flux measurements might be highly uncertain,  
%rely on the positions of prior 24\,$\mu$m sources, 
but use a map-based technique keeping the information of the flux distribution in 
the {\it observed} images.
% thus providing a uniform search for candidates of distant starbursts. 
% without any assumption about redshift and prior positions 
%for each 500$\mu$m source, thus provides a complete census of distant star-forming galaxies which 
%are typically faint in the 24$\mu$m.
%Therefore, our method provides a uniform search for distant starbursts which can be identified 
%by their higher $S_{500\,\mu m}$/$S_{24\,\mu m}$ ratios by pixels.
%In particular, our method can identify a missing population of distant galaxies that are faint in the  
%24\,$\mu$m, or so close to nearby brighter sources at 24$\mu$m that 
%their flux densities in longer {\sc Spire} 
%wavelengths are significantly mis-measured because of blending. 

%\subsection{Object identification in Ratio Map}
%Spitzer MIPS 24$\mu$m positions reach fainter fluxes than the Herschel bands (see Figure 4 in 
%Elbaz et al. 2011). 
Our search approach consists of building a 500$\mu$m/24$\mu$m ratio map, which   
%Our map-based search method 
relies on the deep $Spitzer$/MIPS 24$\mu$m data. 
%which reach fainter fluxes than the Herschel bands up to redshift $z\simgt3.5$ 
%which reach fainter in far-IR luminosity than the \herschel bands up
%to redshift $z\sim3$ (see Figure 4 in Elbaz et al. 2011).
% and most \herschel sources \have a 24\um counterpart. 
For the GOODS-{\it North} field of interest here, 
the 24\um data reach fainter far-IR luminosities than the \herschel bands up 
to redshift $z\sim3$ (see Figure 4 in Elbaz et al. 2011). 
To build our \ratio~ratio map, we use the real SPIRE 500\um map that we divide 
by an image containing all 24\um detections down to 21$\mu$Jy (at a level of 3$\sigma$, totally 2704 galaxies), 
which are distributed at their actual positions. We do not use the real $Spitzer$-MIPS 24\um image 
because we found that even after smearing the image to the 500\um beam, negative pixels due to noise 
fluctuations introduced fake sources in the \ratio~ratio map.
%We used all 24$\mu$m sources 
%detected down to 21 $\mu$Jy (at a level of 3$\sigma$, totally 2704 galaxies).
%and most \herschel sources have a 24\um counterpart. 
%Our approach is to make a 500$\mu$m/24$\mu$m ratio map. 
%Firstly, w
%We convolved the flux of each 24\um source using the 500\um point spread function 
%(PSF), and distributed them spatially using the coordinates of the 24\um sources 
%under identical astrometric projection as 
%with an astrometric projection identical to the 500\um map,
In brief, we added point sources using the 500\um point spread function (PSF) at the positions 
24\um sources under identical astrometric projection as the 500\um map, scaling each PSF to the 24\um flux, 
to build an image that matches the actual 500\um map in beam size and pixel scale. 
{  The astrometric accuracy of the 500\um map is verified by stacking at the positions of 241 
24\um sources with $S_{250\mu m}$ in the range 10-35 mJy in the field. The stacked map has an offset from the 
center of the 24\um counterparts of $\sim0.8$ arcsec, which is small compared to the 7.2 arcsec pixel size of the 500\um map}. 
In order to take into account the astrometry uncertainties of the \herschel data, 
a random astrometric error of $\sim$0.5\arcsec~was introduced to the source positions. 
Note that in this beam-smeared image, 24\um flux density for each source is kept, 
and the background is dominated by the smeared faint sources populating the whole 
image. 
%Since the instrumental noise to this image, 
%and all pixels have positive values. 
%Since we haven't introduced any  
%Note that we do not simply PSF-smooth the actual 24\um map to the resolution of 500\um, 
%because in this situation the noise fluctuation  
%the peak flux for each source is 
%In order to take into account the uncertainties in the astrometry solutions, we applied 
%a random astrometric error of $\sim$0.5\arcsec to the source positions from the 24\um. 
% beam-matched map with each source at the central pixel 
%having the 24\um flux.  
%;to match the beam size of 500$\mu$m, but keeping;

We then construct a 500$\mu$m/24$\mu$m ratio map by dividing the actual  
500$\mu$m image by the beam-smeared 24$\mu$m map. 
Since low-$z$ galaxies are expected to have much lower \ratio~ratios 
compared to high-$z$ ones (Figure 1), this procedure thus effectively reduces
the contribution to the map from the bulk of the low-redshift galaxy population, 
while considerably boosting the signal from candidate distant sources.
%Note that we do not simply convolve the actual 24\um map to the resolution of 500\um to 
%construct the \ratio~map, because our tests show that the negative 24\um pixels due 
%to noise fluctuations can introduce too many fake sources in the final \ratio~map 
%and hence the approach is not ideal.  
%which we called as "spectral deconfusion". 
%We then search for sources in the ratio map using Sextractor (??). 

%In Figure~\ref{fig:fakeimg}, we show an example of the ratio map cutout centered on 
In Figure~\ref{fig:contour}, we show an example of a spectroscopically confirmed $z=4.05$ SMG 
(GN10, Daddi et al. 2009a) identified in the \ratio~ratio map (see Section 2.3.1), 
%which turns out to be a spectroscopically confirmed $z=4.05$ SMG (GN10, Daddi et al. 2009a). 
% based on the \ratio~map, 
whose 500\um flux was initially wrongly associated to a nearby source 
 at $z_{\rm phot}=1.44$ due to the strong source blending. 
%As can be seen from the higher-resolution images at shorter wavelengths, 
%a simulated point source at $z=2$, 2.5, and 3 (from top to bottom respectively), 
%along with the observed 24\um and 500\um images.
%high-$z$ source into the 24\um and 500\um, the ratio map cutout constructed using 
%the procedure mentioned above. 
{ It can be seen from Figure~\ref{fig:contour} that 
%while the source appears 
%relatively faint in the actual 500\um map compared to the heavily blended 500\um 
%peak, it is more unambiguously seen in the ratio map.
% due to the increased contrast to nearby sources. 
%Based on the ``effective" noise estimation (defined in Section 2.1), the source is detected 
%at $2.8$, 3.1 and 5.3 $\sigma$ in the ratio map for redshift at $z=2$, 2.5 and 3, 
%indicating an increase of the contrast towards higher redshift, as illustrated in the right panel in Figure~\ref{fig:fakeimg}.  
%using the spectral confusion technique as described above, 
%along with its multi-band images from the MIR to FIR. 
GN10 is faint at 24$\mu$m, and its SPIRE  
emission particularly at 500$\mu$m is heavily blended with nearby 
sources, making it difficult to identify directly. 
In contrast, the galaxy is unambiguously uncovered in the \ratio~ratio map 
at a position coincident with its 24\um counterpart. 
Note that the measured SPIRE 350\um and 500\um flux in the public catalog 
(Elbaz et al. 2011) is associated with the nearby $z_{\rm phot}=1.44$ galaxy. 
However, as shown in Figure 2 (lower panel), SED fittings {  using the whole library of 
CE01 SED templates} to the 24\um to far-IR photometry 
suggest that this source contributes to $<20$\% of the measured 500\um flux. 
{  The use of the starburst SED derived by Magdis et al. (2012) didn't improve 
the fittings and yielded similar results, strongly suggesting that the source   
is likely wrong associated as the counterpart. }
Conversely, the best-fitting SED model at the redshift of GN10 yields a 
much better result ($\Delta \chi^2=$31.9), indicating it {  should 
dominate} the measured 500\um emission.   
%The nearby $z_{\rm phot}=1.44$ galaxy (red cross in the middle panel) 
%Our careful flux decomposition (Appendix B) shows that the $z\sim4$ source has 
%a significant contribution to the 500\um emission.} 
} 
%reveals a significant source 
% which clearly deviates 
%from the center of 500\um emission 
%By checking its 24\ym counterparts 24\um~ within 
%,
%The center of the source in the \ratio~ratio map 
%with a much smaller positional correspondence to the 24\um~ counterpart, 
This demonstrates the effectiveness of our method for uncovering and selecting 
high-$z$ 500\um sources. 
The selection efficiency of our method 
will be investigated in detail on simulated data (Section 2.3.1) and a validation of the 
method {  using SCUBA2 data is presented in Section 2.3.3}. 
\begin{figure}
%\centering{
%\includegraphics[scale=0.5, angle=0]{/Users/xshu/paper/goodsn/submit/1028/simulate/sed/ms12/sex/pdplot_500to24.ps}
%\includegraphics[scale=0.5, angle=0]{/Users/mac/paper/goodsn/submit/revise/0921/simulate/sed/ms12/sex/pdplot_500to24.ps}
\includegraphics[scale=0.5, angle=0]{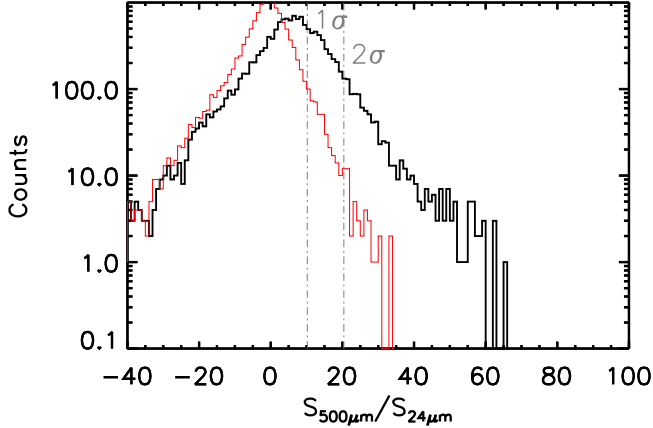}
% \vspace{0.5cm}
\caption{
%$Left:$ Image cutouts centered on a simulated source at $z=2$, 3 and 4 (from top to
%bottom). From left to right, 24\um, 500\um and the ratio map.
The P(D) plot for the $S_{500\mu m}/S_{24\mu m}$ ratio map.
The vertical dot-dashed lines correspond 1$\sigma$ and 2$\sigma$
of the dispersion. 
The red curve represents the $S_{500\mu m}/S_{24\mu m}$ ratio 
after removing all individually detected 500\um sources in the \herschel 
catalog used in Elbaz et al. (2011). 
This illustrates that the false detections caused by map artifacts that 
would pass our selection threshold are negligible (see Section 2.5).   
%Joint 99\% confidence contours of the \fekalfa emission line intensity vs. velocity width FWHM, obtained
%from Gaussian fits to the line observed with the \chandra HEG, as described in the text.
%The vertical dotted lines correspond to the FWHM of H$\beta$ line (Wang et al. 2007).}
}\label{fig:pdplot}
\vspace{0.1cm}
\end{figure}

\subsection{Object identification in the $S_{500\mu m}/S_{24\mu m}$ map}
%Subsequent source extraction was performed in the ratio map using Sextractor 
Given that the $S_{500\mu m}/S_{24\mu m}$ ratio is very sensitive to  
high-$z$ sources, we then perform a systematic search for candidates on 
the ratio map using SExtractor (Bertin \& Arnouts 1996). 
Since the noise distribution is highly non-Gaussian in the final ratio map, 
it is difficult to define a minimum signal-to-noise (S/N) 
requirement for source extraction. 
%We have adopted a minimum S/N of 2, 
We therefore used simulations to determine the minimum S/N threshold in order to 
optimize the ``purity" and completeness of the resulting sample. 
Based on such simulations (see Section 2.3.1), 
we adopted a minimum S/N of 2, corresponding to a minimum \ratio~ratio of 20.4, 
where the noise fluctuation, or defined as ``effective" noise, 
was estimated from the dispersion of pixel distribution in the ratio map, P(D) plot, as 
shown in  Figure~\ref{fig:pdplot} (black curve). 
%As we are searching for a method to identify high-$z$ candidates, not secure ones,
%the 2$\sigma$ threshold is sufficient to detect a $z\simgt3$ source based on simulations 
%(Figure. )
%The details of the simulations to quantify the selection efficiency (and sample incompleteness) 
%are given in Section ??. 
Note that this S/N requirement based on simulations 
cannot be translated to a nominal flux detection limit in a trivial way, 
but it is sufficient for our purpose of identifying candidate high-$z$ galaxies.  
% Figure~\ref{fig:peak} shows the distribution of $S_{500\mu m}/S_{24\mu m}$ values 
%for the simulated sources of the various 500\um fluxes, at $z=2$, 3 and 4, respectively, 
%which are randomly distributed in the ratio map 1000 times.  
%For the adopted threshold of $S_{500\mu m}/S_{24\mu m}$$=$20.4,  or $2\sigma$, we 
%reach detection rates of greater than 50\% for a $S_{500\mu m}=$15 mJy source 
%at $z=2$, and $\sim$85\% for a $z=3$ source with the same 500\um flux. 
A more detailed description on the simulations and the detection efficiency will be given in 
the next section.  

\begin{figure*}[tph!]
\centering
%\includegraphics[scale=0.49, angle=0]{/Users/xshu/paper/goodsn/submit/1028/simulate/sed/ms12/newsed/ce01template/goodsn_efficiency_ce01.ps}
%\includegraphics[scale=0.49, angle=0]{/Users/xshu/paper/goodsn/submit/1028/simulate/sed/ms12/newsed/ce01template/density/z3/goodsn_efficiency_density_z3.ps}
%\includegraphics[scale=0.49, angle=0]{/Users/mac/paper/goodsn/submit/revise/0921/simulate/sed/ms12/newsed/ce01template/goodsn_efficiency_ce01.ps}
%\includegraphics[scale=0.49, angle=0]{fig4_goodsn_efficiency_ce01.ps}
%\includegraphics[scale=0.49,
%angle=0]{/Users/mac/paper/goodsn/submit/revise/0921/simulate/sed/ms12/newsed/ce01template/density/z3/goodsn_efficiency_density_z3.ps}
%\includegraphics[scale=0.49, angle=0]{fig4_goodsn_efficiency_density_z3.ps}
\includegraphics[scale=0.85, angle=0]{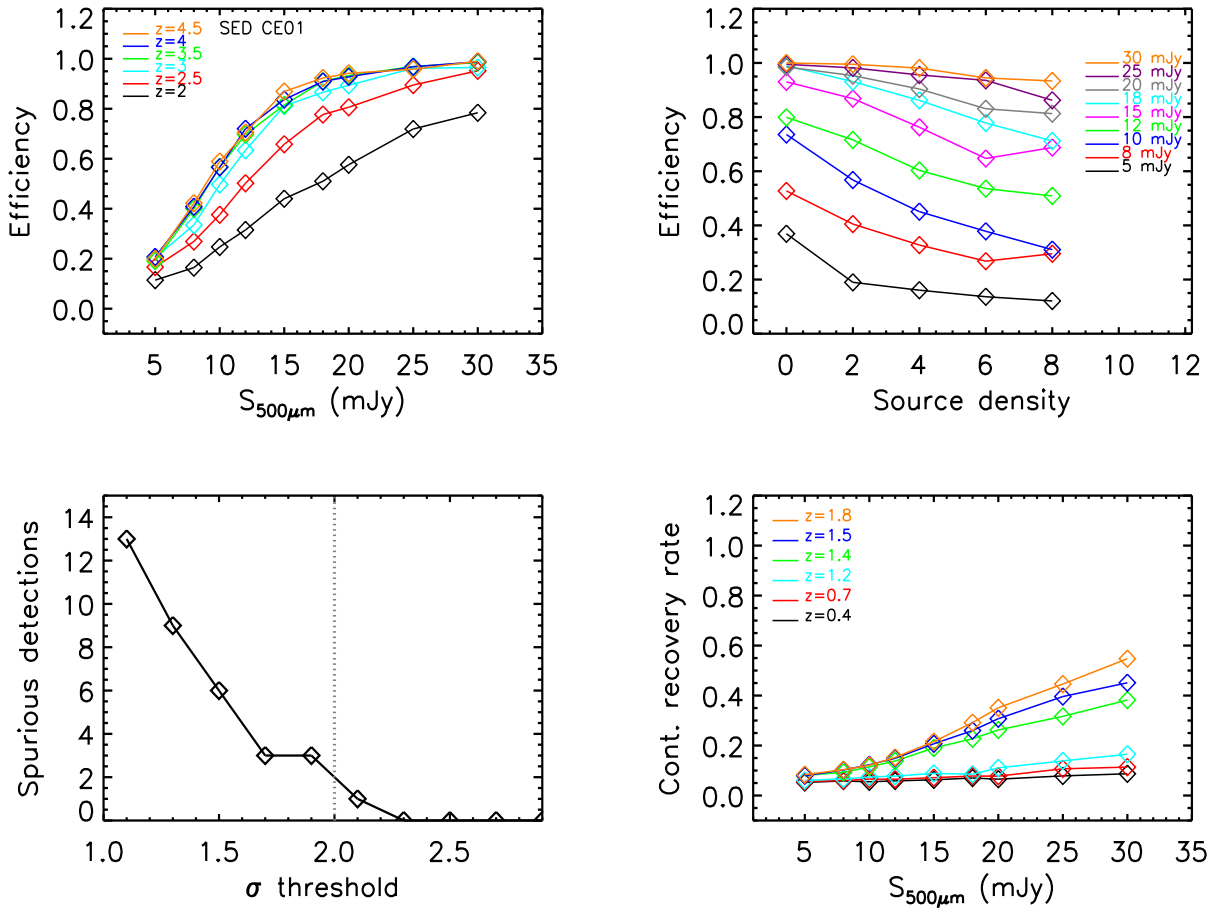}
\caption{
{\it Upper left:} Selection efficiency as a function of 500\um flux density and
redshift, assuming a mean CE01 SED (Chary \& Elbaz 2001). 
Note that using the main-sequence SED from Magdis et al. (2012) in simulations yields similar results. 
%The color ratios between 500\um~ and 24\um~ are strong function of redshifts
%and $S_{500\mu m}$.
For sources with $z>3$, the selection functions appear very similar and
the recovery rate is $\simgt$ 60\% above $10$ mJy. 
{\it Upper right:} Selection efficiency as a function of source density for simulated galaxies at $z=3$. 
The source density is defined as the number of 24\um sources within 20\arcsec radius. 
%Joint 99\% confidence contours of the \fekalfa emission line intensity vs. velocity width FWHM, obtained
%from Gaussian fits to the line observed with the \chandra HEG, as described in the text.
%The vertical dotted lines correspond to the FWHM of H$\beta$ line (Wang et al. 2007).}
%{\it Lower left:} {  The contamination rate of sources at $z<2$ as a function of 500\um flux at 
%different redshift bins}. 
{\it Lower left:} {  Number of expected false detections above a given detection threshold. }
{\it Lower right:} {  The detection rate of contamination sources at $z<2$ as a function of 500\um flux for 
different redshift bins}. 
}
\label{fig:efficiency}
\end{figure*}

%We then impose the minimum flux ratio of 30, resulting in the total 
%detection of 40 sources in a region covered by both the 24$\mu$m and 500$\mu$m. 
%\subsection{Counterpart matching}
%Determining the optical/NIR counterparts to Herschel/SPIRE sources is complex, 
%but thanks to lower positional uncertainty of our sources detected in the ratio maps, 
%it is much more straightforward way than in the original, confusion dominated SPIRE maps. 
%

\subsubsection{Selection efficiency}

%Furthermore, we need to determine how efficient our selection 
%is in finding $z\simgt2$ infrared luminous galaxies and hence the completeness of the 
%sample.
For our sample to be robust and meaningful, we need to determine the efficiency of our selection of $z\simgt2$ infrared-luminous galaxies.
% and hence the completeness of the sample. 
{ To do this we perform Monte-Carlo (MC) simulations by injecting fake sources into 
the actual maps. 
%whose $S_{500\mu m}$/$S_{24\mu m}$ colors match 
%In the simulations, we assumed a mean $S_{500\mu m}$/$S_{24\mu m}$ color evolution which is 
%derived from a suite of CE01 templates (dashed curve in Figure 1).  
% then to determine how many are recovered.} 
The positions of injected sources are randomly distributed but are chosen to avoid overlapping within
known sources detected in the 
ratio map (Section 2.3). 
In the simulations, we assumed a mean $S_{500\mu m}$/$S_{24\mu m}$ color evolution with redshift, 
as shown by the dashed curve in Figure 1. 
The evolutionary trend is derived by averaging the $S_{500\mu m}$/$S_{24\mu m}$ color 
over a suite of CE01 templates for a given redshift bin. 
}%We then determine how many injecting sources can be recovered in the ratio map.} 
%The typical size of known sources is ???, thus the chance of
%the overlapping is low. 
%Note that  the size of known sources????
As we will describe later, we used mainly the CE01 library to fit the mid-to-far-IR photometry and
 derive the integrated far-IR luminosities. 
The CE01 library comprises a total of 105 SEDs with a range of dust temperatures 
and is widely used to characterize the far-IR properties of ULIRGs (e.g. Magnelli et al. 2012; Swinbank et al. 2014).  
%{  Note that the results of simulations are not changed if using  }
% ({  explained why CE01 templates}). 
%Therefore, the SED we assumed in the simulations works closely to, or at least 
%does not deviate too much from the observed ones.  
The simulation samples a $S_{500\mu m}$ range from 5 mJy to 30 mJy in nine steps and 
a redshift range $1.5<z<5$ in six steps.
%  and flux densities at {\sc SPIRE} 
%500\um ($5\rm mJy<S_{500\mu m}<30 \rm mJy$), whose properties 
%({\sc Spire}/{\sc Mips} color ratios) match the measurements of main sequence galaxies 
%in the GOODS-North field (Elbaz et al. 2011). 
We estimated the efficiency of detection at a 
given flux and redshift as the fraction of the injected sources 
{ that were recovered in the ratio map}, 
using the same selection procedure as above\footnote{Sources are considered recovered if they are found within 2 pixels of the injected position 
with a peak S/N above $2\sigma$.}.

We carried out $N_{\rm sim}$=200 independent simulations for each of the flux densities, each of which has $N_{\rm src}$=20 
sources injected. $N_{\rm src}$ was chosen to be small enough to avoid overlapping between the simulated sources. 
% ({too clean, not as real one?}).  
The averaged detection efficiency over the $N_{\rm sim}$ simulations
forms the final detection efficiency, 
which is displayed as a function of redshift and $S_{500\mu m}$ in Figure~\ref{fig:efficiency}. 
This figure shows that, as expected, the recovery efficiency increases  
at high flux densities and redshifts. 
For instance, for a $S_{500\mu m}$=15 mJy (typical $3\sigma$ confusion limit, Nguyen et al. 2010) 
source at redshift $z=2$, the selection efficiency is $\sim$45\%, while it increases to $\simgt80$\% for 
$z>3$ galaxies with the same flux. 
Above $z=3$, the curves of the detection rate as a function of flux 
evolve very similarly, suggesting that our method is effective at selecting 
high redshift sources with a completeness of $\simgt50$\% for a $S_{500\mu m}\simgt$10 mJy source. 
Note that the detection efficiency for each real source depends
on the intrinsic SED, as well as the possible presence of source blending. 

%In the Appendix E, we also tested our method by applying it to the data from COSMOS field where 
%SCUBA-2 450\um data are available.  
%Higher resolution SCUBA-2 450\um 

%We use the estimated efficiencies on the grid of nine flux densities and six redshifts 
%and then interpolate for each source to correct for the incompleteness when computing the SFRD. 

Due to the contamination of bright nearby sources, which is particularly strong  
in the SPIRE 500\um band, the selection could introduce a bias against the areas where the projected 
number density of 24\um sources is higher. 
Our random distribution of sources in the real $500$\um map allows for checking 
the effect of the ``local" confusion noise on source detections. 
We find that the recovery rate decreases with the increase of projected 24\um source 
density ($N_{\rm neib}$)\footnote{For a given galaxy of interest, the source density,
  $N_{\rm neib}$, is
  defined as the number of 24\um sources within a 20\arcsec radius. }. 
As shown in Figure~\ref{fig:efficiency} (right), for a galaxy
%($N_{\rm neib}\simlt2$) 
at $z=3$ with $S_{500\mu m}$=10 mJy, 
the probability that the source with $N_{\rm neib}\simlt2$ is detected in the ratio map is
$\sim$60\%, while it decreases to less than 40\% 
in regions where there are more than six 24\um sources within a radius of 20\arcsec. 
%Figure~\ref{fig:densitymap} shows the spatial distribution of our {\sc Spire} sources, 
%overlaid on the map representing the projected number density of underlying 24\um~ galaxy 
%population (down to 20$\mu$Jy).
Correcting for such bias, the detection efficiency would be higher. 
Visual inspection of source spatial distribution suggests that  
sources appear to be found in areas where the projected 24\um source 
density is relatively low. 
Note that sources which are significantly fainter than their close neighbours 
in the {SPIRE}/500\um would not pass our S/N selection. 
%On the other hand, t
There are (rare) $z\simgt4$ ULIRGs detected at submm and/or mm that are too faint to be 
detected in the SPIRE data, such as the one at $z=5.2$ (HDF850.1, Walter et al. 2012).  
Such sources would not be selected by our method and hence would be missing from our sample.  
{  In addition, our minimum S/N requirement in detecting sources in the ratio map 
discriminates against galaxies with M82-like SED unless they are at $z>3$ (Figure 1). 
}

\subsubsection{Counterpart Identification}
 We describe in this Section the procedure used to attribute shorter-wavelength 
counterparts to the \herschel 500\um sources detected in the \ratio\ ratio map. 
It is important to note that the use of this ratio map constructed from a list of 24\um sources does 
not preclude the identification of 500\um sources with no 24\um counterparts. Such sources 
will naturally emerge as objects with a high ratio. In the following, we show how 
probabilities of association are determined for potential 24\um counterparts. We also define a quantitative threshold 
below which no 24\um counterpart is considered reliable, hence identifying 24\um dropouts. 
%Once a source is detected in the ratio map, we have searched for potential counterparts in the 
%igher resolution 24Œºm image out to a radius of 15 arcsec from the S500Œºm/S24Œºm centroid
Once a source was detected in the ratio map, we 
%For a given 500\um~source in the ratio map, we 
%identify its potential counterpart i
searched for potential counterparts in the higher-resolution 24\um image 
out to a radius of 15 arcsec from the $S_{500\mu m}/S_{24\mu m}$ centroid.  
%by searching for objects in the higher resolution 24\um image within a radius $r=14.4$\arcsec
%~($\sim$0.3$\times$FWHM of 500\ums, or $\sim$1.5 pixels) of the $S_{500\mu m}/S_{24\mu m}$ peak. 
%Note that for 24 \um sources separated by less than $r=10$\arcsec, 
%it has been proposed ({  reference?}) 
%there are large uncertainties in deblending their far-IR fluxes at 500$\mu$m 
%(e.g., Leiton et al. 2014). 
%Note that when sources are separated by less than $\sim$0.3$\times$$FWHM$ of 500\um,
%there are large uncertainties in deblending  
%This search radius 
%was determined empirically through simulations by injecting artificial 
%500\um sources 
%into a noise map with known flux densities and positions}\footnote
%{We generated simulated \herschel 500\um map to be as close as possible to the real image 
%in a statistical sense, e.g., number counts, the photometric and confusion noise. 
%The methodology is described in detail in Schreiber et al. (2015).}, 
%{  and measuring the distribution of positional offsets between a source 
%detected in the ratio map and the input $z$\simgt2 source with 
%highest \ratio~ratio. 
%We found that there is a $\sim$90 per cent probability that the probable 
%counterpart will be within 15 arcsec of the source centroid in the ratio map.
%Note that 
%Using our simulated \herschel 500\um map (Schreiber et al. 2015),
{  The search radius was determined from our simulations (Section 2.3.1), 
where we found that for more than 90 per cent of sources, the extracted positions 
are less than 15 arcsec from the input positions, i.e., 
the probability of missing potential counterparts is less than 10 per cent}\footnote{This search radius also 
represents a 2$\sigma$ positional uncertainty, 
with $\sigma\sim0.6\times$FWHM/(S/N) (Ivison et al. 2007), where FWHM = 36 arcsec, 
for a detection in the 500\um map with S/N=3.}.
Note that at the extreme depth reached by the 24\um imaging reported here, the cumulative 
surface density of 24\um sources yields 1-8 sources per search area with a median of 3 
sources. 
{  For larger radii, one will find more counterparts but the probability of 
chance associations will also increase. }

%it is impossible to have a meaningful deconvolution of sources 
%that are separated by less than $\sim$0.3$\times$FWHM of 500\um (Wang
%et al. 2014, in preparation).

\begin{figure}[th]
\centering
\includegraphics[angle=0,scale=.70550]{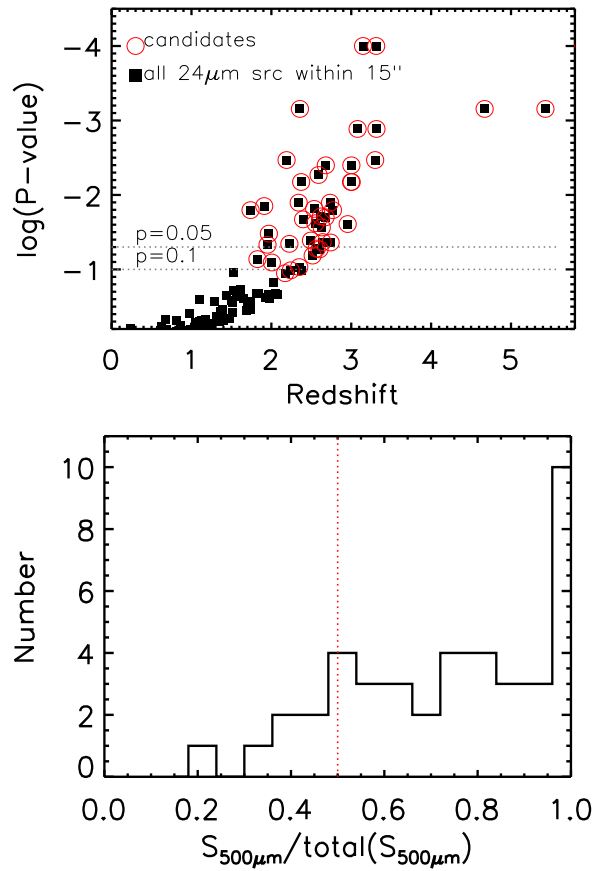}
\caption{
{\it Upper:} 
$P$-values calculated as $P=1-$exp$(-\pi n \theta^2)$, where $n$ is $n(>$$S_{500\mu m}^{\rm Pred}$,$>z$) and $\theta$ 
is the search radius of 15\arcsec, are plotted against redshifts for all 24\um sources within the 
search radius (filled squares). { Candidate counterparts identified with $P\simlt0.1$ are shown 
in red circles}. 
For sources with multiple candidate counterparts (11/42), the one with the least $P$-value is plotted. 
{The $S_{500\mu m}^{\rm Pred}$ is predicted 500\um flux for each source (see Section 2.3.2)}. 
%\DIFdelbegin \DIFdel{based on its 24\um flux and redshift, assuming a mean \ratio~color evolution of CE01 SEDs.}\DIFdelend  
%This is a test on the simulated data from c.s. 
%We found that at a threshold of $p\simlt0.1$, $\sim$90\% sources detected in the ratio map have at 
%least one 24\um counterpart at $z\simgt2$. 
%Our sample seems to peak at higher redshift than those selected at $\sim$850\um, 
%with a significant excess at $z>3$. 
{\it Lower:}
{ The distribution of $ S_{500\mu m}/{\rm total}(S_{500\mu m})$, where
$ S_{500\mu m}$ is the 500\um flux density of the counterpart identified with $P\simlt0.1$, and ${\rm total}(S_{500\mu m}$) 
is the summed 500\um flux of all $input$ 24\um sources within the search radius}. 
{ Here we used the $input$ $ S_{500\mu m}$ of each source in the simulated map.}
%ratio of 500\um flux density of counterpart to the total 500\um flux of all 24\um sources within
%the search radius. %The plot used the same sources as in Figure (red circles) from {\it the simulated data}. 
%For those with multiple 24\um counterparts ($P\simlt0.1$, $\sim$31\%), the one with least $P$-value 
%is plotted. 
Note that 36 out of 42 sources have one 24\um galaxy contributing more than 50\% of the total 
500\um flux (a majority contribution to the 500\um emission), suggesting a high efficiency of the $P$-value 
in identifying the {\it best} counterpart. For the remaining six objects with $S_{500\mu m}$/total($S_{500\mu m}$)$<$0.5, 
five have multiple counterparts ($P$-values between 0.0027 and 0.0864). The summed 500\um flux density of those counterparts to the total($S_{500\mu m}$) is greater than 0.5.  %The 
The red dashed vertical line represents $S_{500\mu m}/{\rm total}(S_{500\mu m})$=0.5. 
%median trend of $z<2$ \herschel/SPIRE sources 
%from Casey et al. 2012(a). 
%Filled squares are $z\simgt2$ sources identified from the ratio map, with 
%spectroscopically confirmed ones highlighted by red circles.  
}
\label{fig:csimulate}
\end{figure}

{  The primary method for identifying the most likely counterparts to each 500\um source is `$P$-statistics' ( 
Downes et al. 1986), by calculating the Poissonian probability of sources in a high-resolution catalog 
(24\um and/or radio) that lie within the search radius, $\theta$. 
Given a potential counterpart with flux density $S$ in the high-resolution imaging, 
the probability of finding at least one object within $\theta$ of at least that flux density is 
$P=1-$exp$(-\pi n_{(>S)}\theta^2)$, where $n_{(>\rm S)}$ is the surface density of sources above flux density $S$. 
$P$ gives the likelihood that a {  500\um source is associated  
with a counterpart in a higher-resolution map by chance}. 
%$P=1-$exp$(-\pi n\theta^2)$. 
%This depends on $n_{\rm (>S)}$, the surface density of sources in the high-resolution catalog above the flux density $S$ of the given candidate, and the search radius, $\theta$. 
The lower the $P$-value, the less likely it is that the candidate is associated with 
a 500\um source by chance, and the higher the probability of a genuine match.  
However, as has been pointed out in the literature (e.g., Yun et al. 2012),} 
the often adopted high-resolution 24\um or radio continuum data for the counterpart identification 
may suffer from a systematic bias against high-$z$ sources. 
%Since the traditional $P$-statistics can be biasing the identifications towards the 
%brightest 24\um sources which could result in more misidentifications with low-redshift 
%galaxies, i
Since the main goal of the ratio map is to favor {  higher-$z$ sources which 
have statistically higher \ratio~ratios}, 
we replace the surface density of $n_{(>\rm S)}$ in the computation of $P$-value 
by $n_{(>\rm z, >S)}$ to take into account the redshift information.
{  For a given counterpart, $n_{(>\rm z, >S)}$ is the 
surface density of objects brighter than $S$ which have redshifts above $z$. 
This means that in a catalog that is cut at flux density $S$ for a given counterpart, we considered only 
those with redshifts above $z$ in calculating the probability of it being a chance 
association. 
}
{
%, this being a function of both the flux and the redshift of counterparts.  
%Given a potential counterpart with flux density $S$ and redshift $z$, we can 
%calculate a Poissonian probability, $P$, of finding at least one object (within the search 
%radius) of at least that flux density and redshift, 
%which is given by $P=1-$exp$(-\pi n\theta^2)$. 

\begin{figure}[t]
\centering
\includegraphics[scale=0.5, angle=0]{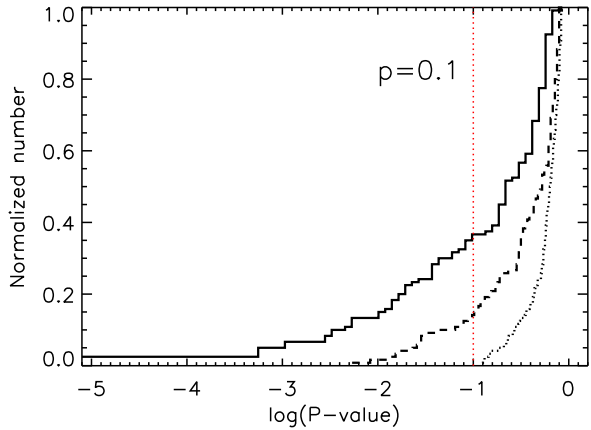}
\caption{
{  Cumulative distribution of $P$-values for all 24\um counterparts
within the search radius. 
The solid lines shows the distribution of the modified $P$-values when
taking into account the redshift information. 
Also shown is the distribution of $P$-values which were calculated
using only the 24\um flux, as often adopted in literature (dotted line), and those 
using only the predicted 500\um flux (dashed line). 
}
}
\label{fig:pchist}
\end{figure}

\begin{figure*}[thb]
\centering
\includegraphics[scale=0.7, angle=0]{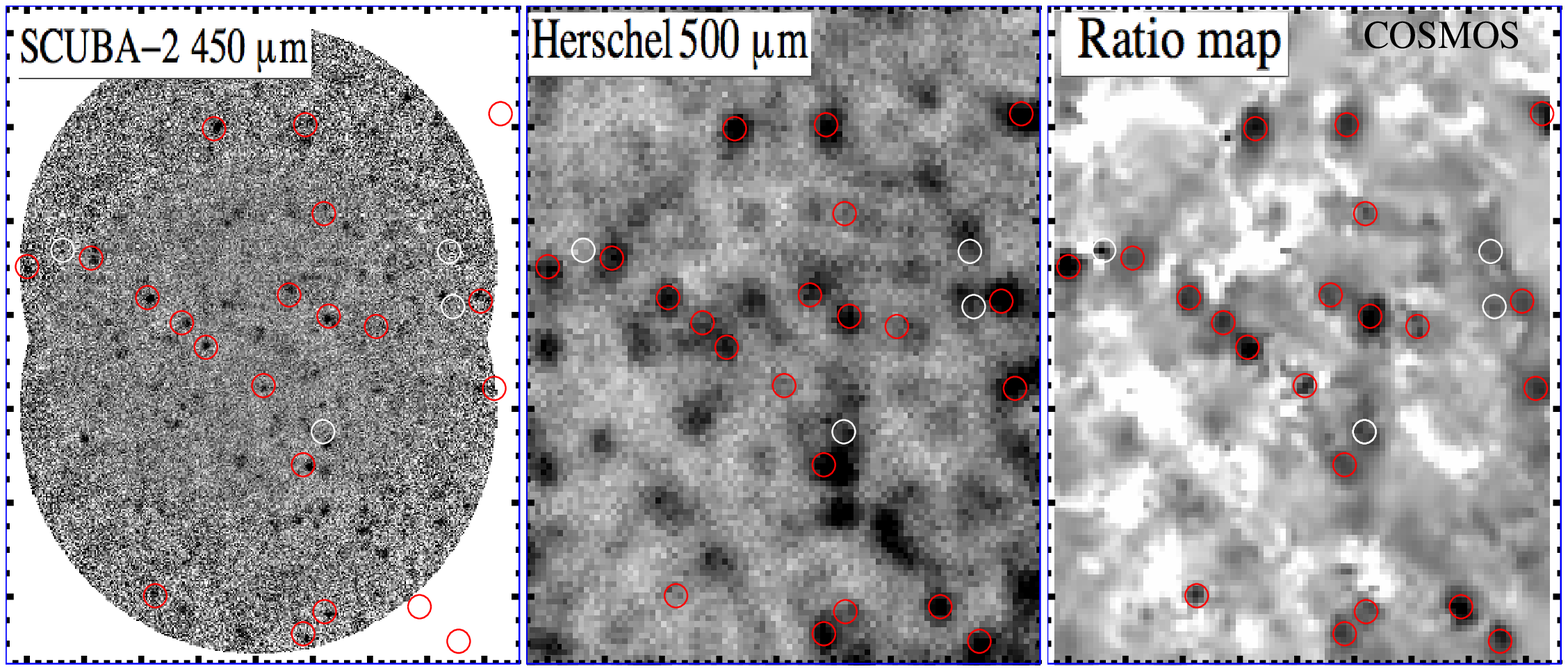}
\caption{
SCUBA-2 450\um image (left), \herschel 500\um image (middle) and the \herschel 500\um to $Spitzer$ 24\um ratio map (right). 
The data are taken from observations of the COSMOS field.  
Sources that are detected in the ratio map are shown in red circles with a radius of 15\arcsec, 
while white shows those without counterparts in the 450\um map. 
%Joint 99\% confidence contours of the \fekalfa emission line intensity vs. velocity width FWHM, obtained
%from Gaussian fits to the line observed with the \chandra HEG, as described in the text.
%The vertical dotted lines correspond to the FWHM of H$\beta$ line (Wang et al. 2007).}
{  Note that while many bright 500\um sources disappear in the ratio map, 
there are few faint sources being more unambiguously revealed in the ratio map.} 
}
\label{fig:scuba2_rmap}
\end{figure*}

%In order to avoid bias towards identifications to the brightest 24\um sources, 
%unlike most previous works,  
%we use the number density of a list of 24\um sources with $S_{500\mu m}^{\rm Pred}$ above 
%the flux value of the source of interest.   
%in computing $P$-statistics for the counterparts, 
%Here, the $S_{500\mu m}^{\rm Pred}$ is the predicted 
%500\um flux density for each source based on its 24\um emission and redshift 
%assuming a mean \ratio~color evolution of CE01 SEDs (dashed line in Figure 1, see also Section 2.3.1).} 
% we employed a modified Poisson probability, 
%$P_{\rm eff}$, based on the detection efficiency of 500\um sources in the ratio map from simulations (Section 2.4).  
%We denote a $P_{\rm eff}=P_{\rm 0}\ast(1-\epsilon)$ for each source, where $P_{\rm 0}$ is 
%the nominal $P$-value calculated using the 24\um catalog, and $\epsilon$ is the probability 
%that a given 500\um source at a given redshift can be detected in the ratio map, 
%i.e., a `probability of detection' for each source.  
%Accordingly, we are able to assign a `probability of detection' for each source. 
%A higher detection probability $(\epsilon)$, a lower $P_{\rm eff}$ and probability of 
%chance association. 
Unlike what happens at shorter wavelengths, the best counterpart for a 500\um detection 
is not the brightest 24\um prior but instead results from a combination
of source luminosity and redshift. To account for this, 
{  For a given potential 24\um counterpart, we do not use the number density of sources above flux $S$ 
at the 24\um in 
calculating the Poissonian probability of chance associations, but of 24\um counterparts brighter than a 
``predicted" 500\um flux density}, $S_{\rm 500\mu m}^{\rm pred}$}.
%accounting for both 24\um flux and redshift} 
This is obviously only possible in a field
such as GOODS-{\it North} with a high redshift completeness. 
To estimate $S_{\rm 500\mu m}^{\rm pred}$ 
for each 24\um counterpart and then determine a density criterion based on it, we assume
a characteristic 500/24\um flux density ratio that depends on redshift following the trend 
shown with a dashed line in Figure 1. 
{  Such a redshift dependence of the 500/24\um color represents 
the evolutionary trend of mean \ratio~ratio of CE01 SEDs}.
Since high-redshift 500\um sources are expected to present a larger \ratio~ratio and hence a 
higher probability of being detected in the ratio map, 
%radio catalogs undoubtedly contain a number of spurious sources near the detection thresholds, 
this procedure will naturally select the corresponding 24\um sources as potential counterparts 
 and reduce the probability of chance associations with lower-redshift sources. 
{  However, one has to treat this statistic with caution, as we assumed that 
lower-$z$ galaxies with abnormal IR SEDs that would have similarly high \ratio~ratios 
are rare (e.g., Rowan-Robinson et al. 2010).  
We will comment on later in this paper, $\sim$17 per cent of 
identifications might be due to the contaminations from 
some of the lower-$z$ populations.  
%evidence does exist for  some of the populations . 
Note that using the observed 500\um flux
to predict 24\um flux is an alternative way to identify counterparts, but maybe problematic
(regardless of the uncertainty in choosing IR SEDs), since without a perfect knowledge
of priors, the precise measurements of the 500\um flux are impossible due to the effect of
blending, resulting in the predictions of 24\um flux with high uncertainty.
}

%Note that the precision of the method relies on 

%To identify the most likely counterparts for our 500\um sources detected in the ratio map, 
%we need to choose a threshold of $P_{\rm 24\mu m}^C$ and ensure that the fraction of 
%incorrect counterparts remains low. 

%Our procedure to do this is using the simulated      
%\herschel data which are described in detail in Schreiber et al. (2015). 
In order to define a threshold for our modified $P$-value {  calculated for the 24\um catalog}, 
$P_{\rm 24\mu m}^C$, that will correspond to a realistic 
association probability, we use the realistic mock \herschel data that are described in detail 
in Schreiber et al. (2015).
We construct a \ratio~map for the mock \herschel data and detect candidate high-$z$ sources using the same procedure as 
we described above. 
%We found that at a threshold of $p\simlt0.1$, $\sim$90\% sources detected in the ratio map will 
%have at least one 24\um counterpart at $z\simgt2$. 
In Figure 5 (upper), we plot the $P$-statistics calculated using the formulation above, against 
the redshifts for all input 24\um sources within the search radius (filled squares).
% centered on the source detected in the ratio map. . 
We found that at a threshold of $P\simlt0.1$, $\sim$90\% sources detected in the ratio map 
would have at least one 24\um counterpart at $z\simgt2$ (red circles), and no source below $z\sim1.5$ 
has a $P$-value greater than 0.1. 
Figure 5 (lower) shows the distribution of the ratio of 500\um flux of candidate counterparts ($P\simlt0.1$) 
to the total 500\um flux of all 24\um sources within the search radius. 
This ratio provides a measure of the importance, in terms of 500\um flux, of the identifications. 
A ratio close to 1 indicates that the 500\um emission is dominated by a single (high-$z$) 24\um counterpart.  
Only $\sim$14\% (6/42) 500\um sources detected in the ratio map have more than one 24\um galaxy with additional source(s) 
contributing up to $50$\% of the total 500\um flux, and five of them in fact have multiple $z\simgt2$ counterparts. 
Therefore $\sim$86\% (36/42) of the 500\um sources have been associated with a single $z\simgt2$ source, 
which is main contribution to the 500\um emission ($>$50\%), suggesting a high efficiency of the identifications. 

{  In Figure~\ref{fig:pchist}, we plot the cumulative distribution
  of our modified $P$-values when taking into account the redshift information for all 24\um counterparts
within the search radius (solid line). 
Also shown is the distribution of the traditional $P$-values which were calculated
using only the 24\um flux, as often adopted in literature (dotted
line).  Based on the resulted Poisson probabilities, almost none of
the associations would be judged to be significant at $\ge$90\% confidence
                                (i.e., $P\le0.1$). 
Using the predicted 500\um flux instead (dashed line) would result in slightly more
sources with $P\le0.1$, but still not statistically high enough to
quantify the formal significance of the identifications.  
}

%Note that adopting $P\simlt0.1$, 
\subsubsection{Validation of the method with SCUBA2}

{  
In order to further test our method and its efficiency in uncovering candidates of high-$z$ 
\herschel 500\um sources, we build a \ratio~map using the data from
the COSMOS field for which we have 
a SCUBA-2 450\um image for direct comparison. The higher resolution of SCUBA-2 (7.5\arcsec~FWHM, Geach et al. 2013) at 
450\ums, compared to 36\arcsec~at 500\ums, 
offers a unique opportunity to assess the reliability of the high-$z$ 500\um 
sources identified using our \ratio~method. 
Figure~\ref{fig:scuba2_rmap} shows the SCUBA-2 450\um image of the
COSMOS field (left). The corresponding \herschel 500\um image of
similar size is shown in the middle. 
On the right, we show the \herschel 500\um to $Spitzer$ 24\um flux ratio map, 
overlaid with detected sources (red circles) using the same search 
criteria as described in Section 2.1. 
It can be seen that many bright 500\um sources in the actual map (middle) 
disappear in the \ratio~map. 
This is likely due to the fact that many of them are dominated by 
low-$z$ galaxies where the \ratio~ratios are low. 
Conversely, while some sources appear relatively faint and undetectable in the actual 
500\um map (close to the confusion limit $\sim$6mJy), they are more 
unambiguously seen in the ratio map (e.g., the last two panels in Figure 9). 
% due to an increase in the contrast of signal.  
These fainter 500\um sources are in fact clearly seen in the deeper SCUBA-2 450\um 
imaging (rms$\sim$2.5-4.0 mJy/beam) of COSMOS.
%??Jim, whether we need to quote the PI of SCUBA-2 here?).    

When compared to the field covered by the SCUBA-2 450\um map, a
 total of 22 sources are detected in the ratio map, and 18 of them have 450\um counterparts 
within {15\arcsec}.  
For the remaining four sources without SCUBA-2 counterparts, three are located close to the 
edge of the SCUBA-2 map 
%and hence their detections suffer from a higher noise 
where the local noise level is relatively high (white circles in Figure~\ref{fig:scuba2_rmap}). 
%In Figure E2, we shows t
We also checked the positional offsets between the sources identified in the 
ratio map and the SCUBA-2 450\um sources. We found that the positional offsets for most sources are less than 
10\arcsec. Considering the SCUBA-2 beam size of $\sim$7\arcsec~ which is close to the pixel size 
of \herschel 500\um map (7.2\arcsec), this strongly suggests 
a physical association between the sources detected in the ratio map 
and those in the SCUBA-2 map.

}

{  With the same $P$-statistics and cut ($P\simlt0.1$), 
we present in Figure~\ref{fig:scuba2_pvsz} the counterpart identifications to high-$z$ 500\um sources in the COSMOS field, 
where the higher resolution SCUBA-2 450\um imaging allows for 
more precisely localizing the 24\um counterparts. 
%for testing 
%the realism of the 500\um source identifications.  
We found a similar trend of $P$-value against redshift as in Figure~\ref{fig:csimulate} . 
At a $P$-value cut of $\simlt0.1$, most, if not all, 500\um sources detected 
in the ratio map have at least one 24\um counterpart at $z\simgt$2. 
Most high-$z$ 500\um sources that we identified are also detected in the SCUBA-2 450\um image  
with positions coincident with the 24\um counterpart,
suggesting a low probability for chance associations.  
These results further support the reliability of our approach for counterpart identification. 
%using the $P\simlt0.1$    
%suggesting that the identifications to high-$z$ 500\um sources are reliable. 
%In order to test this and the modified $P$-statistic procedure, we have used the simulated 
%\herschel data as described in detail in Schreiber et al. (2015). 
%The results are that 
%We found that there is a $\sim$90\% probability that a simulated $z$\simgt2 \herschel 500\um source 
%with flux density above 6 mJy ($\sim2\sigma$) detected in the ratio map  
%has a counterpart with $P_{\rm eff}\simlt0.5$ within a search radius of 15\arcsec. 
We therefore adopt a cut on the Poisson probabilities of $P\simlt0.1$ for the counterpart 
identifications to our sample. 
%Note that 500\um detections that do not present 24\um counterparts fulfilling this criterion 
%will therefore be considered as 24\um dropouts. 

\begin{figure}
\centering
\includegraphics[scale=0.5, angle=0]{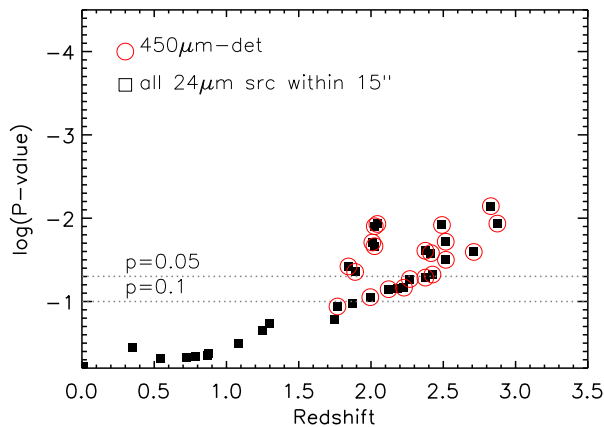}
\caption{
%The same as Figure 1, but from the test with SCUBA-2-450\um data, to p
Plot of redshift vs. $P$-value for all 24\um sources
(filled squares) within the search radius of 15\arcsec, and counterparts to 500\um sources (red circles). Most high-$z$ candidate 500\um sources (red circles)
identified with $P\simlt0.1$ (at $\simgt$90\% confidence) are also
detected in the SCUBA-2 450\um imaging with higher resolution, suggesting that
the rate of chance associations is low.
%the counterpart identifications are reliable. 
%The distribution of positional offsets between the SCUBA-2 sources and those identified in 
%the ratio map. The red dotted line represents the pixel size of 7.2\arcsec~for the \herschel 500\um image. 
%Joint 99\% confidence contours of the \fekalfa emission line intensity vs. velocity width FWHM, obtained
%from Gaussian fits to the line observed with the \chandra HEG, as described in the text.
%The vertical dotted lines correspond to the FWHM of H$\beta$ line (Wang et al. 2007).}
}
\label{fig:scuba2_pvsz}
\end{figure}

In Figure~\ref{fig:imgshow}, we show examples of the identifications
of two relatively clean sources from our catalog 
where the blending effect is small, as well as four sources in the COSMOS field. 
Though they could be also identified in the original 500\um map, this demonstrates that 
the detection in the ratio map is without any doubt coincident with a high-z source: 
one is spectroscopically confirmed at $z=3.19$ and another has a photometric redshift at $z=4.15$.  
Further support comes from the comparison with the SCUBA2 data at the 450\ums, 
showing that candidate high-$z$ 500\um sources have been unambiguously identified.
}

 \begin{figure*}[]
 \centering
 \includegraphics[scale=0.55, angle=0]{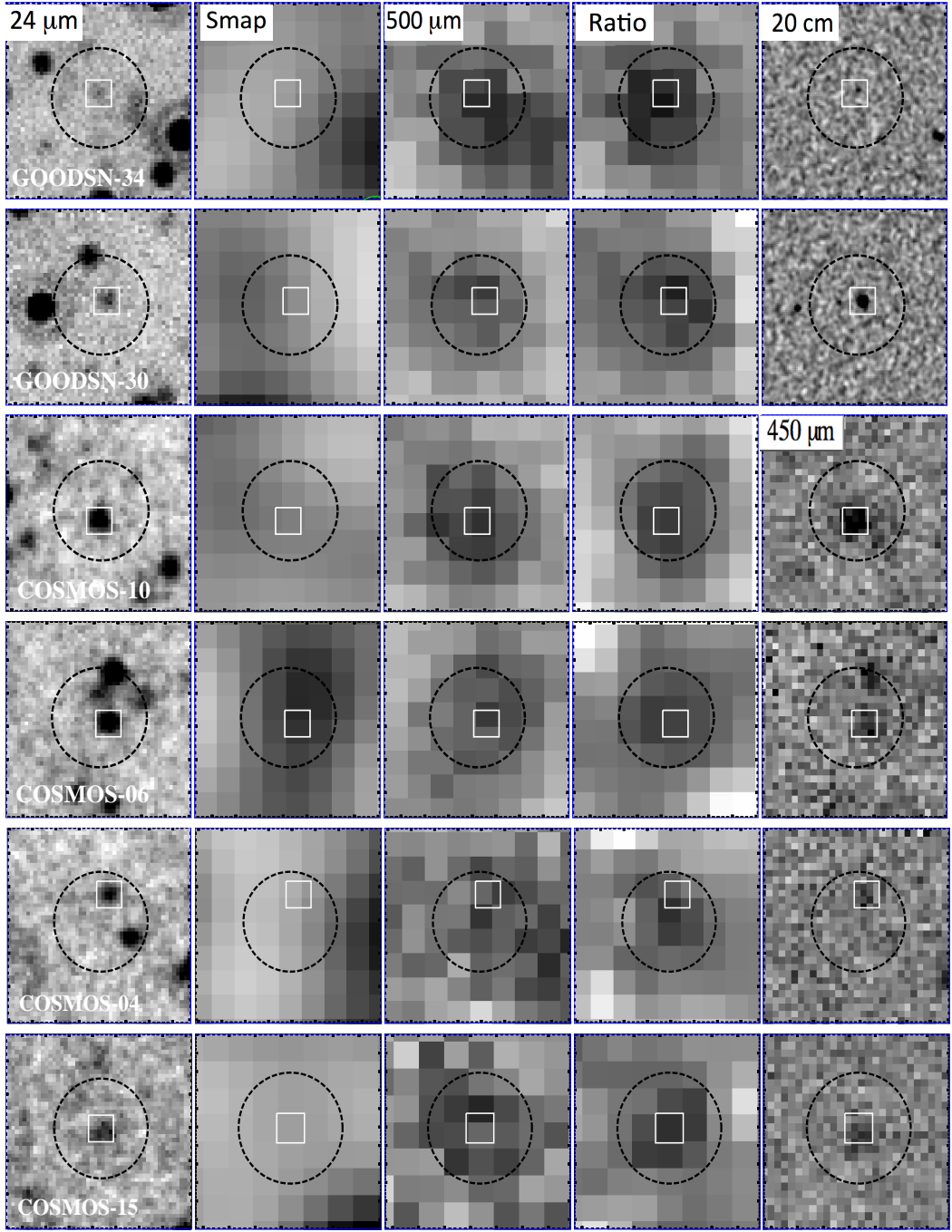}
 \caption{
{  Examples of ``clean" sources detected in the ratio map.
Image cutouts from left to right: the 24\ums, the ``beam-smoothed"
24\um (Section 2.2), the 500\ums, the \ratio~ratio and the radio 1.4 GHz map.
%The upper panel shows the two sources in the GOODS-North field, 
%while 
In the lower four panels, we also show example sources in the COSMOS field where the SCUBA2-450\um 
image is available for direct comparison (last column).
%The lower two rows represent two faint 500\um sources revealed in the ratio map, which 
%are close (or below) to the confusion limit at 500\um and hence may not be detectable in the actual 500\um image. 
}
}\DIFaddend
 %Joint 99\% confidence contours of the \fekalfa emission line intensity vs. velocity width FWHM, obtained
 %from Gaussian fits to the line observed with the \chandra HEG, as described in the text.
 %The vertical dotted lines correspond to the FWHM of H$\beta$ line (Wang et al. 2007).}
 \label{fig:imgshow}
 \end{figure*}

\begin{figure}[t]
\centering
\includegraphics[scale=0.35, angle=-90]{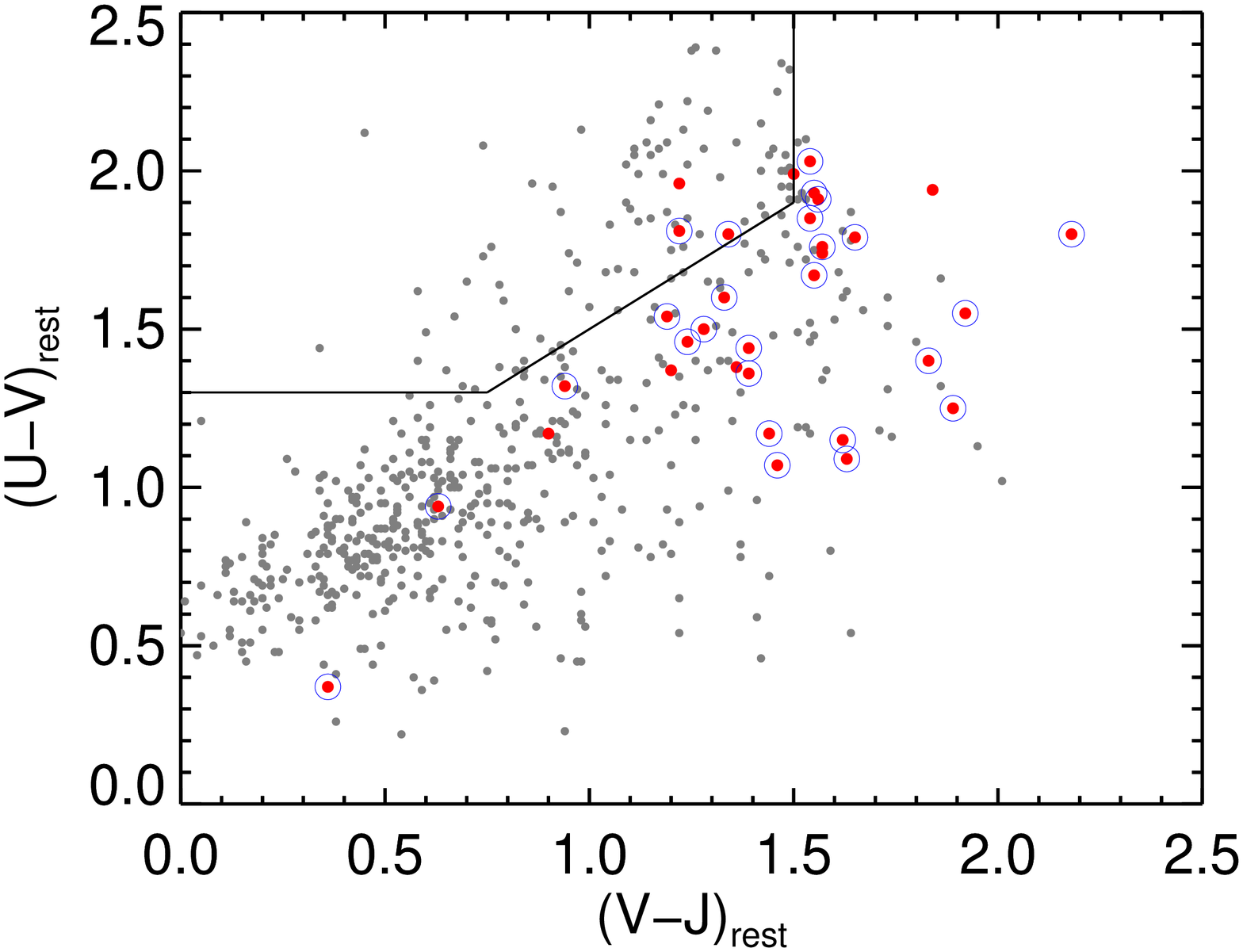}
\caption{ The rest-frame (U-V) vs. (V-J) color diagram for all IRAC sources within the search radii of 15" (grey dots). 
Red solid circles represent secure Spitzer (radio)-identified 500\um counterparts (Table 2 in the text), and
those satisfying IRAC color selection for SMGs are highlighted in blue (Biggs et al. 2011).
}
\label{fig:uvj}
\end{figure}

{  In order to considering alternative counterparts for the few sources with no 24\um counterpart, which may be an IRAC source with
a predicted 500\um flux density greater than the one associated with the 24\um counterparts, we in fact
considered two extrapolations of $S_{\rm 500\mu m}$, the one derived from the observed 24\um and the one
derived from star-formation main-sequence relation (Column 8 in Table 1).
The latter approach is based on redshift and stellar mass measurements, associating an SFR to each IRAC source 
using the main-sequence relation (Schreiber et al. 2015). 
The SFR was converted into a total IR luminosity 
using the Kennicutt (1998) relation\footnote{  
We used a mass-dependent dust extinction, i.e., IR excess IRX = $\rm log_{10}(L_{IR}/L_{UV})$, 
to decompose the SFR into a dust-obscured component seen in the FIR and a dust-free component 
which emerges in the UV (e.g., Heinis et al. 2014).
}. 
%relative contribution of unobscured and obscured SFR. }.  
%We then assigned a favorable CE01 SED to
%each galaxy following the observed SED evolution trends with redshift to predict $S_{\rm 500\mu m}$.
We then assigned a favorable CE01 SED to each galaxy based on the IR luminosity to predict $S_{\rm 500\mu m}$.  
We used the UVJ selection technique (Whitaker et al. 2012) to isolated passive galaxies and excluded 
them from the counterpart catalog.
Although the main-sequence approach provides only a lower limit to SFR and hence estimate of $S_{\rm 500\mu m}$,
it will help to discriminate against candidates of high$-z$ passive galaxies which are undetected at 24\ums.
Note that 500\um detections that do not present 24\um counterparts fulfilling the $P$-value criterion 
will therefore be considered as 24\um dropouts. 
%robustness of proposed faint where the surface density is much greater.
}

Using the above procedure, we have found at least one 24\um counterpart for {  28 (out of 36)} 
sources detected in the ratio map (Table 1).  
%first search for counterparts in the 24\um catalog 
%with $P_{\rm eff}\lt0.5$. 
In {  eight} cases where no such counterparts can be found, 
we first proceed to search in the IRAC 3.6\um catalog using the same radius and $P-$value cut. 
% on $P_{\rm 24\mu m}^C$. 
%In calculating the $P_{\rm 24\mu m}^C$ for the IRAC sources, we used the main-sequence relation and 
%a mass-dependent dust extinction to predict the 500\um flux for them
%(see the details in Schreiber et al. 2015).  
% which are not detected in the 24\ums, 
%we have assumed a 3$\sigma$ upper limit of 21$\mu$Jy for them. 
This results in two further identifications in the IRAC band {  (GH500.01 and GH500.15).
%, one of which are also detected in the radio. 
For the remaining six sources}, we used the radio data for counterpart identifications. 
%In total, this procedure yields at least one counterpart within 15 arcsec for  
%sources, ?? of which are robust with $P_{\rm eff}\simlt0.5$. 
Out of a total of 36 sources, {  seven} have more than one potential 24\um counterpart, and 
{  seven} are found to have additional counterpart(s) from the IRAC identifications.
Such multiple statistical associations could be partly due to the source clustering at similar redshifts (e.g., GH500.26, GH500.27).  
Therefore, a unique counterpart is identifiable in about 60 per cent of the cases,            
while two or more candidates are present in others. %, requiring a more quantitative analysis.  
For the latter case, we choose the one(s) detected in the radio (if available) as the most probable 
counterpart for further analysis (Table 2)\footnote{For GH500.08, 
we choose the one of not being detected in the radio as best counterpart (see Appendix A.).}.
%where the radio-detected 
%24\um counterpart has little contribution to the 500\um emission from our detailed analysis.}.   
Candidate $Spitzer$ and radio counterparts and their computed $P$-statistics\footnote{
In computing the $P$-statistics for the radio catalogue, we used the nominal method 
without taking into account the redshift information (e.g., Pope et al. 2006). } 
are given in Table 1.  
%A unique counterpart is identifiable in about 60 per cent of the cases, 
%while  
%For eight sources which have multiple counterparts, we choose the one(s) detected in 
%the radio as the most probable counterparts, which are highlighted in boldface 
%letters, though a more quantitative analysis is required. 
%Note that when using the IRAC 3.6\um catalog for counterpart identifications, 
%Five sources (all at z$>$4) are found to have additional counterpart(s) from the IRAC identifications. 

%Of totally 51 identifications, 34 are also detected in the radio.  
%?of the ?? sources which have more than one potential counterparts? 
%The probable interpretations of such multiple statistical associations are 
%either gravitational lensing, or clustering of star-forming galaxies at the source 
%redshift. 

{
We note that some bright silicate absorption galaxies at $z\sim0.4$ and $z\sim1.3$ which have 
depressed 24\um emission (Magdis et al. 2011), 
may have elevated $S_{500\mu m}$/$S_{24\mu m}$ ratios meeting our selection criteria. 
%We screened against such objects b
By inspecting the photometric redshift and far-IR SED, 
we found one such lower-$z$ galaxy (GH500.06) that
may coincidently have a large $S_{500\mu m}$/$S_{24\mu m}$ ratio 
and we excluded it from our sample. 
In addition, there are {  five} sources 
%which are detected in the ratio map and have counterparts
%in both the radio and 24\ums (GH500.03, GH500.11, GH500.22), 
at redshifts less than $2$ (GH500.03, GH500.11, GH500.22, GH500.24, {  GH500.36}), 
%among which two are detected in the radio, 
%The second can arise from the SED variations. 
% 
among which three are spectroscopically confirmed at $z_{\rm spec}=1.6$, $z_{\rm spec}=1.76$ and 
$z_{\rm spec}=0.8$, and another {  two} has a photometric redshift of $z_{\rm phot}=1.61$ {  and $z_{\rm phot}=0.60$}, 
respectively.
% derived from a $K_{\rm s}$-based
%multiwavelength photometric catalog (Pannella et al. 2014).
As we are mainly interested in the $z\simgt2$ population of \herschel 500\um sources,
we exclude them from the following analysis. 
%Finally, we removed one source at $z_{\rm spec}=0.8$
%hat is heavily blended (GH500.24): there are five bright 24\um sources 
%within the searching radius, and three of them have photometric redshifts at $z$\simgt2.
% greater than 2 and detection in the radio. 
%They are so closely separated (less than 5\arcsec~from each other) that deblending of 500\um is impossible.    
%in the $S_{500\mu m}$/$S_{24\mu m}$ ratio map 
%Our final sample includes 34 sources which meet all selection criteria,
%which are listed in Table 1.

For the remaining {  35} sources (including four with more than one candidate counterpart), 
we provide the coordinates, photometric 
redshift (or spectroscopic redshift if available), mid-IR to radio flux densities, and the derived 
{ far-IR} 8--1000\um luminosity and dust temperature in Table 2. 
{  Their distribution in the rest-frame (U-V) vs. (V-J) color digram
  is shown in Figure~\ref{fig:uvj}. 
It can be seen that most objects (32/36, $\sim$90\%) can be classified as
dusty starburst galaxies (Whitaker et al. 2012), consistent with
their  high dust attenuation of $A_{\rm V}$$\sim$2 mag 
inferred from the UV-MIR SED fittings.  
%???need more descriptions, two sources are classified as passive, which may have unreliable identifications.??
We note that there are two sources (GH500.10a and GH500.25) falling
into the regime for quiescent galaxies. However, both are
significantly detected in the \herschel PACS 160\um ($\sim4-6$ mJy, Table 2), suggesting the
presence of ongoing
obscured star-formation. 
The possibility of wrong identifications seems low, as both are relatively isolated 
and well detected in the 24\um and radio.   
Since the two sources have relatively high photometric redshift at
$z\simgt4$ where the UVJ selection is still poorly explored (e.g.,
Straatman et al. 2013), whether they can be considered quenched
remains further investigations. Indeed, their specific SFRs (8-40$\times10^{-9}$/yr) are already $>$2x higher than
the sSFR= $5\times10^{-9}$/yr of similarly massive star-forming galaxies
at $z>$3 (Schreiber et al. 2015). 
% those typical values for quiscent galaxies (e.g., 
} 

The {\it Herschel}/SPIRE fluxes are from the GOODS-\herschel catalog (Elbaz et al. 2011), 
or derived from point-spread function (PSF) fitting using GALFIT (Peng et al. 2002) 
for cases where the flux could be wrongly attributed to nearby 24\um sources 
due to blending, or is not measured in the catalog. 
A description of the {\it Herschel}/SPIRE flux measurements is presented in Appendix B. 
{  Note that due to the poor resolution of the \herschel data, measuring reliable 
SPIRE flux densities for individual galaxies, especially those at 500\ums, is a 
challenging task. The de-blending approach of using 24\um data of better resolution 
would be still affected by the degeneracy between the positions of the sources and 
flux uncertainties attributed by faint sources.  
Given the case of heavily blending, the SPIRE flux measurements 
should be treated with some caution.}
The thumbnail images for each identified source are presented in Appendix C.
%Among 35 sources, ten have multiple (two) $z\simgt2$ radio/24\um
%counterparts. T
Among 14 sources that have two or more candidate radio/24\um 
counterparts, four (GH500.13, GH500.26, GH500.28, GH500.35) have counterparts separated by only $\sim$5\arcsec or less. 
%Since it is impossible to have a meaningful flux deblending, we report the
%combined SPIRE fluxes in Table 1. 
Three of these cases are treated as single systems because 
the counterparts are consistent with being at the same redshift and are assumed to be 
interacting galaxies. 
%Two of these (GH500.26 and GH500.28) 
%
These three sources are so closely separated that deblending the {\it Herschel}/SPIRE 
flux is impossible, hence 
%Since it is impossible to have a meaningful flux deblending, 
we report the combined fluxes of the two counterparts in Table 2. 
For GH500.35, we cannot assess whether the three counterpart galaxies are
interacting systems {  or spatially clustered}, and 
we assume that the source with the brightest 24\um flux is the only counterpart, 
which should be treated with caution. }

\subsubsection{Contaminants}
%There are three classes of contaminants with potentially   
%high $S_{500\mu m}$/$S_{24\mu m}$ ratios thus meeting our selection criteria:  
%(a) sources slightly below our S/N cut, (b) low-$z$ sources with depressed 24\um emission or extremely 
%cold dust temperature (enhanced 500\um emission), and (c) noise artifacts in the ratio map. 
%The former, also known as ``Eddington bias", can be partly caused by blending effects, 
%while the second can arise from the SED variations. 

%We find three sources which are detected in the ratio map and have counterparts 
%both in the radio and 24\ums, at slightly lower redshift than $2$.
%Two are spectroscopically confirmed at $z_{\rm spec}=1.72$ and $z_{\rm spec}=1.89$, and another 
%has a photometric redshift of $z_{\rm phot}=1.61$ derived from a $K_{\rm s}$-based 
%multiwavelength photometric catalog (Pannella et al. 2014). 
%As we are interested mainly in the $z\simgt2$ population of \herschel sources, 
%we exclude them in the following analysis.   

%The blending or boosting effects would be an issue, though we used a 
%beam-smeared 24\um image to construct the ratio map. 
% where source confusion, in some sense, is reduced ({  not clear}).  
We have visually inspected all sources which pass our S/N cut in order to search for 
noise artifacts and blends of bright sources. 
%Only for one source that is found in the ratio map hasn't radio or 24\um counterpart at $z>$
%In general, such false sources are easily to be identified in the higher resolution 
%{\sc SPIRE} 350\um and 250\um real map, and found to be negligible fraction. 
%To check for false detections caused by map noise artifacts, we run 
%our search procedure on the \ratio~map having removed all individually detected 500\um 
%sources in the catalog used in Elbaz et al. (2011). 
{  To check for false detections caused by map noise artifacts and/or from 
sources well below the nominal ``flux" detection limit, we run 
our search procedure on the \ratio~map by removing all individually detected 500\um 
sources in the catalog used in Elbaz et al. (2011). 
%We find only two detections which are possibly false sources arising from noise fluctuations or 
%very faint objects below the detection limit (at a 3$\sigma$ level). 
%In fact, one of them has a potential counterpart in the radio 
%and $Spitzer$-IRAC bands, but is not detected in the deep $H$ and $K_s$ band imaging. 
%We performed flux extraction by using PSF fitting at the position of radio counterpart, 
%and found a 500\um flux of $\sim$3 mJy, consistent with the 1$\sigma$ noise level (Elbaz et al. 2011).  
%We cannot tell whether these 500\um sources are real or artifacts. 
The P(D) plot of the \ratio~ratio map omitting all 500\um detections in the existing 
catalog is shown in  Figure~\ref{fig:pdplot} (red curve). 
It can be seen that few pixels have values above our S/N cut of \ratio=20.4, 
suggesting that the contamination from noise artifacts is negligible. 
%consistent with the two false sources we found.  
Figure 4 (lower panel) shows the number of spurious sources as a function of detection threshold.  
As one would expect, the number of false detections depends strongly on the chosen S/N 
threshold. 
At a 2$\sigma$ cut, the result suggests that two detections will be spurious in our 
ratio map, corresponding to a spurious source rate of $\sim6$\% (2/36). 
Note that while increasing the S/N requirement would decrease the number 
of false sources, it would also remove a fraction of real but slightly fainter sources from the map. 
Therefore, we adopt 2$\sigma$ as the detection S/N cut for sources in the ratio map, because 
it provides a good compromise between the catalogue size and source reliability. 
}

{  The above estimate, however, provides only an lower limit to the number
of contaminants. 
%It is possible that some lower redshift sources with abnormal high \ratio~ratios 
As shown in Figure 1, the diversity in IR SED of galaxies means that 
lower redshift sources would have probabilities of being detected in the 
ratio map if they have similarly high \ratio~ratios, 
contaminating the selection of $z\simgt2$ sources. 
We account for this effect statistically by performing simulations 
to measure the recovery rate for injected low-$z$ sources of varying flux density, 
using the similar procedure described in Section 2.3.1.  
For each injected source we assumed an IR SED randomly selected from the 
CE01 libraries to predict its 24\um flux density at a given redshift. 
Note that we didn't assume any peculiar IR SED for injected sources 
to avoid bias in detections. 
Although very ``cold" galaxies have more chance of 
being detected, 
%we haven't chosen peculiar IR SED of this kind to avoid 
%bias, as 
they are expected rare given the small area covered by GOODS-North. 
%During the injection process, we assume sources have equal probability of 
%being selected any SED from CE01 libraries.  
Since lower redshift sources are not expected to be detected in our data except those with abnormal high 
\ratio~ratios, any source recovered in the map would be considered as a contaminant.
The average recovery rate for low$-z$ contaminants as a function of flux density is shown in Figure 4 (lower right panel).
It can be seen that sources below $z\sim$1.2 have a low probability to be detected, but the detection rate
increases with redshift above $z\sim$1.4 due to the increased fraction of sources with higher \ratio~ratios. 
%The major contaminants could be due to some bright silicate absorption galaxies which have depressed 24\um emission
%at $z\sim0.5$ and $z\sim1.5$, hence a boosting of S500/S24 ratios.
The low$-z$ contaminants become more significant at higher flux density, making them non-negligible 
effect in selecting $z\simgt2$ galaxies. 
In fact, such low$-z$ sources have been identified in the ratio map.  
As we mentioned above, we found 6 out of 36 sources at $z<2$ ($\sim17$\%) that could be contaminants
in our catalog, which is consistent with the simulation result of $\sim$20\% detection rate at $\sim15$ mJy, 
a median flux for catalog sources.  
}

%\begin{figure}
%\includegraphics[angle=-90,scale=.37]{/Users/xshu/paper/goodsn/submit/0625/figure/sedvsredshift/sedvsz_withdata.ps}
%\includegraphics[angle=-90,scale=.37]{/Users/mac/paper/goodsn/submit/revise/0921/figure/sedvsredshift/sedvsz_withdata_new.ps}
%\caption{Animation still frame taken from \citet{kim03}.
%This figure is also available as an mpeg
%animation in the electronic edition of the
%{\it Astrophysical Journal}.}
%\end{figure}
%\subsection{Purity and Efficiency}
%\begin{figure*}[th]
\section{Basic observed source properties}
\subsection{Redshift distribution}
%While the spectroscopic redshift coverage of  field is less than 10\% for both 24\,$\mu$m and Herschel 
%detected sources, 
%better than XX\,\%  and YY\,\% respectively for {\it Herschel} sources and 
%The unique multi-wavelength coverage of the GOODS-North field provides 
%well constrains on the photometric redshifts (e.g., Schreiber et al. 2014), 
%with 85\% of 24$\mu$m sources have a redshift photometrically measured 
%(and 48\% of spectroscopic redshift).
% 
%The photometric redshift estimation for candidate $z\simgt2$ \herschel 500\um sources 
%comes from the ground-based $Ks$-band catalog (Pannella et al. 2014), 
%and when available, spectroscopic redshift is used either from optical spectroscopy (Barger et al. 2008) 
%or CO emission lines (Daddi et al. 2009a, b).   
In this section, we present the redshift distribution for our 
%$S_{500\mu m}$/$S_{24\mu m}$-selected 
$z\simgt2$ \herschel 500\um sources. 
To derive the redshift distribution, we take spectroscopic redshifts if available, based 
on optical and/or CO spectroscopic observations (Barger et al. 2008; Daddi et al. 2009a, b), 
and otherwise photometric redshifts based on a $K_s$-selected multi-wavelength photometric catalog (Pannella et al. 2015). 
Spectroscopic redshifts are available for 12 out of the {  35 \herschel sources (33\%)} in the 
sample. {  Note that the photo$-z$ for one source (GH500.15) is not available, as it is not detected in the $K_s$ band. 
Although SED fittings in the far-IR suggests it likely being at $z\simgt4$ (Section 4), 
we exclude the source in the following analysis due to the relatively poor constraint on the photo$-z$}. 
%Note that for 1 source (Table 1, ), we use the redshifts derived from the best-fit far-IR spectral templates, 
%as the source is either not detected in the $K_s$-band, or has a photometric redshift significantly 
%inconsistent with the one derived from far-IR SED fittings ({ the $\chi^2$ difference for the best-fitted SED 
%is $>$10}). 

\begin{figure}[th]
\centering
\includegraphics[angle=0,scale=.69550]{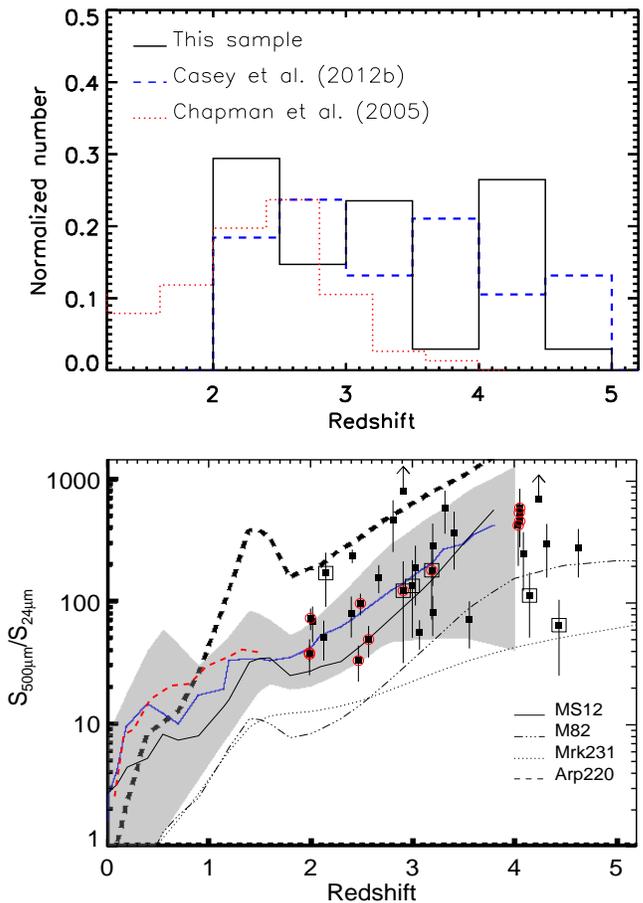}
\caption{
{\it Upper:} 
The redshift distribution of the \ratio-selected
sources in the GOODS-North field (black line, normalized by the total). 
The red dotted histogram is the redshift distribution of SCUBA 850\um sources 
from Chapman et al. (2005).
% where only $z>2$ SMGs are shown. 
For comparison, the blue dashed histogram shows the distribution of spectroscopic redshifts for 
$2<z<5$ SPIRE sources from Casey et al. (2012b).  
%For each sample, the distribution is normalized to the total number of sources. 
The excess number of objects at $4.0<z<4.5$ may be due to cosmic variance and 
the presence of a high-redshift proto-cluster (Daddi et al. 2009b).
%Our sample seems to peak at higher redshift than those selected at $\sim$850\um, 
%with a significant excess at $z>3$. 
{\it Lower:}
\ratio~flux ratio as a function of redshift. 
Black curves are model SEDs of Arp220 (dashed), 
a main-sequence galaxy template (solid) from Magdis et al. (2012), M82 (dot-dashed) and Mrk231 (dotted). 
The shaded region represents the range of the CE01 models, and the blue solid is 
its mean trend. The red dashed line is the median trend of $z<$2 {\it Herschel}/SPIRE 
sources from Casey et al. 2012(a). Filled squares are $z\simgt$2 sources identified 
from the ratio map, with spectroscopically confirmed ones highlighted by red circles. 
Large open squares denote galaxies whose 500\um emission is less than $3\sigma$. 
%The red dashed line is the median trend of $z<2$ \herschel/SPIRE sources 
%from Casey et al. 2012(a). 
%Filled squares are $z\simgt2$ sources identified from the ratio map, with 
%spectroscopically confirmed ones highlighted by red circles.  
}
\label{fig:redshift}
\end{figure}

%??? have good spectroscopic redshift.   
The redshift distribution for the {  34} $z\simgt2$ \herschel sources identified from the 
ratio map is shown in the upper panel of Figure~\ref{fig:redshift}. 
For comparison, we also plot the redshift distribution for SMGs (red) from Chapman et al. (2005).
%where only $z\simgt2$ sources are shown, and 
%and \herschel/SPIRE-selected galaxies at $z>2$ (blue) from Casey et al. (2012a). 
Each distribution is normalized to the total number of sources in the sample. 
The redshift distribution derived by Chapman et al. (2005) is based on a sample 
of SMGs drawn from various SCUBA/850\um surveys with radio counterparts targeted for spectroscopic 
redshift follow-ups.
%, and is renormalized in counts in Figure~\ref{fig:redshift} (left) to match ours for comparison.   
It can be seen that the distribution of our $S_{500\mu m}$/$S_{24\mu m}$-selected sample 
is in sharp contrast to that SCUBA/850\um SMGs.   
%clearly shows two peaks, one at $z\sim2-3$, similar to known SMGs, and another 
%significant population (11/34, $\sim$36\%) at $z\simgt4$. 
%, which is 
%span a broad range, from 1.99 to 4.7, with a non-ignorable fraction of sources at $z>3$. 
The median redshift for our sample {  is $z_{\rm med}$=3.07}, 
with a significant fraction of sources {  (10/34, 29\%) at} $z\simgt4$. 
The radio-identified SCUBA SMGs have a lower median redshift of $z_{\rm med}$=2.6 (considering only sources 
at $z\simgt2$), for which only 18\% of the population is at $z>3$. 
%?? Wardlow et al. 2011).    ??
% (Chapman et al. 2005).
% Wardlow et al. 2011).    
%We compare to the redshift distribution of spectroscopic sample from 
%\citep{chapman05}, median $z=2.2\pm0.1$.
Part of this difference could be due to the identification bias in the sample of Chapman et al. 
The redshift distribution of SCUBA SMGs includes only sources with robust radio detections;
the highest-redshift sources may fall below the depth of radio data and are likely to be missed.  
%the identifications of submm counterparts require robust radio detections.  
On the other hand, the sample considered here is confined to {\sc 500}\um 
sources significantly detected in the ratio map which efficiently selects 
sources at the highest redshifts. 
Lower-redshift submm sources are %rejected as a result 
%of our pre-exclusion of some of the less luminous sources in 
down-weighted in 
the ratio map  
in an effort to obtain a homogenous sample of galaxies at $z\simgt2$ as possible. 

%A recently published survey of 1.1mm-selected SMGs with optical counterparts 
%determined from high-resolution millimetre imaging found a median redshift of 
%$z_{\rm med}=2.8$ (Somolcic et al. 2012), which 

\begin{figure}[th]
\centering
\includegraphics[angle=0,scale=.550]{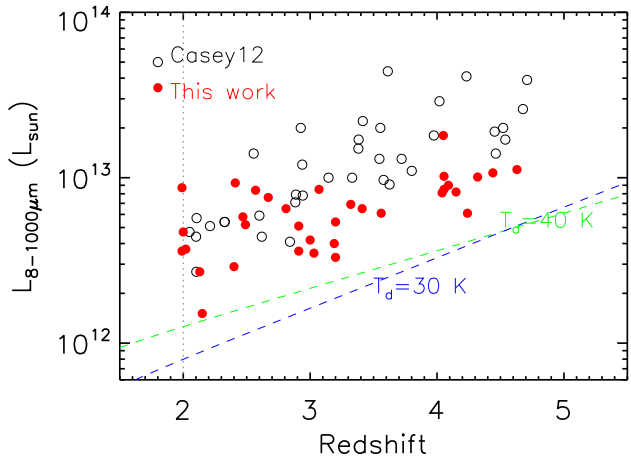}
\caption{
Infrared luminosity (8-1000\ums) against redshift for the \ratio-selected $z\simgt2$ ULIRGs (filled red circles) 
in the GOODS-North field. 
The {\it Herschel}/SPIRE-selected, spectroscopically confirmed 
sources at $2<z<5$ (open circles; Casey et al. 2012b) are overplotted for comparison. 
The detection boundaries as a function of redshift for $T_d=30$ K 
(blue dashed line) and $T_d=40$ K (green dashed line) are illustrated, 
assuming a flux limit of 9 mJy at 500\ums.   
}
\label{fig:lfir}
\vspace{0.1cm}
\end{figure}

%Comparing the distribution of \herschel/{\sc SPIRE} sources 
%derived from the large spectroscopic survey of 250--500\um\ selected galaxies at $2<z<5$ \citep{casey12b}, 
A recently published spectroscopic survey of \herschel/{\sc SPIRE} sources with 
optical counterparts determined from high-resolution radio/24\um imaging found a population 
of ULIRGs at $2<z<5$ \citep{casey12b}, and a median redshift of $z_{\rm med}=3.38$ (blue 
dashed line in Figure~\ref{fig:redshift}, upper).
It should be noted that this population is only a tail of spectroscopically confirmed \herschel sources in
their survey, the majority of them ($\sim$95\%) being at $z<2$ (Casey et al. 2012a).
%If we only compare sources at $z>2$, 
%, which is not significantly different from our sample. 
Though the sample of $z>2$ \herschel sources in Casey et al. (2012b) is likely 
biased towards optically bright population for which spectroscopic identifications are made, 
the redshift distribution appears similar to that derived from our sample. 
%The probability that these objects and the \ratio-selected sources are drawn 
A Kolmogorov-Smirnov test shows that the probability that both samples are drawn 
from the same underlying parent distribution {  is 0.43}.
Note that while the Casey et al. (2012b) redshift distribution is relatively flat from z=2 to 5, 
our sample shows a tentative excess at $z\simgt4$. 
% by the Kolmogorov-Smirnov test, indicating no significant difference between the two samples.  
%By comparing only $z>2$ sources with the Casey et al. (2012b) sample, there appears to be a discrepancy 
%in the shape of the distribution. 
%The Casey et al. (2012b) distribution is relatively flat from z=2 to 5. 
%There is a deficit of sources in our sample at $z\sim3.5$, 
%and an additional peak at $z\simgt4$ that is not observed in the Casey et al. sample. 
%({  similar distribution??})
This could be due to cosmic variance, since our sample probes a small volume in a single field, 
while the Casey et al. data cover several deep fields over a much larger ($\simgt$5 times) sky area\footnote
{Casey et al. (2012b) sample comes from a spectroscopic survey of bright 
SPIRE sources ($>$10-12 mJy) in four HerMES legacy fields, including the Lockman Hole, COSMOS, ELAIS-N1 and ECDFS (see 
their Table 1).}. 
In fact, Daddi et al. (2009b) have reported a proto-cluster structure
at $z=4.05$ in the  
GOODS-North field (see their Figure 13) which includes three
CO-detected SMGs within $\sim$25\arcsec~at the same redshift. 

Using a \ratio~ratio method, we find {  10} \herschel 500\um sources at $z\simgt4$
% ($S_{500\mu m}\simgt$10 mJy) 
 in the GOODS-North field, corresponding {  to $\sim29$\%} of the whole sample. 
{  Eight} of these $z\simgt4$ candidates have $S_{500\mu m}>10$ mJy.   
This suggests significantly more ULIRGs at the highest-redshift end compared to other surveys 
(e.g., Smol\v{c}i\'{c} et al. 2012 and references therein). 
This will enrich our understanding of obscured star
formation in the early Universe.
%As our selection is not complete, 
Excluding four {spectroscopically confirmed $z>4$ sources at the same redshift} which are likely associated with a protocluster at $z\sim4.05$ (Daddi et al. 2009a, b), 
we can place a lower limit on the surface density of $z>4$ ULIRGs  
at $\simgt6/0.05=120$~deg$^{-2}$.
%, much higher than that derived from 
%the Casey et al. (2012) sample ($\sim$30/deg$^2$). 
%significantly higher than that of any other surveys. 
%Though the GOODS-North field is affected by cosmic variance,
% (of at least a factor of ??? overdensity), as suggested by ...
This value is more than an order of magnitude higher than what is expected in cosmological models 
(Baugh et al. 2005; Swinbank et al. 2008; Coppin et al. 2009; B\'{e}thermin et al. 2012). 
%though the GOODS-North field we probed is affected by cosmic variance. 
%{  so our source density, hence efficiency, is larger. This should said, emphasized and quantified.} 
%a possible discrepancy is that the redshift shift of these sources is relatively flat from z=2 to 5, 
%whereas there is a deficit of our sources at $z\sim3.5$. 
%Interestingly, as shown in Figure~\ref{fig:redshift}, the redshift distribution 
%of our sample is basically identical to that derived by 
%Casey et al. (2012a) 
%\citep{casey12a}  
%for their large spectroscopic survey of 
%250--500\um\ selected galaxies at $2<z<5$. 
%large spectroscopic redshift survey of \herschel sources in the HerMES fields \citep{casey12b}.
%Though the sample from Casey et al. (2012b) is incomplete and not 
%representative of the whole population {\sc SPIRE}-bright sources at $z>2$, 
%the consistency between the redshift distribution of these objects and the 
%{\sc Spire}-selected sources in present sample 
%While we cannot draw conclusion from the difference between these two redshift distributions, 
%given the limited number of sources in the comparison, both 
%Our results confirm that 
%luminous, dusty starburst galaxies  in the early Universe were more abundant than previously thought 
%({  Why? Any number, ref for that}).
%a population of intense starburst galaxies at the high-redshift previously 
%unidentified (or unknown).  
%This will significantly enrich our understanding of obscured star 
%formation in the early Universe. 
However, given the very small area covered by GOODS-N 
and lack of the spectroscopic confirmation for most candidates, 
these values should be taken with caution.  
%However, as only four out of ten $z>4$ candidates have been spectroscopically confirmed, 
%the GOODS-North field we probed is affected by cosmic variance, 
%more conclusive results will have to wait for the study of $z>4$ ULIRGs 
%in a larger field. 
A complete view of the properties of the $z>4$ ULIRG population will require 
a detailed submm follow-up and robustly determined redshifts. 
%a more conclusive result will have to await for 
%detailed submm follow-up and robustly determined redshifts, 
%and search for 
%which is crucial to further constrain the SPIRE population at $z>4$. 
% will require detailed submm follow-up. 
Notably, recent ALMA follow-up of a bright sample of lensed SMGs discovered 
with the SPT yielded at least 10  
spectroscopically confirmed sources at $z>4$ (Vieira et al. 2013; Wei{\ss} et al. 2013), 
challenging the current models for galaxy formation and evolution. 
%redshift distribution which apparently peaks at $z>3$ 
%and at least 10 of the sources are found to lie at $z>4$ 
%(Vieira et al. 2013; Weiss et al. 2013). 
%Moreover, the pre-selection of red {\sc SPIRE} 
%sources from \herschel has resulted in a total of $\sim$40 extreme starburst candidates 
%at $z>4$ ($\sim2$ per deg$^2$ at $S_{500\mu m}>30$ mJy, Dowell et al. 2014), 
%challenging the current models for galaxy formation and evolution. 
%previously thought. 
%
%our map-based 
%method has discovered a new population of high-redshift starbursts, which will 
%enrich our understanding of obscured star formation in the early Universe. 

In Figure~\ref{fig:redshift} (lower panel), we show the measured
$S_{500\mu m}$/$S_{24\mu m}$ ratio (Table 2) for each of our $z\simgt2$ sources
%identified from the ratio map 
as a function of redshift, along with the various model SED predictions. 
The \ratio~ratios of our galaxies are all over 30, 
and are in good agreement with the dusty SEDs allowed by current models, albeit with 
a large scatter. 
The large \ratio~ratios therefore support the expectation that most sources,
if not all, in our sample are indeed at $z\simgt2$. 
%\DIFdelbegin \DIFdel{Notwithstanding uncertainties in our knowledge about the physics of galaxy evolution, 
%our results indicate that the IR SED of distant galaxies is on average 
%similar to that of local star-forming galaxies.}\DIFdelend 
In the next section, we will investigate the far-IR properties of individual 
galaxies.  

\subsection{Infrared luminosities and dust properties}
%In Table 1, we list, when available, the MIR 24\um~through radio 1.4 GHz 
%flux measurements for each of the $S_{500\mu m}$/$S_{24\mu m}$-selected galaxies in 
%our sample. 
To derive the far-IR emission and SED characteristics 
of the sample, we fit the photometry with a library of galaxy 
templates including CE01 and the well-studied SEDs of the 
local starburst M82, the ULIRG Arp220 and the dusty AGN Mrk231. 
The SEDs for the latter three local galaxies are taken from the SWIRE template library (Polletta et al. 2007).
Our library consists of a total of 108 templates with characteristic 
dust temperatures ($T_d$) in the range $T_d\sim20-60$K.
The SED fitting results are shown in Figure D1 in the Appendix, and their infrared luminosities and 
dust temperatures are given in Table 1.  

We compute the infrared luminosity of each source by integrating the best-fit SED 
between rest-frame 8 and 1000$\mu$m. 
{  The luminosities for this sample range from 1.5$\times10^{12}-1.8\times
10^{13}$\lsol, implying infrared SFRs of $\sim$260--3100 \Mpy with a Salpeter IMF. 
Note that the IR luminosities can be determined to less than 50\% accuracy when excluding 
SPIRE 350 and 500\um measurements for 60 per cent of sources where the data at 
longer wavelengths ($>$500\um) are available.} 
This suggests that they are among the most extreme star-forming galaxies seen 
in the early Universe (e.g., Casey et al. 2014 and references therein).
Figure~\ref{fig:lfir} shows the infrared luminosity against redshift (filled red circles). 
% relative to a 500\um detection limit of 10mJy for a given dust temperature.
The distribution of the Casey et al. (2012b) spectroscopically confirmed
sample of $2<z<5$ SPIRE sources is shown for comparison (open circles). 
%The blue and green dashed line represents the detection limit in the 500\um (9mJy at 3$sigma$, 
%Elbaz et al. 2011),  
Our $S_{500\mu m}$/$S_{24\mu m}$--selected population and 
the Casey et al. (2012b) sample each probe different regions of parameter space. 
Though the redshift distribution is similar (Section 3.1), 
our sources are systematically lower in luminosity by $\sim0.1-0.5$ dex at a fixed redshift.
%they probe a $\sim0.1-0.5$ dex fainter population than those in Casey
%et al. (2012b) sample.
%???need to revise to make the reason for the difference clear???
%This may suggest that our $S_{500\mu m}$/$S_{24\mu m}$ ratio method is efficient in detecting 
%more intrinsically fainter objects, which is important for a complete view of the properties 
%of the $z\simgt2$ \herschel population.
%, compared to those in 
%the much larger effective area ($\sim$5 times) probed by Casey et al. (2012b) spectroscopic survey. 
%({  need to compare the projected number densities of $z>3$ sources})  
%In Figure~\ref{fig:lfir}, we also plot the limiting IR luminosity assuming a single modified blackbody with 
%dust temperature $T_{\rm dust}=$30 (blue dashed line) and 40 K (green dashed line) as a function of redshift. 
%The flux limit is chosen to be 9 mJy at 500\ums, 
%representative of the sample in this 
%paper (see Table 1), and is 
%which is the 3$\sigma$ detection limit reported in Elbaz et al. (2011). 
%It is clear that both samples are above the detection boundaries, suggesting 
%It is clear that 
While the Casey et al. (2012b) $z>2$ sample may not be representative of all 
high-$z$ 500\um sources due to their selection of brighter objects for spectroscopy follow-ups,  
our selection is offering a unique probe of intrinsically fainter objects, 
which is important for a complete view of the properties of the $z\simgt2$ \herschel population.  
{Note that, as we mentioned above, it is important to understand the different volumes probed 
for a more meaningful comparison of the typical luminosities in the two samples.}
%at a given detection limit of the \herschel data. 
%under the current detection limit.  
 
\begin{figure}[ht]
\centering
\includegraphics[scale=0.4, angle=-90]{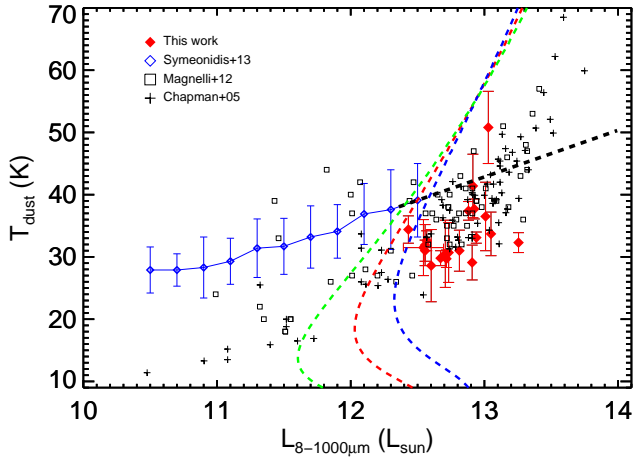}
\caption{
%The {  effective} dust temperature ($T_d$) versus infrared luminosity ($L_{8-1000\mu m}$),
%for the \ratio-selected galaxies discussed here (red solid diamonds). 
%color coded by redshifts. 
%The $T_d$-$L_{\rm IR}$ relation for $z=0-1$ SPIRE-selected LIRGs and ULIRGs (blue diamonds)
%from Symeonidis et al. (2013), and SMGs (black crosses) from Chapman et al. (2005) are plotted for comparison. 
%Recent measurements based on the \herschel data for SMGs from Magnelli et al (2012) are also included (open squares). 
%The large, filled circles show the median temperature in bins of IR luminosity, 
%It appears that the high-$z$ galaxies in our sample have cooler temperatures at fixed luminosity than the
%extrapolation (dashed line) of the $T_d$-$L_{\rm IR}$ relation at $z=0-1$. This 
%may indicate more extended star formation at higher redshifts, as proposed by Hwang et al. (2010). 
%The dotted lines illustrate the luminosity detection limits corresponding to $S_{500\mu\rm m}=9$ mJy as 
%a function of dust temperature, for a galaxy at z=2 (green), z=3 (red) and z=4 (blue), respectively.  
%Joint 99\% confidence contours of the \fekalfa emission line intensity vs. velocity width FWHM, obtained
%from Gaussian fits to the line observed with the \chandra HEG, as described in the text.
%The vertical dotted lines correspond to the FWHM of H$\beta$ line (Wang et al. 2007).}
 Effective dust temperature ($T_d$) versus total far-IR luminosity ($L_{8-1000\mu m}$), for $z<$1 galaxies (blue line and diamonds: SPIRE sample of Symeonidis et al. 2013; black crosses: SMG sample of Chapman et al. 2005), $z$ = 1--3 SMGs$+$Herschel galaxies (open squares, Magnelli et al. 2012) and our $z\simgt$2 \ratio--selected sample (red filled diamonds). Colored dashed lines: transposition of our detection limit of $S_{\rm 500\mu m}$ = 9 mJy for three redshifts, $z$=2 (green), $z$=3 (red) and $z$=4 (blue). 
Black dashed line: extrapolation of the $T_d$-$L_{\rm IR}$ relation for 
$z<$1 SPIRE sample of Symeonidis et al. (2013). 
}
\label{fig:lirvsbb}
\end{figure}

The dust temperature is derived using the code of Casey (2012), 
% which performs a modified blackbody fitting to the far-IR SED, and determines 
 which fits the far-IR photometry with 
an SED consisting of a modified blackbody spectrum and a power-law component in 
the mid-IR portion ($\simlt40$\um). 
The combination of a greybody and mid-IR power-law takes into account 
both galaxy-wide cold dust emission and smaller-scale warm dust emission 
(e.g., Kov\'{a}cs et al. 2006; Casey 2012). 
%The details of the fitting method are described in Casey et al. (2012c).   
%Since the number of far-IR photometric data points is limited, 
To reduce the number of free parameters, 
we fix the slope of the mid-IR power-law to $\alpha=2$ 
for sources which are not detected in {\sc PACS}. 
%, we fix the slope of mid-IR 
%power-law to $\alpha=2$ to reduce the number of free parameters. 
%Since the number of photometric data points sitting on the Rayleigh-Jeans 
%tail in the mm is limited, for consistency in our fittings,
In addition, the emissivity index of the blackbody spectrum on the Rayleigh-Jeans portion 
is fixed to $\beta=1.5$ (a commonly chosen value of $\beta$ in the literature, e.g., 
Chapman et al. 2005; Pope et al. 2005; Younger et al. 2009), 
since {we have not enough data points at longer wavelengths ($\simgt$850\um) 
to constrain $\beta$ in a meaningful way}.  
However, we note that there is very little change in the derived $T_d$ 
by fixing $\beta$ within the range of $1-2$. 

%The dust temperature for each galaxy was measured by determining 
%the peak of the SED via Wien's law, which are less model dependent, 
%so is more easily compared to dust temperatures measured with other 
%alternate techniques (e.g., Chapman et al. 2005; Casey et al. 2012b). 

%The luminosities for this sample range from 1.9?$\times10^{12}-2.2\times
%10^{13}$\lsol, implying infrared SFRs of $\sim$300-4000 \Mpy, suggesting they 
%are among the most extreme star-forming galaxies seen in the early universe (e.g., 
%Bridge et al. 2012). 
%Figure 4 shows the infrared luminosities against redshift relative to the 
%detection limit of a given dust temperature. 
%The distribution of the Casey et al. (2012b) spectroscopically confirmed 
%sample of {\sc SPIRE}-selected sources at $2<z<5$ is shown for comparison. 
%The parameter space probed by our SPIRE/MIPS--selected population is 
%different. Though the redshift distribution is similar, the sources in the 
%present sample are systematically lower in luminosities, i.e., at a given redshift, 
%they probe $\sim0.1-0.5$ dex fainter population than those in Casey et al. (2012b). 

%Figure~\ref{fig:lirbb} 

%Figure~\ref{fig:lirvsbb} plots the infrared luminosity against dust temperature 
{ We present in Figure~\ref{fig:lirvsbb} the dust temperature against infrared luminosity 
}for the $z\simgt2$ sources in our sample which have a meaningful constraint on the dust temperature
from the submm and/or mm measurements.
%at least one measurement in the (sub)-mi
%, color coded by redshifts. 
We include in the comparison the measurements for SCUBA SMGs from Chapman et al. (2005) 
as well as the $z=0-1$ {\sc SPIRE} galaxies from Symeonidis et al. (2013), 
which appear to closely follow the relation that the dust temperature increases with 
infrared luminosity, as observed in the local IRAS 60\ums-selected sample (e.g., Chapman et al. 2003). 
To examine whether our high-$z$ sample has temperatures similar to those of local 
samples and at $z<1$, we divided our sample into two bins of $L_{IR}$ with roughly equal numbers of 
{  sources: 12.2$\leq$log10($L_{\rm IR}$/\lsol)$\leq$12.8 and 12.8$\leq$log10($L_{\rm IR}$/\lsol)$\leq$13.4. 
%The luminosity bins of division are 10$^{(12.2-12.8)}$\lsun} and 10$^{(12.8-13.4)}$\lsun}. 
The derived median temperatures and luminosities are $T_d=31.0\pm1.6$ and 36.5$\pm6.3$~K for $L_{\rm IR}=4.0$ and 
8.7$\times10^{12}$\lsol, respectively. 
}As Figure~\ref{fig:lirvsbb} shows, at a fixed luminosity the high-$z$ galaxies 
in our sample tend to have cooler dust temperatures ($\delta T_d\sim5$K) compared to 
that expected from the 
$L_{IR}-T_d$ relation of $z=0-1$ SPIRE galaxies (Symeonidis et al. 2013). 
%Note that including galaxies without sub(mm) measurements 
%yields similar results. 
This could be partially due to the selection in the 500\um band, 
which is biased against the warmer sources, even at these high redshifts. 
%The dashed lines illustrate the $S_{500\mu m}$=10 mJy  luminosity detection limits  
%as a function of dust temperature, for a galaxy at z=2 (green), z=3 (red) and z=4 (blue).      
%though the errors on dust temperature are large.

%, we find a tendency that 
%dust temperature increases with luminosity, which is  
% ({  need to add Magnelli and Magdis samples for comparison?}).  

%On the other hand, s
%Such a shift of $T_d$ with increasing redshift but similar IR luminosity has also been 
%reported by Hwang et al. (2010), who found modest changes in the $L_{IR}-T_d$ relation 
%as function of the redshift 
%Using a sample of galaxies selected with {\it AKARI} and \herschel, 
By comparing the dust properties between local galaxies ($z<0.1$) 
observed with {\it AKARI}  
and galaxies in the redshift range $0.1\simlt z \simlt2.8$ selected by \herschel,   
Hwang et al. (2010) found modest changes in the $L_{IR}-T_d$ relation 
as a function of redshift. 
\herschel-selected galaxies appear to be 2-5 K colder than that of {\it AKARI}-selected 
local galaxies with similar luminosities.  
%the shift of $T_d$ with increasing 
%redshift but similar IR luminosity 
% by comparing a sample of galaxies selected with \herschel 
%with local galaxies from {\it Akari}. Hwang et al. (2010) also found 
%and a large dispersion in $T_d$.  
%This conclusion is further strengthened
Such evolution of the $L_{IR}-T_d$ relation with redshift has also been reported 
by the studies of Rex et al. (2010) and 
Chapin et al. (2011), and recently by Symeonidis et al. (2013) and Swinbank et al. (2014) 
with a much larger \herschel dataset. 
%Since $T_d$ is closely related to the dust emissivity and the geometry 
%of star-forming regions, the offset in the $L_{IR}-T_d$ relation may imply that 
The offset in the $L_{IR}-T_d$ relation may imply that
high-$z$ ULIRGs have a weaker average interstellar radiation field than 
 local analogs of a similar luminosity, 
which could be due to a more extended distribution of dust and gas (e.g., Hwang et al. 2010). 
{ 
However, we note that dust temperatures derived in this work and others in the literature suppose that the emission-weighted sum of all the dust components could be well fitted by a single modified blackbody model. In reality, the far-IR emission of a galaxy results from the combination of a series of blackbodies associated to a temperature distribution depending on a number of factors, including dust emission spectral index, dust grain distribution and geometry (e.g., Casey 2012, Magnelli et al. 2012), that is often summarized by an effective dust temperature associated to the peak emission of the far-IR SED.
%However, we note the dust temperature derived using a single modified blackbody model in
%this work and literature  
% assumes that the emission-weighted sum of all the dust components could be well fitted. 
%The absolute dust temperature of a galaxy is in fact nontrivial to calculate, as it depends on a 
%number of factors, including dust emission spectral index, dust grain distribution and geometry 
%(e.g., Casey et al. 2012c, Magnelli et al. 2012).  
}

\begin{figure}[ht]
\centering
\includegraphics[scale=0.51, angle=0]{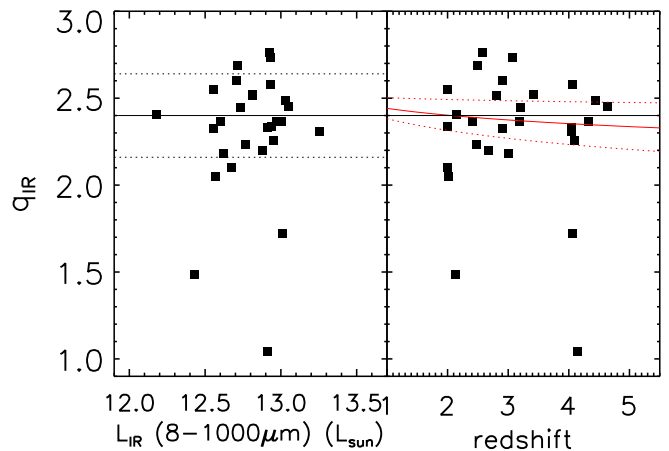}
\caption{
Radio-to-far-IR correlation coefficient ($q_{\rm IR}$) as a function of far-IR luminosity (left) and redshift (right).
The solid and dashed lines represent the median and $2\sigma$ values for
SPIRE sources from Ivison et al. (2010b).
Good agreement is seen with the Ivison et al. (2010b) results, and no
significant evolution with redshift is observed (right). 
The red line shows the moderate redshift evolution of $q\propto(1+z)^{-0.04\pm0.03}$ 
obtained by Ivison et al. (2010b) for comparison.
%Joint 99\% confidence contours of the \fekalfa emission line intensity vs. velocity width FWHM, obtained
%from Gaussian fits to the line observed with the \chandra HEG, as described in the text.
%The vertical dotted lines correspond to the FWHM of H$\beta$ line (Wang et al. 2007).}
}
\label{fig:radiolfir}
\end{figure}

\subsection{High-$z$ Radio/far-IR relation}

%Since our IR luminosities were derived independently of radio luminosity, we measure
For the {  28} radio-detected sources, we measured the far-IR/radio ratio, $q_{\rm IR}$, to investigate the 
far-IR/radio correlation for ULIRGs at $z>2$.  
The tightness of the far-IR/radio relation is very useful in many aspects in the 
study of galaxy evolution, e.g., estimate of the SFR of dusty starbursts in the absence of far-IR data 
(Condon 1992; Barger et al. 2012), 
%crude redshift estimator (Carilli \& Yun 1999; Roseboom et al. 2010) 
and search for radio-excess AGN (Del Moro et al. 2013). 
However, controversial results have been obtained in literature studies on the evolution of such relation. 
%Sargent et al. (2010) showed that, by taking into account the sample selection effect, 
%the FIR/radio relation is roughly invariant up to redshift $z=1$, and possibly up to higher redshift 
%(e.g., Barger et al. 2012; Panella et al. 2014). 
Recent studies based on \herschel data claimed a redshift evolution of the correlation 
(e.g., Ivison et al. 2010a; Casey et al. 2012a, b; Magnelli et
al. 2010, 2012, 2014; Thomson et al. 2014). 
On the contrary, Sargent et al. (2010) showed that, by taking into account the sample selection effect, 
the far-IR/radio relation is roughly invariant up to redshift $z=1$, and possibly up to higher redshift
(in agreement with Bourne et al. 2011; Barger et al. 2012; Pannella et al. 2014). 
Both results are inconsistent with the theoretical prediction that this tight relationship should break down
at high redshift due to rapid Compton cooling of the relativistic electrons (e.g., Condon 1992).
  
We use the ratio ($q_{\rm IR}$) of rest-frame 8--1000\um~flux to 1.4 GHz radio flux as 
defined in Ivison et al. (2010a). The rest-frame radio power was computed assuming 
$S_{\nu}\,\propto\,\nu^{-\alpha}$ and a radio spectral index of $\alpha=0.8$ (e.g., Thomson et al. 2014). 
For these radio detected sources, the measured 
$q_{\rm IR}$ ranges from 1.04 to 2.76 with a median value of 2.37. 
%Only one galaxy in our sample is "radio-loud" and indicative of an AGN (the source at $z=4.4$). 
In Figure~\ref{fig:radiolfir}, we plot the $q_{\rm IR}$ versus (left) their 
far-IR luminosities and (right) their redshifts.
Using the \herschel data in the GOODS North field, Ivison et al. (2010b) performed a stacking analysis of 
24$\mu$m-selected galaxies at $z=0-2$, and found a median value of $q_{\rm IR}=2.4$ 
(solid line in Figure~\ref{fig:radiolfir}) and a scatter of 0.24 
(dashed lines). Our $z\simgt2$ galaxies have $q_{\rm IR}$ values 
in good agreement with this range.
%, which may reflect that our sample is not biased to radio-bright counterparts. 
Only three galaxies (GH500.21, GH500.27c, and GH500.30) in Figure~\ref{fig:radiolfir} lie off the far-IR/radio
correlation. 
This is likely due to the existence of a strong AGN, which contributes
significantly to the radio emission hence lowers the $q_{\rm IR}$ (e.g., Del Moro et al. 2013). 
%with one being
%???Any idea as to why it is outlier to the usual qIR???
%``radio-loud" and indicative of existence of a strong AGN (GN500.22 at $z=4.15$). 
This is consistent with the 
observed excess of 24\um emission in GH500.30 (see Figure D1 in the Appendix).
%This negligible fraction of AGN in the sample is likely due to 
%the fact that AGN are expected bright at the 24\um, our selection, the \ratio ratio, 
We note that our selection is not sensitive to high-$z$ AGN,
which are expected to be bright at 24\um and thus have depressed \ratio~ratios unless they 
are extremely luminous in the far-IR.  

Since the radio continuum of galaxies without strong AGN 
is predominantly produced by supernova remnants from young stellar populations, our 
results suggest that the far-IR luminosity for high-$z$ \herschel sources 
is dominated by heated dust emission due to star-forming process in massive galaxies, 
with little contribution from old stars. 
Parameterizing the $q_{\rm IR}$ evolution in the form $q_{\rm IR}\propto(1+z)^\gamma$, 
and excluding {three} galaxies with a strong AGN contribution to the radio emission,  
we {  find $\gamma=0.05\pm0.05$}, which is consistent with no evolution. 
For comparison, Ivison et al. (2010b) 
%performed a stacking analysis using \herschel data of the GOODS-North 
%for galaxies at $z=0-2$ and 
found a slight evolution with $\gamma=-0.04\pm0.03$ 
(red solid curve with dotted $\pm1\sigma$ uncertainties). 
By extending to higher redshifts (out to $z=4.6$) than most other studies,  
our data suggest no evolution of $q_{\rm IR}$. 
 This agrees with the recent finding by Pannella et al. (2014), 
who studied the far-IR/radio correlation up to $z\simeq4$ in 
a stacking analysis of \herschel data for a mass-selected sample 
of star-forming galaxies. 

\begin{figure}[th]
\centering
\includegraphics[scale=0.61, angle=0]{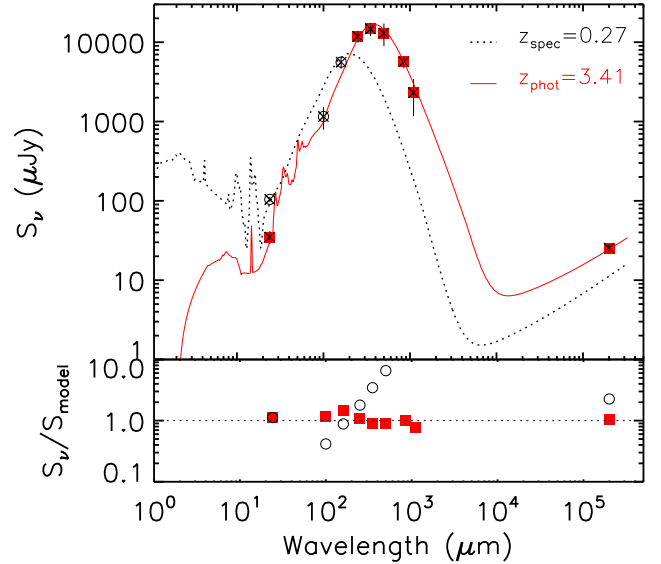}
\caption{
Far-IR SED of a mis-identified $z=3.4$ \herschel source GH500.19 (red squares).   
The best-fit CE01 template at $z=3.4$ is shown by the red line, while the black dotted 
line is the best fit at $z_{\rm spec}=0.27$ 
to $all$ data points from 24\um to radio.   
It is clear that the best-fit SED at $z_{\rm spec}=0.27$ ($\chi^2=51.2$) is much poorer 
than that at $z=3.4$ ($\chi^2=1.38$). 
The $z_{\rm spec}=0.27$ source can only have a significant contribution to the \herschel 
fluxes at shorter wavelengths (black open circles).   
%The 500\um/24\um~ flux ratio as a function of redshift.
%The black line is the SED from from Magdis et al. (2012) for different redshifts and
%the cyan dot-dashed line is the main sequences SED from Elbaz et al. (2011).
%For comparison, we overplot the SED from the local starburst galaxy M82 (purple dashed line).
%The grey shaded region shows the selection of high-$z$ sources, $z\simgt2$ and $S_{500\mu m}$
%/$S_{24\mu m}$\simgt30.
%Joint 99\% confidence contours of the \fekalfa emission line intensity vs. velocity width FWHM, obtained
%from Gaussian fits to the line observed with the \chandra HEG, as described in the text.
%The vertical dotted lines correspond to the FWHM of H$\beta$ line (Wang et al. 2007).}
}
\label{fig:obj4sed}
\end{figure}

\begin{figure*}[th]
\centering
\includegraphics[scale=0.55, angle=0]{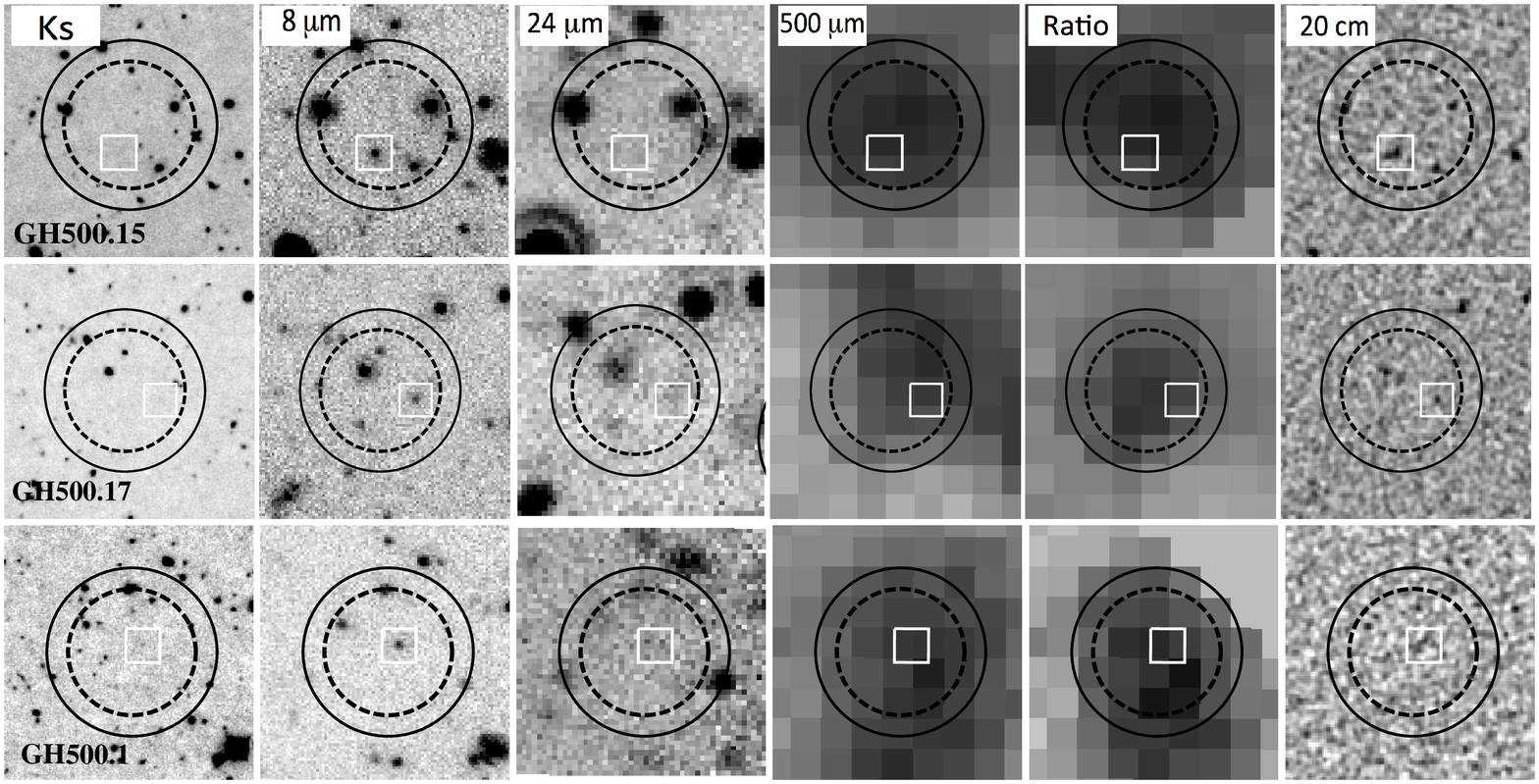}
\caption{
Cutout images for three 500\um sources which are not detected at 24\ums, from left to right: 
ground-based $K_s$-band, IRAC/8\ums, 
MIPS/24\ums, SPIRE/500\ums, \ratio, and VLA 1.4GHz. 
The smaller white square 
%\DIFdelbegin \DIFdel{(8\arcsec\ on one side)} \DIFdelend 
indicates the counterpart identified from the radio or IRAC.
The dashed circle represents the search radius of $15$\arcsec, centered on the
source position determined by SExtractor, while the solid circle indicates the 
beam FWHM of 500\um with a radius of $18$\arcsec.  
The non-detections at 24\um and the red colors  ($K_s$--[3.6\ums]$>$2)
suggest that they are likely to be extremely dust-obscured galaxies at $z\simgt3$.    
%, while the large circle of 20\arcsec, 
The images are 1 arcmin on a side. 
%The far-IR photometry with the best fit CE01 template for each source is shown 
%on the right panel. 
%Submm flux densities as function of the redshift and FIR luminosity.
%Solid and dashed lines show the linear fit to the $S_{\rm 850\mu~m}-L_{\rm IR}$ relation for the
%galaxies detected in submm.
%Radio-to-far-IR correlation ($q_{\rm IR}$) as a function of redshift (left).
%The solid and dashed lines represent the median and $2\sigma$ values for
%SPIRE sources from Ivison et al. (2010b).
%Good agreement is seen with the Ivison et al. (2010b)'s results, and not
%significant evolution with redshift is observed (right), as compared to
%the moderate redshift evolution of $q\propto(1+z)^{-0.04\pm0.03}$ (red lines)
%obtained by Ivison et al. (2010b).
%Joint 99\% confidence contours of the \fekalfa emission line intensity vs. velocity width FWHM, obtained
%from Gaussian fits to the line observed with the \chandra HEG, as described in the text.
%The vertical dotted lines correspond to the FWHM of H$\beta$ line (Wang et al. 2007).}
}
\label{fig:24umdrop}
\end{figure*}

\section{Discussion}
\subsection{Misidentified high-redshift sources in the \herschel catalog}
The large beam sizes (FWHM$\sim$20-40\arcsec) of the {\it Herschel}/{SPIRE} 
images mean that obtaining robust multi-wavelength 
identifications and photometry for SPIRE sources is challenging. 
\herschel point source photometry is frequently performed 
by flux extraction at positions of known 24\um sources or 
1.4~GHz radio sources (e.g., Elbaz et al. 2011; Roseboom et al. 2010, 2012), 
as these wavelengths have much better resolution ($\sim1-5$\arcsec)
and, like {\it Herschel}, are good tracers of star-formation activity (e.g., Ivison 
%are believed to trace star-formation activity as \herschel (e.g., Ivison 
et al. 2010a). 
However, this is particularly problematic for 
 high-redshift sources that are faint or undetected in 24\um and radio surveys, 
%due to the strong $k$-correction with increasing redshift at 24\um, 
resulting in ambiguous and/or incorrect counterpart identifications 
(Roseboom et al. 2010; Yan et al. 2014).  

%In Figure 11, we present an example of the mis-identification 
%of counterpart to the $z\sim3$ \herschel source. 

Our map-based approach has the advantage of identifying 
the correct counterparts to high-$z$ \herschel sources which are bright at 500\um 
but relatively faint at 24$\mu$m. %As mentioned in Section 1, 
%the standard procedure to make the \herschel catalog, if an \herschel 
When using 24\um priors for a \herschel catalog, if the SPIRE
beam contains several blended 24\um sources that cannot be deblended, the brightest 
24\um source 
is usually chosen as the counterpart.
%\footnote{This is the standard
%  procedure used to make the \herschel catalog in the GOODS-North field.}.   
This can lead to misidentifications for high-$z$ sources as we show in 
Figure~\ref{fig:obj4sed} for the case of GH500.19. 
%being misidentified due to blending. 
%We performed far-IR 
%population 
%of high-$z$ 500-\um bright but 24\um faint sources, which would otherwise be missed 
%via the methodology presented above. 
Within the 500\um beam, there are two closely-separated ($<$10\arcsec) 24\um sources, 
one spectroscopically confirmed at $z_{\rm spec}=0.27$ and another at $z_{\rm phot}=3.41$, 
both possibly contributing to the \herschel flux. While the far-IR emission is taken 
to be associated with the galaxy at $z_{\rm spec}=0.27$, the far-IR SED is inconsistent with the low-redshift 
solution. 
%The fit of the CE01 templates to the IR photometry has $\chi^2>50$ for
At $z_{\rm spec}=0.27$, the fit to the IR photometry has $\chi^2>50$ for all the 
templates in the library, with a minimized $\chi^2=51.2$ (dotted line). 
In contrast, the best-fitting template at $z_{\rm phot}=3.41$ yielded 
a much lower $\chi^2=1.38$ (red solid line). 
We therefore conclude that the $z_{\rm phot}=3.41$ galaxy is likely to be 
the true counterpart to the \herschel source, and the low-redshift candidate 
at $z_{\rm spec}=0.27$ 
is ruled out. 
Our method correctly identified the $z_{\rm phot}=3.41$ object as the counterpart, 
while in this case the usual radio identification would favor the low-redshift counterpart, 
which is the brightest radio source within the \herschel beam. 
%(Morrison et al. 2010). 

{  By effectively reducing the confusion from low-$z$ sources using the map-based \ratio~ratio, 
we have identified three potential high-$z$ 500\um sources that are not detected at 24\ums, 
of which two are likely to be at $z>4$.   
In Figure~\ref{fig:24umdrop}, we present the images in the
%the $HST$ $H$-band image from CANDELS, 
ground-based $K_s$ band, 
IRAC 8\ums, MIPS 24\ums, \herschel 500\ums, the \ratio~map, 
and the VLA radio image. 
%The white box denote the position for counterpart which was identified from the radio image 
%The smaller white square indicates the counterpart identified from the radio/IRAC. 
%The dashed circle represents the searching radius of $10$\arcsec, centered on the 
%source position determined by SExtractor.
%has radius of 20\arcsec, roughly the beam FWHM at the 500\um. 
There is no 24\um source within 5\arcsec~of the radio counterpart, 
while a galaxy is clearly detected by {\it Spitzer} in all four IRAC bands. 
All three sources are relatively faint in the $K_s$ band, and the extremely red colors of 
$K_s-[3.6\mu\rm m]>2$, if due to the strong, redshifted Balmer/4000\AA~break of an 
evolved stellar population, imply that they are likely at $z>3$. 
Indeed, %using the $K_s$-based multiwavelength photometric catalog, 
the photometric redshift analysis suggests that two of them are at $z=2.9$ and $z=4.2$ 
respectively. 
The remaining source, which is not seen at $K_s$ and shorter wavelengths, 
is a bright submm source (Pope et al. 2006; Perera et al. 2007; Barger et al. 2014).
%significantly detected in the four {\it Spitzer}/IRAC bands and three \herschel/SPIRE bands. 
Since there is no photometric redshift available for this galaxy,
% based on the $K_s-band$-selected multiwavelength catalog,
we fitted its far-IR to radio photometry with a suite of CE01 templates, 
and found a best-fit redshift of $z=4.4$. 
The high redshifts of 24\ums-dropout \herschel sources which have properties
very similar to the well-studied SMG GN10, if confirmed,
would have important 
implications for galaxy formation and evolution.
}

\begin{figure}[h]
\centering
\includegraphics[scale=0.75, angle=0]{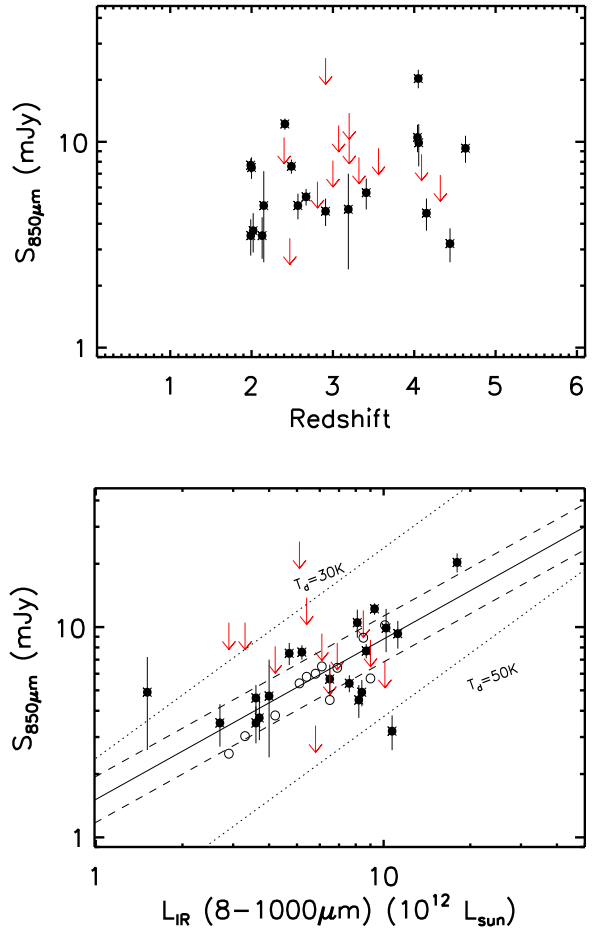}
\caption{
Submm flux densities as a function of redshift (upper panel) and far-IR luminosity (lower panel).
For sources not detected in SCUBA 850\ums, we plot the 3$\sigma$ upper limits on flux (downward arrows) 
and {  IR} SED-predicted 850\um flux densities (open circles). 
Solid and dashed lines show the linear fit (in log space) to the $S_{\rm 850\mu m}-L_{\rm IR}$ relation for the
galaxies detected in the submm.
Dotted lines represent the $S_{\rm 850\mu m}-L_{\rm IR}$ relation for a galaxy at $z=2$, 
with a single modified blackbody at 30 and 50 K, respectively. 
%Radio-to-far-IR correlation ($q_{\rm IR}$) as a function of redshift (left).
%The solid and dashed lines represent the median and $2\sigma$ values for
%SPIRE sources from Ivison et al. (2010b).
%Good agreement is seen with the Ivison et al. (2010b)'s results, and not
%significant evolution with redshift is observed (right), as compared to
%the moderate redshift evolution of $q\propto(1+z)^{-0.04\pm0.03}$ (red lines)
%obtained by Ivison et al. (2010b).
%Joint 99\% confidence contours of the \fekalfa emission line intensity vs. velocity width FWHM, obtained
%from Gaussian fits to the line observed with the \chandra HEG, as described in the text.
%The vertical dotted lines correspond to the FWHM of H$\beta$ line (Wang et al. 2007).}
}
\label{fig:lir_submm}
\end{figure}

%In addition, our approach reduces the confusion noise in a sense of   
%recovering in the ratio map the 500-\um~sources slightly below the nominal 
%confusion limit. 
To summarize, among 36 unique $z\simgt2$ 500\um sources identified in this paper, 
{we find five cases ($\sim$14\%) where the 500\um flux is incorrectly associated with 
a brighter 24\um prior in the official \herschel catalog. 
%The reason for this is that, in cases where two 24\um priors are closer than the FWHM/3 of any given \herschel band, 
%previous studies have systematically favored the brightest of the two as counterpart.  
% have wrong identifications of 
%24\um prior ($\sim$14\%) because of choosing the brightest
%24\um 
%association with {SPIRE}, 
Three high-$z$ 500\um sources that are not detected at 24\um 
are revealed in the \ratio map ($\sim$8\%, see Section 4.2). %and three 500\um sources 
%are recovered in the ratio map whose 
In three further cases, the \herschel 500\um flux in the catalog is probably given to 
nearby sources due to strong source blending ($\sim$8\%)}.
Therefore, using the deepest 24\um and 500\um imaging 
in the GOODS-N field, we estimate that {at least 20\% of $z>2$ dust obscured galaxies in our sample} 
were mis-identified or missed in the current catalog.  
{  This missing fraction of high-$z$ ULIRGs population 
should be taken into account in studies of starburst 
galaxies in the early Universe. }
%for studies of high-$z$ star-forming galaxies.   
%we roughly estimate the mis-identification  
%fraction among the high-$z$ ($z\simgt2$) dust obscured galaxies 
%from the catalog to be $\sim$30\%. 
%these values are subject to large uncertainties. 
Due to the small volume of the GOODS-N field, it is important to apply our method to 
larger survey fields with comparable depths (e.g. HerMES) 
to build a statistically significant and more complete sample of $z\simgt2$ \herschel sources. 
%which is critical to constrain the origin of ULIRGs in the early 
%universe and thus test models for galaxy evolution.
However, detailed, high-resolution submm  
interferometric follow-ups, e.g., with NOEMA or ALMA, are required to 
{\it unambiguously} identify the correct {SPIRE} source counterparts, 
{  and resolve their multiplicities}. 
{ 
%While SMGs at $z\sim1-2$ have been frequently detected by {\it Herschel}, 
%the detection of SMGs by {\it Herschel} at higher redshifts (or vice versa) is not obvious. 
%While \herschel is more sensitive to detect galaxies at their SED peak at $z\sim$1-2, 
%submm (850\um and 1mm) bands have the advantage of detecting higher-redshift galaxies. At the survey limit of HerMES, 
%Casey et al. (2012) estimated that roughly 1/3 of SMGs are not detected by SPIRE at $z>$2. It is therefore 
%not obvious the exact relevance between source detectability at SPIRE and (SCUBA) 850\ums. 
%On the other hand, a fraction of SPIRE galaxies would also not be detectable at a given 850\um flux limit (e.g., $<$5 mJy),  
%due to the well-known dust temperature selection effect at 850\um (e..g, Chapman 2010).  }
While \herschel is highly sensitive to $z\sim$1-2 galaxies whose SED peaks in the far-IR, the long-wavelength submm 
bands (such as 850\um and 1mm) are optimized for higher-redshift
galaxies. 
%At the detection limit of the HerMES \herschel-SPIRE survey, Casey et al. (2012a) estimated that roughly 1/3 of SMGs at $z>$2 are not detected by SPIRE. 
It is thus interesting to compare the \herschel galaxies with submm
sources detected by  SCUBA (for example). 
%not obvious that a submm source detected by SCUBA (for example) should be detectable at 
%SPIRE wavelengths. 
%Conversely, 
%a fraction of SPIRE galaxies would also not be detectable at a given 850\um flux limit (e.g., $<$5 mJy),  
%due to the well-known dust temperature selection effect at 850\um (e.g., Chapman 2010, Magnelli et al. 2012, Magdis et al. 2012). 
%The detection thresholds of SPIRE and (SCUBA) 850\um are therefore not straightforward to compare.}
%submm surveys, e.g. above an 850\um flux density limit of 5 mJy, may miss galaxies with warm dust temperature due to the well-known temperature selection effect at the submm bands (e..g, Chapman 2010, Magnelli et al. 2012, Magdis et al. 2012). The detection thresholds of SPIRE and (SCUBA) 850\um are therefore not straightforward to compare.}
Among our {  34} \herschel sources, 20 have been previously reported as SMGs selected at 850\um 
(Pope et al. 2006; Wall et al. 2008), although three are outside of the SCUBA coverage. 
Therefore, 
%\DIFdelbegin \DIFdel{$\sim$33\%} \DIFdelend \DIFaddbegin \DIFadd{twelve }\DIFaddend 
twelve of our sources are not listed in the 850\um catalog. 
%This could be due to the intrinsic faintness 
In Figure~\ref{fig:lir_submm}, we show the locations of sources with and without 850\um detections 
on the S$_{850\mu m}$-$z$ and S$_{850\mu m}$-$L_{IR}$ planes. 
%\DIFdelbegin \DIFdel{Black filled squares are sources with 850$\mu$m detections, while for non-detections, 3$\sigma$ 
%upper limits (red downward arrows) are shown.} \DIFdelend
We find that there is a correlation between S$_{850\mu m}$ and $L_{IR}$, 
although it is subject to possible selection effects, as submm observations 
could be biased against galaxies with 
warm dust temperatures and therefore relatively faint 850\um flux densities for { \it the same total luminosity} (e.g., Chapman et al. 2010).
}
%\DIFdelbegin \DIFdel{However,} \DIFdelend 

{Using the best-fitting dust SED, the predicted 850\um flux densities (open circles in the 
lower panel of Figure~\ref{fig:lir_submm}) for our \herschel sample appear to be consistent with the 
S$_{850\mu m}$-$L_{IR}$ relation. 
These sources have fitted SED dust temperatures in the range of 30--40 K (Table 1), hence 
their non-detections in 850\um cannot be simply attributed to the selection effect.  
%In fact, most of the non-detections 
%Though the noise of the SCUBA 850\um map is non-uniform across the entire field (Pope et al. 2006), 
%galaxies that are undetectable tend to have 850\um fluxes lower than those detected ones. 
In fact, most of galaxies which have not been detected in 850\um lie at the edge of 
GOODS-North field where the local noise level of SCUBA map is relatively high (e.g., Pope et al. 2006).  
%Since the depths in the SCUBA map are very non-uniform (e.g., Pope et al. 2006), 
%the non-detections are possibly associated with sources with higher local noise (at 
%a given flux). 
The median noise value for the non-detections is 3.1 mJy, compared to the 0.9 mJy for those detected 
by SCUBA in our sample.   
As illustrated in Figure~\ref{fig:lir_submm} (lower panel), 
most of the non-detections could be due to the relatively low S/N given their expected 850\um flux 
densities.  
More sensitive submm imaging of the GOODS-North field, e.g. from SCUBA-2 (Geach et al. 2013), may 
detect the 850\um emission in these high-$z$ \herschel sources and enable a more detailed study of their dust properties. 
}

%most of the have a 3$\sigma$ noise values above 5 mJy. 
% low luminosity galaxies undetectable 
%The dust temperature bias may cause 850\um non-detections for only few brighter sources.  
 
%predicted 850\um flux densities lower than 3$\sigma$ local noise level.    

% where the 850\um observation is less sensitive and has 
%The non-detections for half of sources could be due to that they are at lower redshift, 
%and therefore lower luminosities. 
%The dust temperature bias may cause 850\um non-detections for only few brighter sources.  
% though the dust temperature bias cannot be ruled out.  
%without 850\um detection are located at lower luminosity end, suggesting their 
%intrinsic faintness. 
%The non-detections for sources with higher infrared luminosities 
%There sources are either intrinsically fainter or have hotter dust tempe
%It is of interest to know why the detectability at 850\um of our \ratio~-selected high-$z$ 
%\herschel sources. 
 
%it is well known that the selection of SMGs is subject to strong biases. 

%\subsection{Comparison with {\sc Scuba2} and other {\sc Spire} populations}
%\subsection{How many are missed}
%\subsection{Contribution to Universal star formation}
\subsection{Implications for Star-formation-rate density}

The star-formation-rate density (SFRD) represents the total star formation 
occurring per unit comoving volume.  
The importance of ULIRGs to the buildup of stellar mass 
can be determined by comparing their SFRD contribution to other galaxy populations 
(i.e. UV-selected). This is particularly important for understanding the role 
of dust obscuration at high redshifts. 
At lower redshifts ($z\simlt1$), ULIRGs are rare and their contribution to the SFRD is 
negligible. 
However, the importance of ULIRGs %($L_{IR}>10^{11}$ \lsol) 
grows towards $z\sim2$, where they could contribute half of the total SFRD (e.g., Le Floc'h et al. 2005; 
Magnelli et al. 2013). 
At $z\simgt2$, the contribution of ULIRGs to the SFRD is highly uncertain 
due to their limited numbers (Chapman et al. 2005; Wardlow et al. 2011; Micha{\l}owski et al. 2010a). 
%Our $z\simgt2$ galaxies in this paper selected based on large \ratio~ratios 
Our $z\simgt2$ \herschel sample therefore provides an important opportunity to investigate the SFRD contribution from the high-$z$ ULIRG population.

%\subsubsection{Star formation rate density implications}

To estimate the contribution of our $z\simgt2$ \herschel sources to the cosmic SFRD, we used 
the Kennicutt (1998) relation to convert the far-IR (8-1000\um) luminosity into a dust-obscured SFR, 
assuming a non-evolving Salpeter IMF (Section 3.2). 
{  We split the individual SFR measurements into two redshift bins: $1.9<z<3.1$
and $3.1<z<4.7$ with 18 sources in the former and 16 in the latter.} 
%The SFRD in each redshift bin is measured by dividing SFR for each source 
%in that bin by the accessible volume and summed them. 
The SFRD in each redshift bin was measured by dividing the SFR of each galaxy in that bin 
by the comoving volume and summing them.  
We correct this number by the ratio $V/V_{\rm max}$ for each individual source, where 
%The maximum accessible volume 
$V_{\rm max}$ is the maximum accessible volume within which a source could reside 
and still be detectable by our survey. 
$V_{\rm max}$ is calculated using the survey area of GOODS-North and the 
luminosity detection limit which is determined by setting a flux limit of 9 mJy at 
500\um ($3\sigma$ detection threshold, Elbaz et al. 2011).
%We exclude the two individual sources which have luminosities (log$L_{\rm IR}\simgt13$), 
%as they may be contaminated by an AGN (see Section 3.3). 
%Lensing effect?

\begin{figure}[h]
\centering
\includegraphics[scale=0.49, angle=0]{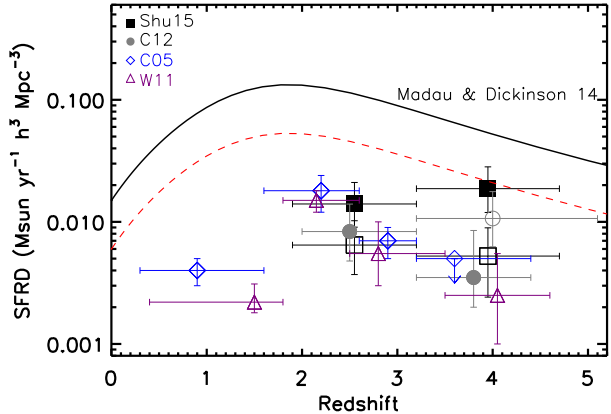}
\caption{
The estimated contribution of \ratio-selected galaxies at $z\simgt$2
(black filled squares) to the cosmic SFRD. 
The open squares are for the spectroscopically confirmed subsample. 
The data points are shown at the median redshift in each redshift bin, and
error bars on SFRD are a combination of $1\sigma$ Poissonian error 
on the number of sources and the typical IR luminosity error of $\sim$0.15 dex.
The black curve shows the UV-based SFRD from a compilation by Madau \& Dickinson (2014),
while the red curve is its rescaling by a factor of 0.4.
The literature SFRD values for comparison are from
Chapman et al. (2005) 850\ums--selected SMGs (blue diamonds), Wardlow et al. (2011)
870\ums--selected SMGs (purple triangles), and {\it Herschel}-{SPIRE} spectroscopically
confirmed $2<z<5$ sources (filled grey circles) from Casey (2012b).
We observe an increase in the ULIRGs' contribution to the SFRD from $z\sim2.5$
to $z\sim4$, consistent with the trend seen in Casey (2012b), when taking into account
the sources with less confident spectroscopic redshifts in their sample (open circle).
%Joint 99\% confidence contours of the \fekalfa emission line intensity vs. velocity width FWHM, obtained
%from Gaussian fits to the line observed with the \chandra HEG, as described in the text.
%The vertical dotted lines correspond to the FWHM of H$\beta$ line (Wang et al. 2007).}
}
\label{fig:sfrd}
\vspace{0.25cm}
\end{figure}

We note that the SFRs of individual sources could be affected by gravitational lensing, 
which would make the observed IR luminosities artificially large, thus overestimate 
the SFRs. 
%The brightest source in our sample (GN500.19, also known as GN20) has $S_{500}=46.6$ mJy, 
%and only three sources are above $\sim$30 mJy (see Table 1). 
%The median value of $S_{500}$ is 14.2 mJy. 
{ The probability of lensing as a function of 500\um flux density (see Wardlow et al. 2013, Figure 9) 
suggests that galaxies with $S_{500}<80$ mJy have a low probability 
%\DIFdelbegin \DIFdel{($<$1\%) of}\DIFdelend \DIFaddbegin \DIFadd 
{of $\lesssim 1\%$ }being lensed by a factor of 
%\DIFdelbegin \DIFdel{$\simgt2$} \DIFdelend \DIFaddbegin \DIFadd
{$\geq 2$}.  
Wardlow et al. (2013) also show that there is little dependence of the conditional magnification probability on redshift at $z>2$. 
Note that the brightest source in our sample (GN500.19, also known as GN20) has $S_{500}=46.6$ mJy, 
and only three sources are above $\sim$30 mJy (see Table 1). 
For the median value of $S_{500}=13.9$ mJy for $z\simgt2$ SPIRE sources in this sample, 
the predicted lensed factor would be less than 2. 
Therefore, the gravitational lensing effect (if there is any) would not change significantly 
the \herschel source contribution to the SFRD. } 
% 
%{  Given the median value of $S_{500}=13.9$ mJy, this model predicts a mean lensing factor for this sample of 
%???, which would change the SFR estimation by a factor of ???. This part is TBD. }
%At the median value of $S_{500}=13.9$ mJy for this sample, 
%Therefore, the lensing is not a significant effect in this sample, given the median value of $S_{500}=13.9$ mJy, 
%and would not change the inferred SFRD.  
%The possibility of source clustering that would affect the SFRD estimate is also low, 
%as there is not obvious trend that sources are strongly clustered on the sky.  
%Note that 
%Schreiber et al. (2014) do measure a flux boosting at the 500\um due to the clustering, 
%of the order of 20\% on average.
Source clustering can boost the flux at the 500\um by $\sim$20\% on average (Schreiber et al. 2015), but %this should have little effect on our SFRD estimation because 
since such clustering only affects sources 
at the same redshift, it does not change our SFRD estimation for the given redshift bin.    
%\subsubsection{Star formation rate density implications}

The contribution of our sample to the SFRD is shown in Figure~\ref{fig:sfrd} as black filled squares. 
We compare our results with previous works based on the SMG samples of 
Chapman et al. (2005, blue diamonds) and Wardlow et al. (2011, purple triangles). 
Using our new technique to explore the 
dusty, high-redshift star-forming galaxies seen 
by {\it Herschel}, 
we recover more of the obscured cosmic star 
formation as shown in Figure~\ref{fig:sfrd}.
% which are not observed before. 
In particular, we observe an increase of the ULIRGs' contribution 
to the SFRD from $z\sim2.5$ to $z\sim4$ (filled squares). 
Note that Casey et al. (2012b) found a similar increase in the SFRD contribution 
from their $z>2$ SPIRE--selected galaxies (open circle in their Figure 9), when
taking into account the sources with less confident spectroscopic redshifts in the bin $3.2<z<5.0$. 
The results imply that extremely dusty starbursts may contribute 
significantly to the build-up of stellar mass in the early Universe 
(about 40\% of extinction-corrected UV samples, e.g., Madau \& Dickinson 2014)\footnote{Our SFRD is strictly a lower limit to the contribution from the overall ULIRG 
population, since we have not corrected for incompleteness in the sample.}. 
%of the SFR density compiled by 
%Madau \& Dickinson (2014) from extinction-corrected UV samples. 
%and dust-obscured IR samples. 
This result may be affected by cosmic variance in 
the GOODS-North field where the number of high-$z$ ULIRGs  
is systematically high compared to other surveys (Section 3.1).  
Conversely, if we consider only the spectroscopically-confirmed subsample, 
the SFRD contribution of our {\it Herschel}--selected ULIRGs (open squares) 
is similar to that of SMGs, which peaks at $z\sim2.2$. 
Future spectroscopic observations of the other candidates in our sample, 
in particular those at $z\simgt4$, 
are therefore required to more robustly constrain their SFRD contribution, 
and to understand the importance of dust-obscured star-forming 
activity in ULIRGs in the first few Gyr of the Universe. 

%confirm their importance in the SFRD constribution. 
%early universe potentially had a very substantial amount 
%of star formation in short-lived, intense bursts ($\sim$1000\sfr). 

%In the $z=2-3$ bin, 
%the contribution from our \herschel/500\um sources to the SFRD is comparable to
%that measured from the SMGs \citep{chapman05,wardlow11a}.
% or1.2\,mm-selected MMGs \citep{roseboom12a}.  
%This is made more
%interesting by the observation that the populations (SMG and HSG) only
%Interestingly, as we discussed in Section 4.3, the SMG population only overlaps 
%with 19 out of 32 \herschel galaxies (60\%), though some of them are undetected 
%due to being out of the SCUBA coverage or in regions of relatively high photometric noise. 
% ??Need a comparison of predicted 850 fluxes from 
%FIR SED with the smm detection limit??TBD.  
%Further work which will use the \ratio~method to search for more $z\simgt2$ \herschel 
%sources in other, larger deep fields,  
%confirming redshifts of $z>$2 {\sc Spire} sources, 
%particularly those without 24\,\um\ counterparts or heavily blended with neighbours, 
%is needed to further more robustly constrain the SFRD contribution, 
%and hence understand the importance of dust-obscured intense star-forming 
%activity ($L_{\rm IR}$\simgt10$^{12}$\lsun)\ in the first few Gyr of the Universe. 

\section{Conclusions}

In this paper we {present} a new, map-based method to select  
candidate $z\simgt2$ ULIRGs using 
the \ratio~color. 
%{\sc Spire}/{\sc Mpis} color.
%{  Compared to the nominal method for counterpart identification and point 
%source photometry of {SPIRE} sources which are required being 
%detectable and bright at 24\um}, o
Our approach {offers}  
a relatively unbiased search for ULIRGs at $z\simgt2$, and recovers a fraction 
of 500\um sources which would otherwise be misidentified and/or missed due to 
incorrect 24\um counterpart associations or non-detections at 24\ums. 
We provide a catalog of {  34} $z\simgt2$ dusty galaxies (all ULIRGs) uniformly selected 
in the GOODS-North field, and the main results are as follows: 
\begin{itemize}

\item From a combination of spectroscopic (comprising 34\% of the sample) and photometric redshifts, we found  
%10 of the 32 galaxies are at $z\simgt4$
that our {\it Herschel}--selected galaxies have a median redshift {  of $z=3.1\pm0.8$ 
and 10 of them are at $z\simgt4$. }
The redshift distribution is different from that of 
%is higher 
%that the redshift distribution of \ratio-selected
 %sources peaks at higher redshifts 
the SCUBA 850\ums-selected SMGs ($S_{850\mu \rm m}\simgt$5 mJy),  
%with a median redshift $z=3.2\pm0.8$. The redshift distribution 
but is consistent with {SPIRE}--selected galaxies between $2<z<5$ from the 
spectroscopic survey of Casey et al. (2012b), %suggesting a higher fraction of 
%high-$z$ dusty star-forming galaxies than previously thought , 
though the latter sample has IR luminosities systematically higher 
by $\sim0.6$dex on average.
%, suggesting that our method is efficient at  
%detecting more intrinsically fainter objects. 
The observed \ratio~colors 
are in agreement with the dusty SEDs allowed for normal low-$z$ star-forming galaxies, 
albeit with a large scatter. 
%follow a trend consistent with the SED of main-sequence  galaxies at $z<2$, suggesting that the IR SED of distant galaxies 
%is on average very similar to the one of normal low-$z$ star-forming galaxies. 

\item We observe a correlation between dust temperature and infrared luminosity. 
The median far-IR luminosity is $L_{\rm far-IR}=6.5\times10^{12}$ 
\lsol, which corresponds to a median SFR$=1100$\sfr, {  with a range of $260-3100$\sfr\ (assuming a Salpeter IMF). 
The median dust temperature is $T_{\rm d}=32.3\pm5.2$K.} 
In comparison with $z\simlt1$ {\it Herschel}-selected galaxies from Symeonidis et al. (2013), 
we find that at a fixed luminosity, our sample has cooler dust temperatures than 
that extrapolated from the $L_{\rm IR}-T_d$ relation at $z\simlt1$, which may 
be attributed to the more extended gas and dust distribution in 
the high-$z$ ULIRGs. 
The result could also be partially explained by the selection at 500\ums, even 
at these high-redshifts, which biases towards galaxies with colder dust temperatures for a fixed luminosity.  

\item The radio-detected subset of our sample {  (28/34}, excluding three
likely strong AGN) follows the far-IR/radio correlation from previous work (Ivison et al. 2010a, b) with a median value of $q_{\rm IR}$=2.37. The result is consistent with no evolution of the correlation with redshift 
out to $z\sim5$.

\item By splitting the sample into two redshift bins ($1.9<z<3.1$ and $3.1<z<4.7$), 
we {find} that the contribution from our sample to the overall SFRD at $z\sim2.5$ is comparable 
to 850$\mu \rm m$-selected SMGs and {\it Herschel}/SPIRE-selected ULIRGs. 
{However, in the higher-redshift bin ($z\sim4$), 
the SFRD contribution of our sample appears to be a factor of $5-10$ higher than 
that of SMGs, and is about 40\% of that of UV-selected star-forming galaxies.} 
We caution that this result may be affected by cosmic variance and apparent overdensity of $z\sim4$ ULIRGs 
in the GOODS-North field. 
Future exploration of the $z\simgt4$ ULIRGs population in larger survey fields 
and spectroscopic follow-ups will be 
crucial to set more robust constraints on their contribution to the cosmic SFRD.   
%The difference may be explained by the fact that our selection is more effectively 
%in selecting higher redshift dusty galaxies, while submm selection is subject from 
%bias against $z>3$ sources, though the effect of cosmic variance should be taken into 
%account. 
 
%It should be noted that since our sample probes a small volume in a single field, 
%the results could be affected by cosmic variance and should be taken with caution. 
 
%\item The radio-detected subset of our sample (22/34, excluding two 
%likely AGN) follow the FIR/radio correlation from previous work (Ivison et al. 
%2010b), but show no evolution of the correlation with redshift.     
\end{itemize}

\acknowledgements
%We wish to thank. 
This work is based on observations taken by the GOODS-Herschel
Program, and is supported by the European Union Seventh Framework Programme (FP7/2007-2013) under grant agreement n. 312725. 
X.W.S. thanks the support 
from Chinese NSF through grant 11103017, 11233002, 11421303, and Research Fund for 
the Doctoral Program of Higher Education of China (No. 20123402110030). 
Partial support for this work was provided by National Basic Research Program 
2015CB857004. R.L. acknowledges support from the FONDECYT grant 3130558.     
%\clearpage

\clearpage
\newpage
%\input{table_coor_0414}
%\begin{landscape}
%\begin{landscape}
\begin{deluxetable}{l@{ }c@{\,}c@{\,}c@{ }c@{ }c@{ }c@{ }c@{ }c@{ }c@{ }c@{ }c@{ }}
\tablecolumns{16}
%\rotate
%\begin{deluxetable}{lcccccccccccccccc}
%\rotate
\tabletypesize{\scriptsize}
%%\begin{landscape}
% \setlength{\tabcolsep}{0.01in}
%\rotate
%\tablewidth{0pt}
\tablecaption{24$\mu m$ identifications of  {Herschel} 500$\mu m$ sources in the GOODS-North field}
\tablehead{
\colhead{Name} & \colhead{RA$_{24\mu m}$} & \colhead{Dec$_{24\mu m}$} &\colhead{$z_{\rm phot}$} & \colhead{$z_{\rm spec}$} &
\colhead{Dist.} & \colhead{$P_{24\mu m}^C$} & \colhead{$P_{3.6\mu m}^C$} & \colhead{$P_{\rm 1.4 GHz}$} & \colhead{$S_{\rm 3.6\mu m }$} & \colhead{$S_{\rm 5.8\mu m }$} \\
%\colhead{350$\mu$m}  & \colhead{500$\mu$m}  & \colhead{850$\mu$m} &
%\colhead{1.1mm} & \colhead{1.4GHz} & \colhead{L$_{\rm FIR}$} & \colhead{$T_{\rm dust}$} \\
%&FIR $K$ & $3.6$\um & $4.5$\um & $5.8$\um & $8$\um \\
%     &             &             &  & ($\mu$Jy)  & (mJy)  & (mJy)  & (mJy)  & (mJy)  & (mJy)  & (mJy)  & (mJy)  &  ($\mu$Jy) & $10^{12}L_{\sun}$ & (K) \\ }
%&   (J2000)   & (J2000) &   & & (arcsec)   &   &  &}
&   (J2000)   & (J2000) &   & & (arcsec)   &   &  & & ($\mu$Jy)& ($\mu$Jy)&\\
(1) &   (2)   & (3) & (4)  & (5) & (6)   & (7)   & (8)  & (9) & (10) & (11) } 
\startdata

GH500.01$^\star$ &  12:35:39.50    &  62:13:11.9 &  4.24 & $\dots$  & 2.9   &  $\dots$ & { 0.0098}  & $\dots$ & 6.6\ppm0.3 & 19.9\ppm1.4 \\%&  GN500.1\\
GH500.02 &  12:35:39.54    &  62:12:43.8 &  3.03 &  $\dots$ & 0.9   &  { 0.057}  & { 0.047} & $\dots$ & 7.5\ppm0.2 & 14.4\ppm1.1\\ %&  GN500.2 \\
GH500.03 &  12:35:53.12    &  62:10:37.2 &  1.53  & 1.60    & 9.7   &  { 0.081}  & { 0.079} &  0.13   & 91.5\ppm0.3 & 100.8\ppm1.0 \\
%GH500.04 &  12:35:53.87    &  62:13:37.2 &  1.19  & 0.87    & 4.6   & 0.011 &  0.097   &  \\
GH500.04         &  12:35:54.28    &  62:13:43.5 &  3.2   & $\dots$    & 6.7   & { 0.018} & { 0.022} & { 0.109}  & 9.6\ppm0.1 & 18.0\ppm0.6   \\
                 &  12:35:53.24    &  62:13:37.5 &  2.8   & $\dots$    & 3.4    & { 0.093} & { 0.069} & { 0.068} & 6.6\ppm0.2 & 9.5\ppm0.5\\ 
GH500.05 &  12:36:05.68    &  62:08:38.4 &  3.21  &  $\dots$   & 8.6   &  0.053 & 0.085 & $\dots$ & 4.5\ppm0.1 & 6.3\ppm0.5\\% passive? z=4.4 src& GN500.3 \\
%GH500.05         &  12:36:06.09    &  62:08:56.3 & 4.44    &  $\dots$   & 10.4 &  $\dots$ & { 0.069} & $\dots$ \\
GH500.06$^\dag$ &  12:36:08.81    &  62:11:43.7 &  1.31  & 1.33    &  7.1  & 0.173  & 0.167 & { 0.056} & 79.7\ppm0.4 & 64.2\ppm0.6\\
%GH500.07$^\dag$ %&  12:36:07.77    & 62:14:49.8  &  2.58  &  $\dots$      & 9.7   & 0.120   & $\dots$ & GN500.4\\
GH500.07$^\dag$ & 12:36:08.60 & 62:14:35.37 & 2.15 &  $\dots$  & 6.1 & 0.228 & 0.212 & { 0.086} & 5.1\ppm0.1 & 10.6\ppm0.4  \\ 
%GH500.08 &  12:36:08.84    &  62:08:04.1 &  3.56 &  $\dots$ & 2.4   &  { 0.0087}  & { 0.0088}  & $\dots$ \\%& GN500.5 \\
%GH500.08         & {  12:36:10.59}    & {  62:08:10.7} &  4.54 &  $\dots$  & 11.8  &  { 0.0033}  & { 0.033} & {  0.04}  \\% uvj=0 
GH500.08  &  { 12:36:08.84}    &  { 62:08:04.1} &  3.56 &  $\dots$ & 2.4   &  { 0.0087}  & { 0.0088}  & $\dots$ & 10.0\ppm0.2 & 23.8\ppm0.6 \\%& GN500.5 \\
& { 12:36:10.59}    & { 62:08:10.7} &  4.54 &  $\dots$  & 11.8  &  { 0.0033}  & { 0.033} & { 0.04}  & 2.5\ppm0.2 & 2.4\ppm0.6 \\% uvj=0 
GH500.09 &  12:36:16.11    &  62:15:13.7 & 2.70  & 2.57 & 1.4  &  { 0.028}  & 0.118  & { 0.05} & 10.5\ppm0.2 & 23.4\ppm0.6 \\%& GN500.7\\
%        &  12:36:16.83    &   62:15:14.4 & 1.78  & 1.99 & 6.5  &  0.28  & 0.09 &\\
GH500.10 &  12:36:16.51    & 62:07:03.0  & 3.07  & $\dots$  & 6.7  &  { 0.0076}  & { 0.014} & 0.13 & 24.6\ppm0.6 & 44.1\ppm0.9 \\%& GN500.6\\
        &  12:36:15.61    &  62:06:43.1 & 4.09   & $\dots$  & 14.1 &  { 0.016}   & { 0.023} & { 0.07} & 2.9\ppm0.2 & 5.2\ppm0.6 \\
%GH500.10 &  12:36:16.11    &  62:15:13.7 & 2.70  & 2.57 & 1.4  &  0.03  & 0.05 & GN500.7\\
%        &  12:36:16.83    &   62:15:14.4 & 1.78  & 1.99 & 6.5  &  0.28  & 0.09 &\\
%GH500.11 &  12:36:18.73   &   62:09:00.6  &  1.61  &  $\dots$ & 0.34  &  0.38  &  $\dots$ & \\
%{  GH500.11}  & 12:36:18.73    & 62:09:00.6     & 1.99 & $\dots$  & 3.5  &  0.099  &   $\dots$ & \\ 
%GH500.10 &  12:36:16.11    &  62:15:13.7 & 2.70  & 2.57 & 1.4  &  0.028  & 0.118  & 0.05 \\%& GN500.7\\
GH500.11  & 12:36:18.49  &  62:09:03.3 & 1.61  &  $\dots$ & 0.15  & { 0.061} & {0.061} & $\dots$ & 57.7\ppm0.2 & 56.8\ppm0.7 \\ 
GH500.12 &  12:36:18.39     &   62:15:50.6  & 1.90 & 2.0 & 2.3  &  { 0.079}  & { 0.056} & 0.011 & 16.2\ppm0.2 & 31.3\ppm0.6 \\% & GN500.8 \\
GH500.13 &  12:36:20.98    &  62:17:09.8    & 1.49 &  1.99 & 2.5   &  { 0.067}  & 0.151 &  0.055 & 14.9\ppm0.1 & 23.5\ppm0.3 \\         %  radio from barger, src034, & 0.01      & GN500.9  \\
         &  12:36:21.27    &  62:17:08.2    &  2.32 & $\dots$  & 0.91 &  0.205  & {0.075} &  0.010 & 13.8\ppm0.1 & 24.6\ppm0.4 \\   % &  0.38  &  $\dots$   & \\  
GH500.14 & { 12:36:22.66}    & { 62:16:29.7}  &  1.92  & 2.47  & 7.2       & { 0.024}  & 0.266 &  {0.03}  & 19.9\ppm0.3 & 35.7\ppm0.9 \\%& GN500.10\\
         &  12:36:22.07    &  62:16:15.9   &  4.63  &$\dots$       & 7.4      & 0.0033 & 0.005  & 0.12  & 3.4\ppm0.2 & 11.2\ppm0.6 \\
GH500.15$^\star$ &  12:36:27.19    &  62:06:05.4   & 4.4  &$\dots$ & 7.2   & $\dots$   &  0.014 & 0.11 & 1.9\ppm0.2 & 5.1\ppm0.7 \\ % & GN500.11 \\
GH500.16$^\dag$ &  12:36:31.29    &  62:09:58.0   & 2.02  &$\dots$ &  1.7   &  0.11  & 0.219 &  0.014 & 13.2\ppm0.2 & 20.9\ppm0.5 \\%& GN500.12\\
GH500.17$^\dag$ &  { 12:36:32.65}    & { 62:06:21.1}   & 2.91   &$\dots$  &  9.1    & $\dots$ & 0.226  & 0.13 & 3.6\ppm0.1 & 9.5\ppm0.4 \\  %GN500.13 \\
                &  12:36:34.53    & 62:06:13.1   & 4.31   &$\dots$  &  11.3   &$\dots$  & 0.055  & $\dots$ & 2.1\ppm0.2 & 2.5\ppm0.6 \\%& GN500.13 \\
%GH500.13 &  12:36:34.53    &  62:06:13.1   & 4.31         &   
GH500.18 &  12:36:33.40    &  62:14:08.5   &    $\dots$    & 4.04  &  1.7     & 0.022  & 0.042  & 0.089  &  0.5\ppm0.1 & 1.4\ppm0.5 \\%& GN500.14\\     
GH500.19 &  12:36:44.03    &  62:19:38.4   & 3.41   &$\dots$           &  6.9     &  0.035 & 0.029 & 0.123  & 2.3\ppm0.09 & 6.9\ppm1.1\\% radio from b08$\dots$ & GN500.15 \\
GH500.20 &  12:36:46.07    &   62:14:48.9  & 2.67    &$\dots$        &  2.6     &  0.062 & 0.121 &  0.02 & 7.9\ppm0.2 & 14.1\ppm0.5 \\%& GN500.16 \\  
GH500.21 &  12:37:00.26    &  62:09:09.9   & 2.13          &   $\dots$ & 12.5  &  0.077  & 0.075  & 0.004 & 22.1\ppm0.2 & 40.8\ppm0.7\\% & \\
GH500.22 &  12:37:00.96    &   62:11:45.6 & 2.01          &  1.76     & 3.92   &  0.036  & 0.058  & 0.022 & 61.5\ppm0.2 & 79.5\ppm0.5 \\%& \\
%GH500.23 &  12:37:02.53    &   62:20:21.4  & 2.23  &   2.22   & 7.6     &  0.032  &  $\dots$ & GN500.17 \\
GH500.23         & { 12:37:01.52}    &  { 62:20:24.8}  & 3.00  &   $\dots$ & 0.39   &  0.060  & 0.038  & {0.049} & 8.7\ppm0.2 & 17.4\ppm0.6\\%\\  & GN500.17 \\
                 &  12:37:02.12    &  62:20:24.9   & 5.92  &    $\dots$ & 3.81  &   $\dots$ & 0.001 &  $\dots$ & 1.1\ppm0.2 & 1.2\ppm0.4\\
%GH500.24$^\dag$ &  12:37:01.27    &   62:21:05.7  &  1.07     &  1.45     &  8.3   & 0.127     &  0.107  \\
GH500.24$^\dag$ &  12:37:01.10     &  62:21:09.6   & 0.79    &  0.8   & 7.9 & 0.87  & 0.316 & 0.005 & 239.8\ppm0.6 & 118.0\ppm1.2 \\
GH500.25 & { 12:37:02.55}    &  { 62:13:02.3}  &  4.44      &  $\dots$  & 2.2    &  0.003  & 0.007 & { 0.12}   & 3.9\ppm0.08 & 8.3\ppm0.3 \\ 
         &  12:37:02.90    &   62:13:07.0  &  4.49      &  $\dots$  & 7.5    & $\dots$  & 0.039 & $\dots$ & 1.5\ppm0.08 & 2.4\ppm0.3 \\
GH500.26 &  12:37:07.21    &   62:14:08.2   & 2.45        & 2.48 &  1.1    &  0.048 & 0.059  &  0.12 & 17.6\ppm0.1 & 28.4\ppm0.6 \\% & GN500.18 \\
%         &  12:37:07.58    &  62:14:09.5    &  5.1        &  $\dots$    &  3.9    & 0.0022 & 0.003   & { 0.10}  & \\
         &  12:37:07.58    &  62:14:09.5    &  5.1        &  2.48    &  3.9    & 0.149 & 0.018  & { 0.10}  & 24.1\ppm0.3 & 51.5\ppm0.8 \\
GH500.27 & { 12:37:11.81}     & { 62:22:12.3 }  &  4.16 &  4.04  &  8.6   &  0.0087   & 0.015  & { 0.046}  & 8.3\ppm0.2 & 19.7\ppm1.0\\%& GN500.19 \\
         & { 12:37:09.73 }   &  { 62:22:02.3}   &  4.10 & 4.04  &  9.1   &   0.022   &  0.014  & { 0.139}  & 7.5\ppm0.3 & 16.8\ppm0.9\\
         &  12:37:08.73   &   62:22:02.5    &  4.39  &  $\dots$  &  13.4    &  0.015  & 0.028  &  $\dots$  & 4.4\ppm0.2 & 3.5\ppm0.4 \\
         &  { 12:37:08.78}   &  { 62:22:01.8}     &  4.19  & 4.04 & 14.7      &    0.021  & 0.025  & { 0.014} & 4.0\ppm0.09 & 4.2\ppm0.4\\
         &  12:37:09.42   &  62:22:14.5    & 1.65    &  1.63  & 11.4    &  0.187&  0.049   &  0.127 & 57.2\ppm0.2 & 68.2\ppm0.9 \\
GH500.28 &  12:37:11.37   &  62:13:31.0     & 1.79   & 1.99  &  1.6    &  0.043   & 0.056 &  0.02 & 54.8\ppm0.1 & 72.9\ppm0.6 \\ %& GN500.20\\
%         &  12:37:12.00   &  62:13:25.6     & 1.86   & 1.99 &  6.2    &   0.116   &  0.06  & \\
%GH500.29 &  12:37:14.29    & 62:12:08.5     & 2.2    & 3.15  & 9.3       & 0.013  &  $\dots$ & GN500.21\\
GH500.29  &  { 12:37:12.05}   &  { 62:12:12.0}     &  2.86    & 2.91 &  7.3     &  0.076   & 0.046  & { 0.10}  & 9.2\ppm0.09 & 17.9\ppm0.5\\
          &  12:37:14.29    & 62:12:08.5     & 2.2    & 3.15  & 9.3       & 0.013  & 0.063   &  $\dots$ & 17.5\ppm0.1 & 26.3\ppm0.5\\ %& GN500.21\\
GH500.30 &  12:37:13.86  &  62:18:26.3     &  4.15    & $\dots$      &  2.1     &  0.010   & 0.011 &  0.003 & 4.4\ppm0.09 & 10.6\ppm0.6\\ %& GN500.22\\
GH500.31 &  {12:37:28.08}  &  { 62:19:20.2}     &  4.32    &$\dots$       &  2.3     &  0.0087    & 0.0033 & { 0.11} & 7.7\ppm0.4 & 16.2\ppm1.6\\%& GN500.23\\
         &  12:37:29.24  &  62:19:28.6     &  4.65    &  $\dots$       & 12.89   & $\dots$   & 0.025  & $\dots$ & 2.9\ppm0.2 & 4.2\ppm0.3\\
%GH500.32 &  12:37:28.55  &  62:14:22.8     &  1.49         & $\dots$  & 3.37     &     &  $\dots$ & \\
GH500.32 &  12:37:28.11  & 62:14:21.82     &  3.32         & $\dots$  & 5.06     &  0.047   & 0.071 & $\dots$ & 1.8\ppm0.08 & 4.7\ppm0.6\\
GH500.33  & 12:37:30.78  &  62:12:58.9     &  2.41      & $\dots$   &  0.9     & 0.071  & 0.102   &  0.017  & 17.1\ppm0.9 & 17.7\ppm2.3\\%& GN500.24 \\
GH500.34  &  { 12:37:38.30}  & { 62:17:36.2}     &  3.13      & 3.19   & 1.5     &  0.059    & 0.193  & { 0.13}  & 4.3\ppm0.2 & 8.6\ppm0.8 \\%& GN500.25 \\
%          &  12:37:39.19  & 62:17:36.5     & 4.02       &  
          & 12:37:39.37   & 62:17:35.0     & 4.33       & $\dots$   &  7.69 & $\dots$   & 0.023  &  $\dots$ & 4.1\ppm0.6 & 3.6\ppm0.9\\
GH500.35  &  12:38:00.27  &  62:16:21.2    &  2.4       & $\dots$    &  4.6     &  0.085    & 0.045 & $\dots$ & 14.4\ppm0.3 & 28.2\ppm0.7\\% & GN500.26  \\
             &  12:38:00.84  &  62:16:11.8    &  2.19      & $\dots$    &  6.5     & 0.094  & 0.051  &   $\dots$ & 40.0\ppm0.2 & 48.3\ppm0.8\\
          &  12:37:58.82  &  62:16:11.7    &  1.98      & $\dots$    &  10.6    &  0.096    & 0.119 &  $\dots$ & 10.2\ppm0.1 & 19.7\ppm0.5\\
%          &  12:38:00.84  &  62:16:11.8    &  2.19      & $\dots$    &  6.5     &     &   $\dots$ & \\
GH500.36$^\dag$  &  12:38:12.54 & 62:14:54.3      &  0.60        & $\dots$    &  3.8     &  0.307 & 0.638   &  0.059 & 126.5\ppm0.6 & 116.4\ppm2.7\\% & GN500.27 \\
%         &  
\enddata
\tablecomments{
Column (1): Name; (2) \& (3): coordinates for the 24$\mu$m counterparts; 
 (4): Photometric redshift;
 (5): Spectroscopic redshift from Barger et al. (2008); Daddi et al. (2009a,b);
 (6): Distance of the 24$\mu$m counterparts from the center of source in the ratio map;  
 (7); Probability $P$ 
{  for random associations of 24$\mu$m counterparts is calculated as $P=1-{\rm exp}[-\pi n \theta^2$], 
where $n$ is the surface density of sources brighter than the predicted 500\um flux,     
$S_{500\rm \mu m}^{\rm Pred}$, and with redshifts above $z$ for a given counterpart of interest (see Section 2.3.2). 
% and $\epsilon$ is the detection 
%efficiency for a source at a given redshift and 500$\mu$m flux from simulations; 
(8): $P$-values for candidate counterparts from the IRAC 3.6\um catalog, which are calculated similar to 
the 24\um catalog, but using the main-sequence relation to predict 500\um flux from stellar masses.  
(9): $P$-values for the radio 1.4 GHz catalogue, but are calculated using $n_{(>\rm S)}$ 
(the surface density of sources brighter than radio flux S$_{1.4\rm GHz}$). 
Traditionally, a radio counterpart is considered to be robust with $P<0.05$.  
%Column (9): Sources presented in our analysis
%.  
$^\star$Counterpart(s) identified from the IRAC 3.6$\mu$m. 
$^\dag$Counterpart(s) identified based on the radio 1.4GHz detection. 
}}
\end{deluxetable}
%\clearpage
%\end{landscape}

%& & & & & &  \\
%\begin{tiny}
%\begin{large}
%\begin{tabular}{lccccccccccccccccc}
%\hline
\clearpage
%\rotate
%\LongTables % optionally
\begin{landscape}
%\begin{landscape}
\begin{deluxetable}{l@{ }c@{\,}r@{\,}c@{ }c@{ }c@{ }c@{ }c@{ }c@{ }c@{ }cc@{ }c@{ }c@{ }c}
\tablecolumns{34}
%\rotate
%\begin{deluxetable}{lcccccccccccccccc}
%\rotate
%\tabletypesize{\scriptsize}
%%\begin{landscape}
\tabletypesize{\footnotesize}
 \setlength{\tabcolsep}{0.01in} 
%\rotate
%\tablewidth{0pt}
\tablecaption{Observed photometry for {Herschel}-selected dust obscured star-forming galaxies}
\tablehead{
\colhead{Name} & \colhead{RA} & \colhead{Dec} &\colhead{$z$} & \colhead{24$\mu$m} & 
\colhead{100$\mu$m} & \colhead{160$\mu$m}  & \colhead{250$\mu$m}  & 
\colhead{350$\mu$m}  & \colhead{500$\mu$m}  & \colhead{850$\mu$m} & 
\colhead{1.1mm} & \colhead{1.4GHz} & \colhead{L$_{\rm FIR}$} & \colhead{$T_{\rm dust}$} \\
%&FIR $K$ & $3.6$\um & $4.5$\um & $5.8$\um & $8$\um \\
     &             &             &  & ($\mu$Jy)  & (mJy)  & (mJy)  & (mJy)  & (mJy)  & (mJy)  & (mJy)  & (mJy)  &  ($\mu$Jy) & $10^{12}L_{\sun}$ & (K) \\ }
\startdata
 %              &                      &                  &                  &                   &                   &                    &                          &                      &                          &                    &                  &                      &        &
GH500.1 & 12:35:39.49 & 62:13:11.7 & 4.24          & $\dots$      & $\dots$     & $\dots$      &5.6$\pm$2.3   &9.3$\pm$2.1   &14.9$\pm$3.9  &$\dots$        &$\dots$      & $\dots$       &6.1   &27.8$\pm$8.8\\
%GH500.2 & 12:35:39.60 & 62:12:44.6 & 3.03           & 54.2$\pm$6.8 & $\dots$     &4.2$\pm$1.0   &8.7$\pm$2.1   &13.0$\pm$2.7  &17.7$\pm$4.1  &$\dots$        &$2.0$          & $\dots$       &4.1   &30.7$\pm$4.7\\
GH500.2 & 12:35:39.60 & 62:12:44.6 & 3.03           & 54.2$\pm$6.8 & $\dots$     &4.2$\pm$1.0   &7.8$\pm$2.1   &10.7$\pm$2.7  &10.6$\pm$4.1  &$\dots$        &$2.0$          & $\dots$       &3.5   &31.5$\pm$4.5\\
GH500.4a &12:35:54.28   & 62:13:43.5 & 3.2  & 149.9$\pm$8.1  & 1.1$\pm$0.3   & 2.8$\pm$0.9 & 12.7$\pm$2.2 & 17.3$\pm$2.3 & 12.6$\pm$3.9 & $<$13.8 & 3.1$\pm$1.1&28.9$\pm$12.1 & 5.4 & 30.7$\pm$4.8\\ 
GH500.4b & 12:35:53.28 & 62:13:37.6 & 2.81  & 32.5$\pm$6.6 & $\dots$     & 3.3$\pm$0.9  & 20.6$\pm$2.2  & 23.2$\pm$2.3 & 15.6$\pm$3.9 & $<$6.4 & $\dots$ &
39.8$\pm$4.5 & 6.5 & 34.5$\pm$8.1 \\ 
%GH500.4b\\ 
%GH500.5 &12:36:05.68  & 62:08:38.5  &  3.2          &37.47$\pm$6.6  &$\dots$      &$\dots$       &6.7$\pm$2.1   &10.2$\pm$2.6   &11.0$\pm$3.9  &$<10.5$         &$\dots$     &$\dots$        &4.6   &28.1$\pm$7.8\\
GH500.5 &12:36:05.68  & 62:08:38.5  &  3.2          &37.47$\pm$6.6  &$\dots$      &$\dots$       &6.7$\pm$2.1   &10.2$\pm$2.6   &11.0$\pm$3.9  &$<10.5$         &$\dots$     &$\dots$        &3.3  &28.4$\pm$9.9\\
GH500.7$^{\dag}$  &12:36:08.57 & 62:14:35.3 & 2.15             & 55.1$\pm$6.7 & $\dots$     &3.1$\pm$0.9   &10.1$\pm$2.2  &11.9$\pm$2.4  &9.7$\pm$3.4$^{\ast}$  & 4.9$\pm$2.3    &2.4$\pm$1.2 & 22.1$\pm$7.1   &1.51 & 23.8$\pm$2.9\\
GH500.8 &12:36:08.84  & 62:08:04.0  & 3.56         &151.2$\pm$8.2  &$\dots$      &3.5$\pm$1.0   &9.3$\pm$2.1   &10.2$\pm$2.7  &11.0$\pm$4.0  &$<9.3$         &$\dots$     &$\dots$        &6.1   &35.2$\pm$7.3\\
%GN500.8b?\\
%GN500.9 &12:36:16.116 & 62:15:13.7& 2.57$^{\star}$   & 322.5$\pm$10.& 5.7$\pm$4.1 &12.0$\pm$4.1  &28.4$\pm$2.1  &29.0$\pm$2.4  &16.2$\pm$4.2  &  4.9$\pm$0.7 & 2.9$\pm$1.1 & 35.8$\pm$4.9  &7.9 & 37.7$\pm$1.6\\
GH500.9 &12:36:16.116 & 62:15:13.7& 2.57$^{\star}$   & 322.5$\pm$10.& 5.7$\pm$4.1 &12.0$\pm$4.1  &28.4$\pm$2.1  &29.0$\pm$2.4  &16.2$\pm$4.2  &  4.9$\pm$0.7 & 2.9$\pm$1.1 & 35.8$\pm$4.9  &8.4 & 37.7$\pm$1.6\\
GH500.10a&12:36:15.7 & 62:06:43.1 & 4.09  & 45.3$\pm$6.7 & $\dots$   & 4.2$\pm$1.0  & 9.3$\pm$2.0 &9.8$\pm$2.8&11.5$\pm$4.1 &$<8.7$& $\dots$ & 43.5$\pm$5.8&9.0&41.0$\pm$8.3\\
GH500.10b &12:36:16.51 & 62:07:03.0  & 3.07 &302.4$\pm$10& 3.1$\pm$4.1& $\dots$ &21.6$\pm$2.0&27.5$\pm$2.8 & 17.2$\pm$4.1& $<12$ & $\dots$ &  26$\pm$4.7  &  8.5&33.8$\pm$4.9\\
%GN500.6b&12:36:15.7 & 62:06:43.1 & 4.09  & 45.3$\pm$6.7 & $\dots$   & 4.2$\pm$1.0  & 9.3$\pm$2.0 &9.8$\pm$2.8&11.5$\pm$4.1 &$\dots$& $\dots$ & 43.5$\pm$5.8&9.0&41.0$\pm$8.3\\
%GN500.10$^\dag$ &12:36:16.116 & 62:15:13.7& 2.57$^{\star}$   & 322.5$\pm$10.& 5.7$\pm$4.1 &12.0$\pm$4.1  &28.4$\pm$2.1  &29.0$\pm$2.4  &16.2$\pm$4.2  &  4.9$\pm$0.7 & 2.9$\pm$1.1 & 35.8$\pm$4.9  &7.9 & 37.7$\pm$1.6\\
%GN500.12 &12:36:18.31 & 62:15:50.5 & 2.0$^{\star}$  & 314.5$\pm$10.& 4.3$\pm$0.4 & 23.4$\pm$1.1 & 42.8$\pm$2.3 & 47.1$\pm$2.1 & 23.6$\pm$4.0 & 7.5$\pm$0.9 & 1.87$\pm$1.1 & 164.4$\pm$6.9 & 5.0 &  29.8$\pm$1.2\\
GH500.12 &12:36:18.31 & 62:15:50.5 & 2.0$^{\star}$  & 314.5$\pm$10.& 4.3$\pm$0.4 & 23.4$\pm$1.1 & 42.8$\pm$2.3 & 47.1$\pm$2.1 & 23.6$\pm$4.0 & 7.5$\pm$0.9 & 1.9$\pm$1.1 & 164.4$\pm$6.9 & 4.7 &  29.8$\pm$1.2\\
GH500.13 &12:36:20.96 & 62:17:10.0 & 1.99$^{\star}$ & 361.3$\pm$10 & 3.98$\pm$0.4 & 15.3$\pm$1.0 &21.6$\pm$2.0 & 26.4$\pm$2.6 & 13.7$\pm$4.1  & 3.5$\pm$0.7 & $\dots$    & 44.9$\pm$2.7   & 3.6   & 31.0$\pm$1.7\\
%GN500.13b?\\
%GN500.14a&12:36:22.10 & 62:16:15.9 & 4.63            & 50.4$\pm$6.8 & $\dots$     & $\dots$     &  8.3$\pm$2.0  & 16.0$\pm$2.5 & 14.3$\pm$4.2 & 9.3$\pm$1.4& $\dots$ &26.1$\pm$6.1 & 10.5 & 33.7$\pm$3.5\\
GH500.14a&12:36:22.66 & 62:16:29.70 & 2.47$^{\star}$ & 412.8$\pm$11 & 3.9$\pm$0.4 & 12.3$\pm$0.9 & 23.7$\pm$2.0 & 21.2$\pm$2.5 & 13.9$\pm$4.2 & $<$3.4& $\dots$ & 91.6$\pm$9.9 & 5.8 & 33.3$\pm$2.5\\
GH500.14b&12:36:22.10 & 62:16:15.9 & 4.63            & 50.4$\pm$6.8 & $\dots$     & $\dots$     &  8.3$\pm$2.0  & 16.0$\pm$2.5 & 14.3$\pm$4.2 & 9.3$\pm$1.4& $\dots$ &26.1$\pm$6.1 & 11.2 & 33.7$\pm$3.5\\
%GN500.10b&12:36:22.10 & 62:16:15.9 & 4.63            & 50.4$\pm$6.8 & $\dots$     & $\dots$     &  8.3$\pm$2.0  & 16.0$\pm$2.5 & 14.3$\pm$4.2 & 9.3$\pm$1.4& $\dots$ &26.1$\pm$6.1 & 10.5 & 33.7$\pm$3.5\\
GH500.15 & 12:36:27.23 & 62:06:05.7  & 4.4$^\ddag$           & $\dots$      & $\dots$      & $\dots$     & 7.3$\pm$2.0  &  14.5$\pm$2.8 & 13.7$\pm$4.2 & 9.5$\pm$2.3  & 4.3$\pm$1.1  & 34.3$\pm$5.9 & 10.0 &  35.2$\pm$5.2 \\
GH500.16 & 12:36:31.23  & 62:09:57.7 & 2.02           &209.4$\pm$8.3&4.9$\pm$3.9  &16.1$\pm$1.0  &25.6$\pm$2.0  &23.5$\pm$2.7  &14.8$\pm$4.2  &$3.7\pm0.8$   & $\dots$    &141$\pm$6.1    &3.7    &32.6$\pm$1.6\\
%GN500.17  & 12:36:32.65 & 62:06:21.1  & 2.91    & $\dots$ & $\dots$      &$\dots$  & 5.5$\pm$2.0 & 6.6$\pm$2.7  & 6.5$\pm$3.2$^{\ast}$   & $<25.5$ & $\dots$ & 23.8$\pm$4.3 & 3.0 & 30.3$\pm$16.1 \\     
%GN500.17  & 12:36:34.53 & 62:06:13.1  & 4.31    & $\dots$ & $\dots$      &$\dots$  &  $\dots$  &  $\dots$   & 7.5$\pm$3.9    & $<18.3$ & $\dots$ &  $\dots$  & 4.91 & $\dots$ \\  
GH500.17  & 12:36:32.65 & 62:06:21.1  & 2.91    & $\dots$ & $\dots$      &$\dots$  & 13.6$\pm$2.0 & 19.5$\pm$2.7  & 17.5$\pm$3.2$^{\ast}$   & $<25.5$ & 3.1$\pm$1.1 & 23.8$\pm$4.3 & 5.1 & 30.5$\pm$5.4 \\      
GH500.18 &12:36:33.42 & 62:14:08.7 & 4.04$^{\star}$  & 30.4$\pm$6.6 & $\dots$     & $\dots$      & 8.2$\pm$2.2  &14.3$\pm$2.7  &13.2$\pm$4.2 & 10.5$\pm$1.6   & 5.4$\pm$0.9 & 34.0$\pm$4.0  & 8.1   &29.1$\pm$2.8\\
GH500.19 & 12:36:44.02 & 62:19:38.4 & 3.41          & 34.9$\pm$6.5  & $\dots$      & $\dots$     &11.9$\pm$2.0  & 14.7$\pm$2.8  &13.1$\pm$4.1  &5.66$\pm$0.97  & 2.3$\pm$1.1  & 25.4$\pm$2.9  &  6.5  & 31.0$\pm$3.3\\
GH500.20 &12:36:46.04 & 62:14:48.6 & 2.67  & 129.8$\pm$8.4& 3.9$\pm$0.4 &16.6$\pm$1.0  &25.9$\pm$2.0  & 24.9$\pm$2.6 &21.1$\pm$4.1 & 5.4$\pm$0.5   &$\dots$      &109.6$\pm$5.3  &7.6  &37.4$\pm$1.5\\
GH500.21 & 12:37:00.26 & 62:09:09.9 & 2.13  & 246.9$\pm$9.1 & 2.0$\pm$0.4&6.6$\pm$0.9 & 16.8$\pm$1.9 & 16.6$\pm$2.8 & 13.0$\pm$4.2 & 3.5$\pm$0.8 &$\dots$ &337.5$\pm$4.3 &2.7 & 34.4$\pm$2.2 \\
GH500.23 & 12:37:01.57 & 62:20:24.9 & 3.0 & 55.7$\pm$6.6 & $\dots$ &  3.1$\pm$0.9  & 10.4$\pm$2.0  &  13.8$\pm$2.8  & 7.7$\pm$3.9$^{\ast}$  & $<8.1$ & $\dots$ & 48.1$\pm$5.1 & 4.2 & 29.8$\pm$6.3\\ 
GH500.25 & 12:37:02.55 & 62:13:02.3 & 4.44 & 79.0$\pm$7.6 & 0.9$\pm$0.3 & 6.0$\pm$0.9 & 11.9$\pm$2.0 & 11.6$\pm$2.7 & 5.2$\pm$2.7 & 3.2$\pm$0.6 & $\dots$ & 25.3$\pm$4.0 & 10.7 & 50.8$\pm$5.8 \\
GH500.26 &12:37:07.18 & 62:14:08.2 & 2.49$^{\star}$  & 248.9$\pm$9.1 & 1.6$\pm$0.4 & 8.6$\pm$0.9      &23.9$\pm$2.0  &29.0$\pm$2.4  &24.7$\pm$4.1 & 7.6$\pm$0.6    & $\dots$     & 28.3$\pm$4.2  &5.2   &29.7$\pm$1.4\\
%GHN2 & 12:37:11.90 & 62:22:12.1  & 4.05$^{\star}$ &  &  
GH500.27a& 12:37:11.89 & 62:22:11.8  & 4.05$^{\star}$ & 84.3$\pm$7.8 & $\dots$      & 6.1$\pm$0.9 & 17.3$\pm$1.9 &  30.1$\pm$2.3 & 46.6$\pm$4.1 & 20.3$\pm$2.1 & 10.7$\pm$9.4 & 79.3$\pm$13. & 18.0 &  32.3$\pm$1.6\\
GH500.27b& 12:37:09.73 & 62:22:02.5 & 4.055$^{\star}$ & 30.2$\pm$6.6 & $\dots$     & 2.9$\pm$0.9  & 12.4$\pm$2.0 & 15.2$\pm$2.4 & 18.2$\pm$4.0 & $\dots$   & $\dots$  & 19.9$\pm$4.0 & 8.5  & 33.9$\pm$5.5    \\
GH500.27c& 12:37:08.77 & 62:22:01.7 & 4.055$^{\star}$& 38.8$\pm$6.5  & $\dots$      & $\dots$     & 14.5$\pm$2.0 &  18.0$\pm$2.4 & 18.2$\pm$4.0 & 9.9$\pm$2.3  & $\dots$      & 172.9$\pm$9.0 & 10.2 & 36.5$\pm$5.5\\
%GN500.27b& 12:37:09.73 & 62:22:02.5 & 4.055$^{\star}$ & 30.2$\pm$6.6 & $\dots$     & 2.9$\pm$0.9  & 12.4$\pm$2.0 & 15.2$\pm$2.4 & 18.2$\pm$4.0 & $\dots$   & $\dots$  & 19.9$\pm$4.0 & 8.5  & 33.9$\pm$5.5    \\     
GH500.28 & 12:37:11.37 & 62:13:31.0 & 1.99$^{\star}$ & 748.0$\pm15.9$ & 11.3$\pm$0.6  & 34.9$\pm$1.5  & 54.1$\pm$2.8    & 49.3$\pm$2.1   & 29.2$\pm$3.9  & 7.7$\pm$0.6 & 4.0$\pm$1.0 & 178.9$\pm12.1$ & 8.7 &33.1$\pm$0.9 \\ 
GH500.29 &12:37:12.04  & 62:12:11.7 & 2.91$^{\star}$ &51.5$\pm$6.7 &1.0$\pm$0.3  &3.6$\pm$1.0   &11.8$\pm$2.0   &10.0$\pm$2.6   &6.5$\pm$4.0$^{\ast}$   &4.6$\pm$0.7    &2.0$\pm$1.0  & 31.7$\pm$4.3  &3.6   &31.9$\pm$3.3\\
%GN500.29b&12:37:14.29   & 62:12:08.6 & 3.15$^{\star}$  &218.8$\pm$8.5&3.2$\pm$0.4  &7.6$\pm$0.9   &12.7$\pm$2.1  &13.9$\pm$2.8  &12.0$\pm$4.0  &$<3.6$        &$\dots$      & $\dots$      &6.4    &42.3$\pm$4.4\\
GH500.30  & 12:37:13.86 & 62:18:26.2 & 4.15          & 63.4$\pm$7.0  & 1.08$\pm$0.4 & 3.5$\pm$1.0 & 10.6$\pm$2.0 & 12.5$\pm$2.8  & 7.3$\pm$3.2$^{\ast}$   & 4.5$\pm$0.8 & 2.4$\pm$1.1 & 627$\pm$8      & 7.8   & 41.4$\pm$5.1\\
GH500.31 & 12:37:28.09 & 62:19:20.1 & 4.32          & 44.1$\pm$6.8  & $\dots$      & 2.2$\pm$0.8 & 12.3$\pm$2.0 & 13.8$\pm$2.9  & 13.4$\pm$4.2 &$<6.9$        & $\dots$     & 33.6$\pm$5.4  &  10.1  & 36.7$\pm$7.8\\
GH500.32 & 12:37:28.11  & 62:14:21.82     &  3.32 & 34.3$\pm$6.6   & $\dots$      & 2.0$\pm$0.8 & 21.1$\pm$2.0 & 21.7$\pm$2.7 & 20.7$\pm$4.0 & $<$8.4      &$\dots$      &$\dots$      &6.9 & 29.5$\pm$3.8 \\
GH500.33$^\dag$ &12:37:30.78  &62:12:58.7  & 2.41           & 184$\pm$7.7  & 4.9$\pm$0.4 &23.4$\pm$1.1  &52.3$\pm$2.3  &55.0$\pm$2.1  &44.8$\pm$3.8  &12.21$\pm$0.77 &4.1$\pm$1.1  & 114$\pm$7     &9.3  &28.3$\pm$0.9\\
%GHN2c& 12:37:08.78 & +62:22:01.8 & 
%GHN3 & 12:37:35.93 & 62:20:43.2 & 2.14           & 179.3$\pm$7.7 & 4.3$\pm$0.4  &10.0$\pm$0.9 & 19.7$\pm$1.9 & 18.7$\pm$2.7  & 9.6$\pm$2.9  &   $\dots$     & $\dots$     &  35.1$\pm$6.0  & 2.91  & 34.8$\pm$2.3\\
%GHN5 & 12:37:28.09 & 62:19:20.1 & 4.32          & 44.1$\pm$6.8  & $\dots$      & 2.2$\pm$0.8 & 10.5$\pm$2.0 & 14.8$\pm$2.9  & 14.2$\pm$4.2 &$\dots$        & $\dots$     & 33.6$\pm$5.4  &  10.0  & 37.2$\pm$7.8\\
%GHN6 & 12:37:22.53 & 62:18:38.0 & 1.55          & 248.2$\pm$9.2 & 3.3$\pm$0.4   & 7.3$\pm$0.9 & 13.4$\pm$2.0 & 13.7$\pm$2.8  & 8.4$\pm$3.2  &$\dots$        & $\dots$     &$\dots$         &2.5    & 31.7$\pm$3.3\\
%GHN8 & 12:37:38.15 & 62:17:37.1 & 3.19$^{\star}$ & 39.3$\pm$6.6 & $\dots$       & $\dots$     & 8.3$\pm$2.1  & 11.4$\pm$2.8  & 11.1$\pm$3.2  & 4.7$\pm$2.3  & 3.4$\pm$1.0 & 26$\pm$5.9     &4.0    & 30.2$\pm$7.0\\
GH500.34  & 12:37:38.15 & 62:17:37.1 & 3.19$^{\star}$ & 39.3$\pm$6.6 & $\dots$       & $\dots$     & 9.6$\pm$2.1  & 12.4$\pm$2.8  & 7.3$\pm$3.2$^{\ast}$  & 4.7$\pm$2.3  & 3.4$\pm$1.0 & 26.0$\pm$5.9     &4.0    & 28.6$\pm$5.8\\
%GHN9$^\dag$ &12:36:20.96 & 62:17:10.0 & 1.99$^{\star}$ & 361.3$\pm$10 & 3.98$\pm$0.4 & 15.3$\pm$1.0 &25.7$\pm$2.0 & 28.3$\pm$2.6 & 17.2$\pm$4.8  & $\dots$      & 3.5$\pm$0.7 & 44.9$\pm$2.7   & 3.8   & 31.2$\pm$2.7\\
%GHN9b &3.4$\pm$0.6 & 164.6$\pm$5; smm and radio 
%GHN11 &12:36:18.31 & 62:15:50.5 & 2.0$^{\star}$  & 314.5$\pm$10.& 4.3$\pm$0.4 & 23.4$\pm$1.1 & 39.7$\pm$2.3 & 44.4$\pm$2.1 & 20.3$\pm$4.2 & 7.5$\pm$0.9 & 1.87$\pm$1.1 & 164.4$\pm$6.9 & 11.2 &  34.6$\pm$1.4\\
GH500.35 & 12:38:00.27 & 62:16:21.2 & 2.4           & 140.5$\pm$8.3& 2.1$\pm$0.4 & 6.0$\pm$1.0  & 11.2$\pm$2.1 & 12.7$\pm$2.8 & 11.6$\pm$3.6  &  $<10.5$    & $\dots$      &  $\dots$      & 2.9    &32.4$\pm$4.9\\
\enddata
%\tablenotetext{$^{\dag}$ Blended with a nearby $z\sim2$ source.}
\tablecomments{
%$^{\dag}$ These 500\um sources have two 24\um counterparts very close each other (within $\sim5$\arcsec), 
%both probably contributing to the SPIRE and submm flux. Therefore, the fluxes we list are for both components. 
%Blended with a nearby $z\simgt2$ source within $\sim5$\arcsec.
%, except for GN500.18 which is bleneded with a $z=4.6$ galaxy.   
{  $^{\dag}$ IR SED fittings are very poor, thus the inferred far-IR luminosity and dust temperature are highly uncertain. 
We did not consider the two sources in our analysis. 
$^{\ddag}$ Redshift from the far-IR SED fit. Although it is likely at high$-z$, we did not include this source in our analysis. }
$^{\star}$ Spectroscopic redshift from Barger et al. (2008); Daddi et al. (2009a,b). 
%$^{\ast}$ Tentative 500\um sources, two of them are detected at $>3\sigma$ in the ratio map. 
}
%There are six sources marginally detected at the 500\um ($<3\sigma$), two of them  
%We seperate the group of sources which are detected less than 3$\sigma$ in the ratio map. }
%\tablenotetext{\ddag}{Redshift from the far-IR SED fittings.}
%\tablenotetext{\star}{Spectroscopic redshift.}
%Star LP 608--62 is also known as BD+1\arcdeg 2341p.  We will
%make this footnote extra long so that it extends over two lines.}
\end{deluxetable}
\clearpage
\end{landscape}

\appendix

\section{Notes on individual sources}
Here we present detailed notes on individual \herschel 500\um sources found in the 
\ratio map. The candidate counterparts in the higher-resolution bands are identified with 
(modified) $P$-statistics (Section 2.3.2 and Table 1). \\

GH500.1: This is a single 500\um source ($S_{\rm 500\mu m}\sim15$ mJy) revealed with 
the \ratio~method. There are two 24\um sources within the search radius of 15\arcsec, but niether has a modified $P$-value less than 0.1. We therefore searched the $Spitzer$/IRAC catalog
for candidate counterparts and identified an IRAC source with photometric 
redshift $z_{\rm phot}=4.24$ as the counterpart based on the $P$-statistics. 
%The weak detection in the $K_s$ band but 
The source is weakly detected in the $K_s$ band, but appears relatively bright in the $Spitzer$/IRAC bands, 
strongly suggesting that it is a candidate high-$z$ dusty star-forming galaxy.\\ 
%It is weakly detected in the $K_s$ band, 
%but appears relatively bright in $Spitzer$/IRAC bands, suggesting a possible candidate at $z>4$.
%The photometric redshift analysis based on a $K_s$-based photometric catalog suggests that 
%the source is at $z_{\rm phot}=4.24$.\\  

GH500.2: There is a single unambiguous 24\um counterpart with photometric redshift of $z_{\rm phot}=3.03$.
There is a faint 1.2 mm source (GN 1200.38, Greve et al. 2008) 3 arcsec away from the 24\um position. 
The source is not detected in the radio or other submm bands. \\

GH500.3: There are seven 24\um sources within the search radius, but $P$-statistics suggest that 
a $z_{\rm spec}=1.6$ source is the likely counterpart to the 500\um source that is detected in 
the \ratio~map. The source is also detected in the radio.\\ 

GH500.4: There are five 24\um sources within the search radius, but only two of them, one at $z_{\rm phot}=3.2$ 
and another at $z_{\rm phot}=2.81$, have a modified $P$-value less than 0.1. While both sources are 
detected in the radio, the one at $z_{\rm phot}=3.2$ is also identified as the counterpart to a 1.1mm source (AzGN13, Chapin et al. 2009).   
The \herschel flux in the official GOODS-\herschel catalog (Elbaz et al. 2011) was given to a nearby 
$z_{\rm spec}=0.98$ source. 
By fitting its far-IR photometry with a suite of CE01 templates, we found that there is a significant 
excess of SPIRE flux compared to that predicted from the best template, suggesting that it is likely 
a wrong association. \\   

GH500.5: There are four 24\um sources within the search radius, but only the one at $z_{\rm phot}=3.2$ 
has a high probability to be associated with the 500\um emission. The source is not detected in the 
radio 1.4 GHz and other submm bands.   
%and another at $z_{\rm phot}=2.3$. 
%Neither of these has 
%a radio counterpart above $5\sigma$. Visual inspection on the radio map suggests weak detections 
%for them.  
%Since the high redshift source is likely a major contribution to the 500\um emission, 
%we consider it as the probable counterpart. 
%However, if we fix the redshift at $z_{\rm phot}=2.3$, fits to the FIR SED with CE01 templates are not good.
%Using the FIR SED fitting method, 
%Allowing the redshift to vary in the fit, we obtained a best-fit redshift of $z=3.59$ for the counterpart, 
%which is reported in Table 1.
\\

GH500.6: There is no source in the 24\um and IRAC 3.6\um catalog with a modified $P$-value less than 0.1. 
Therefore, we searched for a candidate counterpart in the radio 1.4 GHz catalog. There are two 
radio sources within the search radius, one at $z_{\rm phot}=1.54$ and another at $z_{\rm phot}=1.33$. 
Since the latter has a much lower probability ($P_{\rm 1.4 GHz}$=0.056) of being a chance association, 
we consider it as the probable counterpart. This source is also identified as the counterpart to the
SCUBA 850\um emission (GN02, Pope et al. 2006). \\ 

GH500.7: Like GH500.6, there is no source in the 24\um and IRAC 3.6\um catalog with a modified 
$P$-value less than 0.1. The candidate counterpart is therefore identified from the radio 1.4 GHz catalog with 
a photometric redshift $z_{\rm phot}=2.15$. 
The source is also identified as the counterpart to the SCUBA source GN 23 by Pope et al. (2006).
%Within the searching radius, there are two radio sources, one having spectroscopic redshift      
%of 0.8 and the other being at $z_{phot}=2.15$. 
%By fitting the 24\um flux to CE01 templates, we find that the 
%contribution of the low redshift galaxy to 500\um flux is at a level of $5$\% or less.
%Therefore, the 500\um source can be unambiguously associated with the distant counterpart.
%However, if we fix the redshift at $z_{phot}=2.15$, fits to the FIR SED with CE01 templates are not good.
Nevertheless, we found that the fit with $z_{\rm phot}=2.15$ to the far-IR photometric data is poor ($\chi^2$=12.5) and 
there is a significant excess of emission above 250\um compared to the best-fit template.  
{  Allowing the redshift to vary in the fit, we found a much better fit at $z=3.2$ ($\chi^2$=0.49). 
%we adopted as the redshift of the source in this paper. 
Either its photometric redshift was wrong, or it has extremely cold SED that was 
not included in the CE01 libraries.
}
%its far-IR to radio SED is significantly different from the 
The source is also listed in the SCUBA-2 catalog with source name CDFN43, and the 1.1 mm catalog (ID AzGN24), 
for which the same identification was proposed. 
Note that the \herschel flux in the official GOODS-\herschel catalog was assigned 
to a nearby $z=1.46$ source, {  which may suggest severe blending of \herschel flux from multiple counterparts. } \\
%as its 500\um flux is estimated to be no more than 1 mJy. \\

GH500.8: There are three 24\um sources within the search radius, and two of them have low 
$P$-values, with one at $z_{\rm phot}=3.56$ and another at $z_{\rm phot}=4.54$. 
Although the latter is significantly detected at 1.4 GHz, it is further away from the 500\um source centroid 
($\sim$10\arcsec). Our careful flux decomposition suggests 
that the source has little contribution to the 500\um emission ($\sim$15\%). The strong radio emission is likely 
associated with an AGN. We therefore considered the $z_{\rm phot}=3.56$ 
as the most probable counterpart to the 500\um source (S$_{500\mu m}$=11 mJy). \\ 

GH500.9: There are two heavily blended 24\um sources within the search radius, one has a
spectroscopic redshift of $z_{\rm spec}=1.99$ and the other has $z_{\rm spec}=2.578$ (CDFN29, Barger et al. 2014).
The latter has a radio detection and a modified $P$-value less than 0.1, thus was chosen as counterpart.
Barger et al. (2014) also identified it as the counterpart to the 850\um emission (CDFN29, also known as GN04).
%Since two sources are at high redshift and bright at 24\um, 
Since the blending is an issue, there may be a large uncertainty in the 500\um flux.
\\
%though it was not detected in the 
%radio. The 24\um counterpart has photometric redshift of $z=3.24$.  \\

%GHN28:  Like GHN27, there is only one MIPS source in the search radius. 
%There is also a weak radio detection of this source, but below 5$\sigma$ signficance. 
%Deeper radio imaging is able to further localize the counterpart. \\

GH500.10: There are five 24\um sources in the search radius, but only two of them have a 
modified $P$-value less than 0.1. While both sources are also detected in the radio, 
neither is detected in the other submm bands.\\ 
% a radio detection. We note that in this case the source blending is an issue, and 
%other 24\um sources are also capable of contributing to the 500\um flux. 
%There is another radio source ($z_{\rm zphot}=4.1$) 
%just beyond the search radius which has a very 
%faint 24\um counterpart. There appears to be a distinct peak in the \ratio 
%map at the position of the radio source, and we therefore assign it as the 
%additional high-$z$ galaxy candidate. \\ 

GH500.11: Like GH500.9, there are two heavily blended 24\um sources within the search radius, one has a
redshift of $z_{\rm phot}=1.99$ and the other has $z_{\rm phot}=1.61$. 
% (CDFN29, Barger et al. 2014).
The latter has a modified $P$-value less than 0.1, thus was chosen as the counterpart.\\

%GN500.7: There are two heavily blended 24\um sources within the searching radius, one has a
%spectroscopy redshift of $z_{\rm spec}=1.99$ and the other is at $z_{\rm spec}=2.578$ (CDFN29, Barger et al. 2014). 
%The latter has a radio detection.  
%Barger et al. (2014) identified it as the counterpart to the 850\um emission (CDFN29, also know as GN04). 
%Since two sources are at high redshift and bright at 24\um, blending is an issue.       
%Therefore, there is a large systematic uncertainty in the 500\um flux. 
%\\

GH500.12: %The counterpart to this source is identified based on the radio detection.
There are five 24\um sources within the search radius, but only one of them has a modified $P$-value less than 0.1, 
with a spectroscopic redshift of $z_{\rm spec}=2.0$. 
%Given its 24\um flux, the low redshift galaxy is not capable of producing much of 500\um emission.
%The another source has a photometric redshift of 1.68 (within 2.7\arcsec). 
%By fitting the 500\um/24\um flux to a suite of CE01 templates, 
%we find that this second source is not capable of contributing more than ???500\um emission.
The source is also identified as the counterpart to the SCUBA-2 850\um emission based on its strong radio emission 
(CDFN14, Barger et al. 2014). The source is also known as a submm source 850-9 (Barger et al. 2012) or GN 19 (Pope et al. 2006). \\
%Note that the source is so heavily blended with a closely separated $z=1.86$ galaxy ($<$3\arcsec)
%that it is impossible to deblend their \herschel measurements.
%We adopted the total flux densities reported in the GOODS-\herschel catalog as the source photometry. \\

GH500.13:  There are five 24\um sources within the search radius, but only two of them 
are identified as candidate counterparts based on the $P$-statistics, with one at $z_{\rm spec}=1.99$ 
and another at $z_{\rm phot}=2.32$. 
Both sources are also detected in the radio, but separated by only 2.7\arcsec, with
the latter (CDFN34, Barger et al. 2014) being four times brighter than the former. 
% (CDFN34, Barger et al. 2014). 
%
%The fainter radio source has a spectroscopic redshift of 1.99, while the photometric redshift
%for the brighter one is 2.22.
%The latter is consistent the millimetric redshift of 2.79 reported in 
%While the sources cannot be resolved by SCUBA-2, Barger et al. reported 
%the higher resolution SMA measurements, separating two into distinct sources.  
%As it is difficult to separate it into two sources, the \herschel flux is given to the spectroscopically
%confirmed z=1.99 source, which is brighter at 24\um.
While the two sources cannot be resolved by SCUBA (GN7, Pope et al. 2006), Barger et al. presented
higher-resolution SMA measurements, separating it into two distinct sources, each with an
860\um flux of $\sim3.5$ mJy. As it is impossible to deconvolve the \herschel flux, we report the total flux of the two. \\ 
% is given to the spectroscopically
%confirmed z=1.99 source, which is brighter at 24\um.

%An upper limit on the 1.2mm and 850\um for this source is given to be 2 mJy and 14.5 mJy, respectively 
%(Greve et al. 2008). \\ 

GH500.14: Like GH500.13, there are five 24\um sources within the search radius, but only two of them 
are considered as candidate counterparts. 
% There are two radio sources within the searching radius of 10\arcsec, 
% of the "spectral-deconfused" SPIRE source, 
Both sources are detected in the radio, 
with one being four times brighter than the other. The 
brighter radio source has a spectroscopic redshift of 2.47, while the fainter one has 
a photometric redshift of $z=4.63$. 
%Both are likely contributing to the \herschel flux. 
% which is sightly higher than the millimetric redshift of 3.6 (CDFN21, Barger et al. 2014)
Barger et al. (2014) measured the SCUBA-2 850\um flux of $5.2\pm0.7$ mJy for the fainter radio source (CDFN21, also known as GN9).     
%Given their close separation ($$)
Because of their close separation ($<$10\arcsec), the SPIRE flux density from the \herschel catalog is  
given to the $z_{\rm spec}=2.46$, brighter 24\um counterpart. \\

GH500.15: This source is a 24\um dropout galaxy, but 
detected in the AzTEC 1.1 mm Survey of the GOODS-N field (AzGN06, Perera et al. 2007). 
%It was not shown in the SCUBA 850\um catalog in Pope et al. 2006, but listed 
It is also listed 
in the SCUBA-2 sample of 850\um sources (Name: CDFN8, $S_{850\mu m}=9.09$ mJy, Barger et al. 2014). 
%It is not detected in the K$_s$ band catalog of Wang et al. (2010), and hence 
This galaxy is not detected in the K$_s$ or bluer bands, but is detected in all four 
$Spitzer$/IRAC bands, suggesting it is a possible candidate at $z>4$.  
Using the 1.4 GHz to 850\um flux ratio, Barger et al. estimated that the source has 
a millimetric photometric redshift of 3.7. 
By fitting its far-IR to radio photometry with a suite of CE01 templates, we found a  
best-fit redshift of $z=4.4$ for the source.
%The best fit of the CE01 templates is at $z=4.4$, though a redshift range of $z\sim4-5.4$ 
%is also   
\\%As the source was not detected in the 24\um, it is not included in the \herschel catalog. \\ 

GH500.16: This object has a single and clear radio counterpart, and is also detected with 
SCUBA-2 at 850\um (Source ID: CDFN39). \\

%GHN3: This source is not detected in any submm bands. There are three 24um sources with 
%photometric redshifts of around 2 within 10 arcsec of the SPIER source.  
%Our counterpart selection based on the radio emission identifies this source with 
%the $z_{\rm phot}=2.06$ galaxy which is the brightest 24\um counterpart. 
%Note that the two other MIPS sources are faint at 24\um ($\sim30$\uJy), 
%by fitting the 500/24-\um flux to a suite of CE01 templates, we find the combination of 
%two can contribute no more than 10 percent of the 500\um flux. 
% \\

GH500.17: There is single 24\um source at $z_{\rm spec}=1.22$ within the search radius, but the 
probability of this source being a chance alignment with the 500\um emission is high. We therefore 
searched for candidate counterparts in the IRAC 3.6\um catalog, and found a counterpart 
at $z_{\rm phot}=4.31$ based on the $P$-statistics. 
% separated by 15\arcsec. 
%One of them is not detected in the 24\um (dropout). The detected 24\um source 
%has a spectroscopic redshift of 1.21, while the 24\um dropout has a photometric redshift of 
%2.91. 
By fitting the far-IR photometry with a range of CE01 templates, we found that the fit with $z_{\rm spec}=1.22$ 
is very poor ($\chi^2=192$), and { the predicted flux based on the best-fitted SED suggests that  
the source contributes no more than 10\% to the measured 500\um emission.} 
%{  based on the predicted flux from the best-fitted SED}. 
Therefore, the 24\um dropout, at a redshift of $z_{\rm zphot}=4.31$, is likely to be 
the dominant 500\um emitter. 
%Note that this source has also an 1.1 mm detection with flux 3.07 mJy (AzTEC 12), 
%as part of the Chapin et al. (2009) catalog. 
\\ 

GH500.18: %There is a single unambiguous radio counterpart. 
This object is part of the Pope et al. (2006) catalog (GN10) and 
Perera et al. (2008) catalog (AzGN3).  
Similar to GN20, Daddi et al. (2009a) have identified that it has a spectroscopic redshift of 4.042 from a CO emission line. 
The source is also listed in the Greve et al. (2008) 1.2 mm catalog and SCUBA-2 catalog.  
We note that the SPIRE flux densities in the catalog are assigned to a nearby, brighter 24\um source at $z_{\rm spec}=1.45$. 
By fitting the far-IR photometry with a range of CE01 templates, we find that the predicted 500\um emission for 
the $z=1.45$ source is less than 20\% of that reported in the catalog, suggesting an incorrect association, and that most of the 500\um flux should be associated with GN10.  \\

GH500.19: There are three 24\um sources within the search radius, but only one at $z_{\rm phot}=3.41$ 
was identified as a candidate counterpart based on the $P$-statistics. 
It is detected in the AzTEC 1.1 mm imaging by Perera et al. 2007 (source ID: AzGN28), but not in the 
SCUBA map. New SCUBA-2 imaging reveals that the source has an 850\um flux of $5.66\pm0.97$ mJy (ID: CDFN19, 
Barger et al. 2014). 
%The redshift for this object is $z_{\rm phot}=3.41$, consistent with the millimetric redshift of 3.1 estimated by 
%Barger et al.
%The source is also reported in the deeper 1.4 GHz to have a flux of 25.4$\pm$2.9 $\mu$Jy. 
The SPIRE fluxes in the catalog of Elbaz et al. (2011) 
were mis-associated with a $z_{\rm spec}=0.27$ galaxy, 
which is  $\sim10$\arcsec~away from the $z_{\rm phot}=3.41$ source.    \\ 

GH500.20: %Also known as GN 12 in the Pope et al. (2006) catalog and CDFN20 in the Barger et al. (2014) catalog.  
Like GH500.9, there are four heavily blended 24\um sources within the search radius. 
%, threeof which are detected in the radio. 
Only one source at $z_{\rm phot}=2.67$ was identified as the candidate counterpart based on the $P$-statistics. 
This sources is also known as GN 12 in the Pope et al. (2006) catalog and CDFN20 in the Barger et al. (2014) catalog.
Chapin et al. (2009) identified this source as the counterpart to the 1.1mm emission (AzTEC 8), though it is $\sim$7 arcsec 
away from the 1.1mm source centroid. \\ 
%One has a spectroscopy redshift: $z_{\rm spec}=0.55$ and another is 
%estimated photometrically at $z_{\rm phot}=2.67$.
%Since the low redshift source has little contribution to the 500\um flux ($<10$\%), we identified the 
%$z=2.67$ source as the counterpart of the 500\um source.    \\
%, though there is a very bright nearby (S$_{24\mu~m}$=425 \uJy) $z_{\rm spec}=0.83$ source.      

GH500.21: There are four 24\um sources within the search radius, two of which are detected in the radio. 
However, only one object at $z_{\rm phot}=2.13$ was chosen as a candidate counterpart based on the modified 
$P$-statistics computed from the 24\um catalog. 
This source is also identified as the counterpart to the 850\um emission (CDFN44 and GN16, Barger et al. 2014).\\

GH500.22: Like GH500.21, there are five 24\um sources within the search radius, but only one ($z_{\rm phot}=2.01$) 
is a probable counterpart to the 500\um emission. 
This source is also detected in the radio and at 850\um (CDFN22 and GN17, Barger et al. 2014). \\ 

GH500.23: Among three 24\um sources, the one at $z_{\rm phot}=3.0$ is identified as a candidate counterpart. 
There is another IRAC 3.6\um source ($z_{\rm phot}=5.9$) separated by only $\sim4$\arcsec, which possibly contributes 
to the 500\um emission, although it is not detected in the 24\um and radio maps. \\ 
%Like GH500.21, the counterpart to the 500\um emission 
   
GH500.24: There are two heavily blended 24\um sources within the search radius. 
Though both are detected in the radio map, neither has a modified $P$-value less than 0.1 
based on the 24\um and IRAC 3.6\um catalog. We chose the brighter radio source ($z_{\rm spec}=0.8$) as 
the candidate counterpart. 
\\

GH500.25: This source has a single and unambiguous 24\ums/radio identification, 
which is also associated with the SCUBA 850\um emission (GN18, Pope et al. 2006). 
\\

%GN500.17: This object has a single and clear radio counterpart. \\ 

GH500.26: 
This 500\um source appears to have two 24\ums/radio counterparts to the 500\um flux. 
Nevertheless, both objects are confirmed to have a
spectroscopic redshift of 2.484 (Pope et al. 2006), thus 
the 500\um emission comes from two interacting galaxies at the same
redshift.  
While Barger et al. (2014) identified a single counterpart to the 850\um emission (CDFN14), 
Pope et al. (2006) reported two radio identifications to the SCUBA 850\um emission (GN19). 
%At this redshift,
%the separation is 24 kpc and, therefore, we conclude that the
%submm emission is coming from two interacting galaxies at the same
%redshift.
%There are two heavily blended 24\um counterparts (separated by only 3\arcsec), 
%both of them are also detected in the radio. %There are two equally bright radio sources within the searching radius,
%separated by only 3.0 arcsec. Both have 24\um and radio detections.  
%One object has a spectroscopic redshift of 2.49 (Barger et al. 2014) while the other
%is estimated at $z_{\rm phot}=5.13$.
%We assign the spectroscopically confirmed source as the counterpart to the 500\um emission,
%though the contribution from the higher redshift one is also singificant. .
%Note that the source is also detected with SCUBA (known as GN19) and SCUBA-2 (known as CDFN14),        
%for which the same counterpart identification was made.
 \\

GH500.27: There are four 24\um identifications to the 500 \um emission, three of them 
are also known as GN20, GN20a and GN20b. They are associated with a $z=4.05$ protocluster
in the GOODS-North field (Daddi et al. 2009b), and have spectroscopic redshifts of 4.055 based on
CO lines. They are among the brightest submm sources detected with SCUBA. 
There is another IRAC 3.6\um source at a lower redshift ($z_{\rm spec}=1.63$) likely 
contributing to the 500\um emission, though the probability of random association is relatively high 
for the 24\um catalog ($P_{\rm 24\mu m}^C$=0.187). 
\\
%GHN6: A clean $z_{\rm phot}=2.1$ galaxy which is not detected in submm and radio. 
%The counterpart is identified from the 24\um source 0.9" from the center of the ratio map. \\

GH500.28: This object is a significant SCUBA source, GN 39, described in Wall et al. (2008), and
has two radio counterparts, both confirmed to lie at the same redshift ($z_1=1.996$ and $z_2=1.992$, Barger et al. 2008). 
As the two sources are separated by only $\sim$6\arcsec, it is impossible to deconvolve
the SPIRE fluxes, and we report the total flux of the two from the \herschel catalog. \\ 

GH500.29: There are two spectroscopically-confirmed 24\um sources identified as candidate counterparts 
%within the         searching radius 
(one at $z=3.2$ and the other at $z=2.91$).       
The $z=2.91$ source has a radio detection (hence the counterpart identification
is secure), and is associated with a submm source at 850\um (CDFN27, Barger et al. 2014) and     
1.2mm (GN1200.29, Greve et al. 2008).
The $z=3.2$ source is, however, not detected in these (sub)mm bands, but
shows a tentative detection in the radio.
Deeper radio imaging (Owen et al. in preparation) will help to confirm the
counterpart.  \\

GH500.30: The source has a robust identification in the 
24\um and radio catalogs, with a photometric redshift of $z_{\rm phot}=4.15$. 
%However, the SPIRE/500\um flux was given to a bright (S$_{24\mu m}$=425 \uJy) nearby source 
%at $z_{\rm spec}=0.83$. %about 20\arcsec~from the GHN7. 
Note that the source is very bright in the radio (S$_{\rm 1.4 GHz}$=626 \uJy) with a strong radio excess ($q<1.5$), 
suggesting a significant contribution from an AGN. 
The source was reported as an additional 850\um source from the Pope et al. (2006) sample with 
S$_{850\mu m}$= 8.8 mJy (Wall et al. 2008), and was also detected at 1.1mm (ID AzGN26, Perera et al. 2007), 
and 1.2mm (GN 1200.6, Greve et al. 2008).  
However, it was later reported in the SCUBA-2 catalog with an 850\um flux of 4.5 mJy 
(Name CDFN28, Barger et al. 2014). \\ 
%Therefore, the previous measurement of source flux from the SCUBA map likely contains the 
%contamination from the nearby source.  \\

GH500.31: A robust and clean $z_{\rm phot}=4.3$ galaxy identified with a 24\ums/radio counterpart, 
which is, however, not detected in any submm bands. 
Although the $P$-statistics for the IRAC 3.6\um catalog revealed another candidate counterpart 
at $z_{\rm phot}=4.6$, the galaxy is not detected in the radio and is far from the 500\um source centroid, 
thus its contribution to the 500\um emission is marginal. 
\\

GH500.32: There are seven heavily blended 24\um sources within the search radius, but only one at $z_{\rm phot}=3.3$ 
was identified as a candidate counterpart based on the $P$-statistics.  
The source is not detected in the radio and other submm bands. 
\\

GH500.33: The source has a robust and unambiguous counterpart identification the 24\um/radio. 
The galaxy does not appear in the 
Pope et al. (2006) catalog (it is at the edge of SCUBA coverage where the sensitivity is low). 
%probably because of the high noise in the source region. 
Barger et al. (2014) reported the significant SCUBA-2 850\um detection of this source 
(ID: CDFN4) at $z_{\rm phot}=2.41$, with flux of $\sim12$ mJy, making it one of the brightest submm sources ($>10$ mJy) in their catalog. 
%Nevertheless, like GHN16, we found that the fit with $z_{\rm phot}=2.41$ to the far-IR to radio photometry is poor. 
%The best fit of the CE01 templates is at $z=2.63$.  
This source also appears in the 1.1mm catalog of Perera et al. (2008), known as AzGN05. \\

GH500.34: The galaxy has a robust identification in the radio and 24\um with a 
spectroscopic redshift $z=3.19$. 
It is also detected at 850\um, as GN37 in the Pope et al. (2006) catalog, and 
at 1.1mm (ID AzGN9, Perera et al. 2007).  
Like GH500.31, the $P$-statistics for the IRAC 3.6\um catalog revealed another 
candidate at $z_{\rm phot}=4.3$, but it is not detected in the radio. 
We chose the $z=3.19$ source detected in both 24\um and radio as the candidate counterpart. \\ 
 
%The SPIRE/500\um flux in the \herschel catalog is likely underestimated.  \\

%GHN9: Like GHN7, this object was unambiguously identified with the radio and 24\um counterpart 
%at redshift $z_{\rm phot}=3.7$, . 
%The source was also detected at submm bands with SCUBA-2 (ID: CDFN 12) and MAMBO 1.2mm (Name: GN 1200.13). \\

GH500.35: There are three 24\um sources identified as candidate counterparts based on the $P$-statistics, 
%  one of them five times fainter than the other, 
but none of them has a radio or submm detection. 
They are estimated to have photometric redshifts of 2.4, 2.2, and 1.98, and 
may contribute equally to the SPIRE flux. 
Therefore, we are unable to obtain a unique and secure identification. 
However, by inspecting the radio image (Morrison et al. 2010), we find a tentative radio detection at 
the position of source at $z=2.4$. As this source is closest to the peak of the \ratio map, 
we consider it as the tentative counterpart of the 500\um source. Using deeper VLA interferometry, 
% (e.g., Owen et al. 2014) 
or mm interferometric follow-up, it will be possible to further localize the 500\um emission. \\

GH500.36: The source is near the edge of the GOODS-N region, and has a photometric redshift 
of $z=0.6$. 
%The source has a radio and 1.1mm detection (ID 41, Penner et al. 2011). 
However, {  similar to GH500.7, the best-fit with a suite of CE01 library templates 
underestimates SPIRE 350 and 500\um flux significantly. 
Although allowing the redshift to vary yields a better fit ($\Delta\chi^2>10$) at a 
redshift of 2.6,  
%The resulting $\chi^2$ is much better than the fit by fixing the redshift at $z=0.6$ ($\Delta\chi^2>10$). 
it is possible that the source has an unusual IR SED, which is much colder than any CE01 templates.   
}

\section{Notes on Herschel flux measurements}

The \herschel flux measurements for the $z\simgt2$ \ratio-selected ULIRGs 
are mainly from the photometry catalog of Elbaz et al. (2011). 
However, as described in Section 4.1, around 30\% of our sources (including three 
24\um dropouts) have no measured 500\um fluxes in the catalog. 
For these objects, {\it Herschel}/SPIRE flux densities are derived from PSF-fitting 
with GALFIT (Peng et al. 2002), using 24\um prior positions.  
We used a very similar source extraction and photometry method to the one described 
in Elbaz et al. (2011), but fitted a stamp image of size 1.5$\times$1.5 arcmin$^2$ 
centered on the source of interest for each of SPIRE bands. 
In brief, our fittings were performed using an ``iterative" approach. 
We fit simultaneously all 24\um priors above the 5$\sigma$ limit within 
each of the 250, 350 and 500\um stamps, but 
{\it those with derived Herschel fluxes much smaller than the associated flux errors were removed 
from the prior list}. 
The simultaneous fitting was then repeated using the refined list of 24\um priors 
until no such low-signal-to-noise objects were left. 
Note that our restriction of the 24\um priors to the 5$\sigma$ limit would bias against some 
fainter 24\um sources, especially those at high redshift contributing to the SPIRE emission. 
%Our measurements are in good agreement with the alternative
%catalogs used in Elbaz et al. (2011). 
We therefore visually checked the residual image to check for residuals at 
the positions of 24\um sources which were excluded from the original input list. 
If these were found, those objects were added back into the prior list and the fit was re-iterated. 
The flux density for each object was then found from the best fit ($\chi^2/d.o.f<2$). 
%As a final check, we compared the 
%To ensure we have recovered 
%We asses the reliability of our SPIRE photometry through a comparison to 
To assess the reliability of our SPIRE photometry, we performed similar GALFIT fitting to 
all $z\simgt2$ \herschel 500\um sources identified in this paper and 
compared the results to the catalog of Elbaz et al. (2011), as shown Figure B1. 
%The red open squares are our measurements for sources without 
Our measurements are in good agreement with the catalog used in Elbaz et al. (2011),  
except for high-redshift sources whose SPIRE photometry may be significantly under-estimated in the catalog
due to their fluxes being assigned to neighbors with brighter 24\um emission.
%for which the catalog may significantly under-estimate SPIRE photometry for 
%high-$z$ sources by assigning their flux densities to their neighbors with brighter 24\um emission, 
%leading to wrong associations.  

%No spectroscopy information is available, and the source

%The radio source within 6.9" of this SPIRE source is considered as the counterpart

\begin{figure}
\centering
\includegraphics[scale=0.6, angle=0]{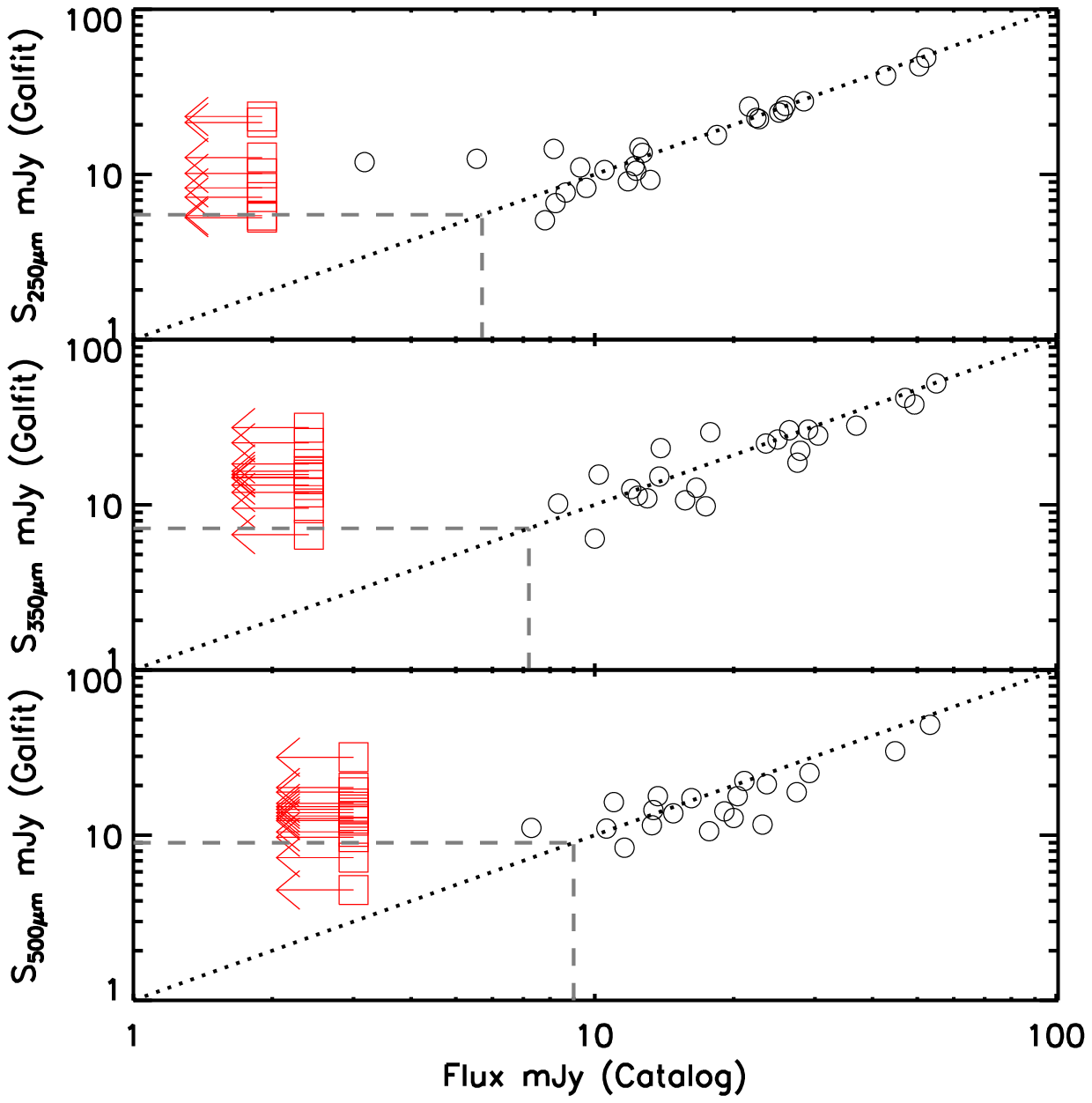}
\caption{
A comparison of flux densities derived from GALFIT with those from the catalog used in Elbaz et al. 
(2011). 
The dotted line in each panel is the one-to-one relation, while the dashed lines 
represent the 3$\sigma$ flux limits of 5.7 mJy, 7.2 mJy and 9 mJy at 250\um, 
350\um and 500\ums, respectively. 
When a source is not detected in the catalog, a 2-$\sigma$ upper limit is shown 
for clarification (red open squares).
%Joint 99\% confidence contours of the \fekalfa emission line intensity vs. velocity width FWHM, obtained
%from Gaussian fits to the line observed with the \chandra HEG, as described in the text.
%The vertical dotted lines correspond to the FWHM of H$\beta$ line (Wang et al. 2007).}
}
\label{fig:galfit}
\end{figure}

%\clearpage
%\begin{figure}
%\centering
%\includegraphics[scale=0.5, angle=0]{/Users/xshu/paper/goodsn/submit/0625/simulate/blending/figure/simulate_dist20.ps}
%\includegraphics[scale=0.5, angle=0]{/Users/xshu/paper/goodsn/submit/0625/simulate/blending/figure/offset_hist.ps}
%\includegraphics[scale=0.5, angle=0]{/Users/mac/paper/goodsn/submit/revise/0921/simulate/blending/figure/simulate_dist20.ps}
%\includegraphics[scale=0.5, angle=0]{/Users/mac/paper/goodsn/submit/revise/0921/simulate/blending/figure/offset_hist.ps}
%\caption{
%Simulations to demonstrate the color doconfusion.  
%Joint 99\% confidence contours of the \fekalfa emission line intensity vs. velocity width FWHM, obtained
%from Gaussian fits to the line observed with the \chandra HEG, as described in the text.
%The vertical dotted lines correspond to the FWHM of H$\beta$ line (Wang et al. 2007).}
%}
%\label{fig:blend}
%\end{figure}

\section{Multiwavelength images of high-redshift Herschel sources}
%We present in Figure C1 the multiwavelength image cutouts of GN10 (Wang et al. 2007, 2009; Daddi et al. 2009a), 
%a $z=4.05$ ULIRG unambiguously revealed with our \ratio ``spectral deconfusion" method. 
%From the left to right: 8\ums, 24\ums, 100\ums, 160\um (top row), 250\ums, 350\ums, 500\ums, and \ratio ratio map (lower row).
%The MIPS 24\um, SPIRE 500\um, \ratio~map and VLA 1.4GHz for all 32 ULIRGs presented in this paper are shown in Figure C2.  %%
%Lower right panel: the false colored 500$\mu$m (red) and 24$\mu$m (blue) image, overlaid with
%the contours of simulated ALMA images with $rms\sim0.1$ mJy.

%\begin{figure}
%\centering
%\includegraphics[scale=0.9, angle=0]{fig3_contour.ps}
%\caption{
%Multiwavelength image cutouts of a high-$z$ source, GN10 (Daddi et al. 2009a) from our sample.
%Each cutout is 30\arcsec on a side. Green square denotes the position for the source, and
%the white circle indicates the beam FWHM of 500\ums.
%The 500\um flux distribution is shown with the red dashed contour, while the red solid
%represents the flux contour in the \ratio ratio map.
%It is evident that the significant 500\um emission of GN10 has been
%unambiguously uncovered in the ratio map, which is otherwise heavily
%blended with nearby sources.
%Joint 99\% confidence contours of the \fekalfa emission line intensity vs. velocity width FWHM, obtained
%from Gaussian fits to the line observed with the \chandra HEG, as described in the text.
%The vertical dotted lines correspond to the FWHM of H$\beta$ line (Wang et al. 2007).}
%}
%\label{fig:contour}
%\end{figure}

\begin{figure}
\centering{
\includegraphics[scale=1.0, angle=0]{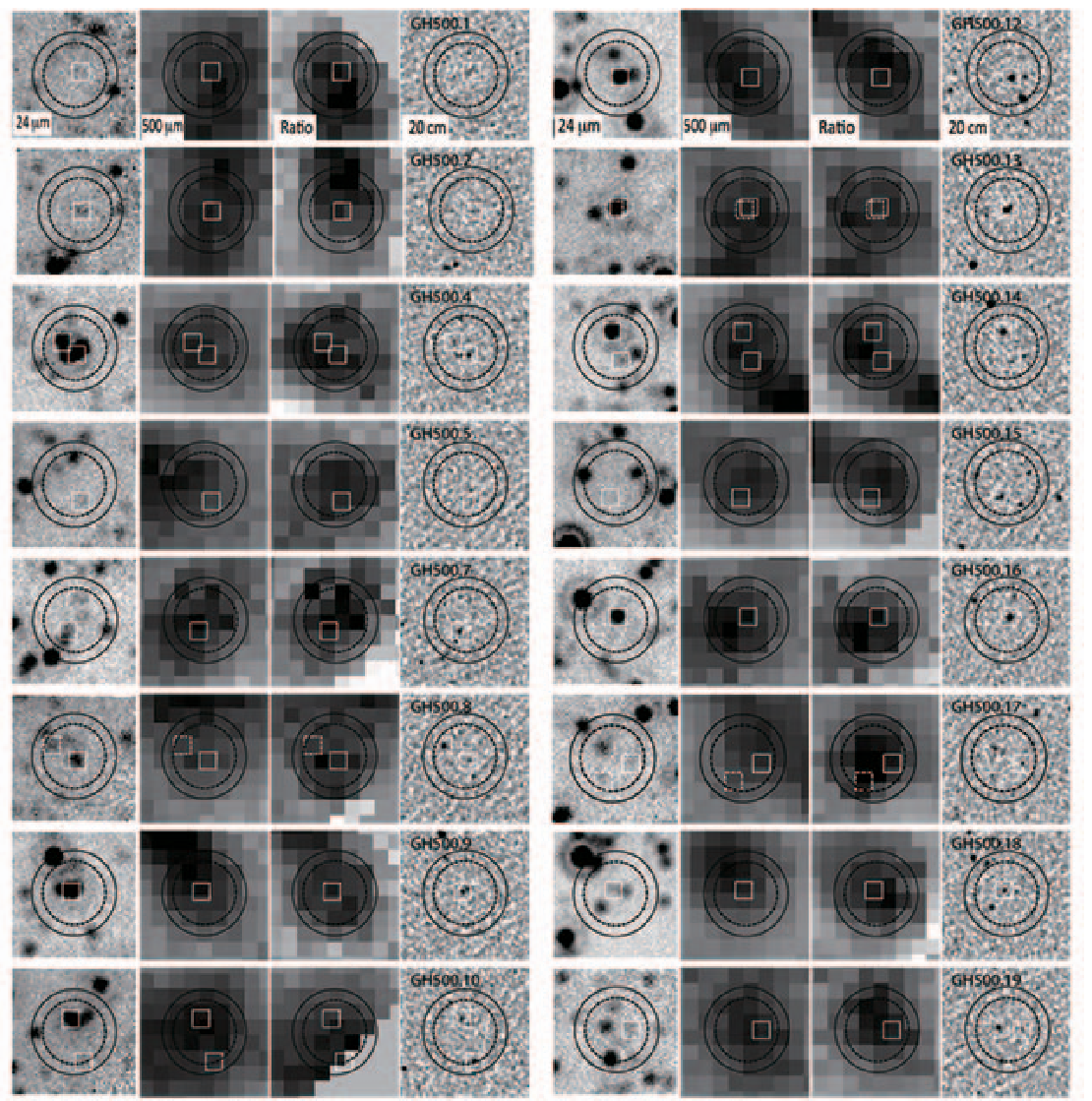}
\caption{
Cutout images of \herschel 500\um counterparts, from left to right: MIPS 24\ums, \herschel 500\ums,
the \ratio\ map, and the VLA 1.4GHz.
All images are 60 arcsec on a side and the black dashed circle indicates the search radius of 15\arcsec.
The smaller white squares are secure counterparts, while the dashed squares represent tentative identifications (Table 1).
%Joint 99\% confidence contours of the \fekalfa emission line intensity vs. velocity width FWHM, obtained
%from Gaussian fits to the line observed with the \chandra HEG, as described in the text.
%The vertical dotted lines correspond to the FWHM of H$\beta$ line (Wang et al. 2007).}
}}\label{fek}
\end{figure}
%\newpage
%\clearpage
\setcounter{figure}{0}
\begin{figure*}[tbh]
\vspace{10pt}
%\centerline{\psfig{file=figure/ratioplot/contour/GHN_img_comb2.eps,width=6.5in,height=7.in,angle=0}}
%\includegraphics[scale=0.4, angle=0]{../figure/ratioplot/contour/newimage/newfigure0428/gh_500_img2.eps}
\includegraphics[scale=1.0, angle=0]{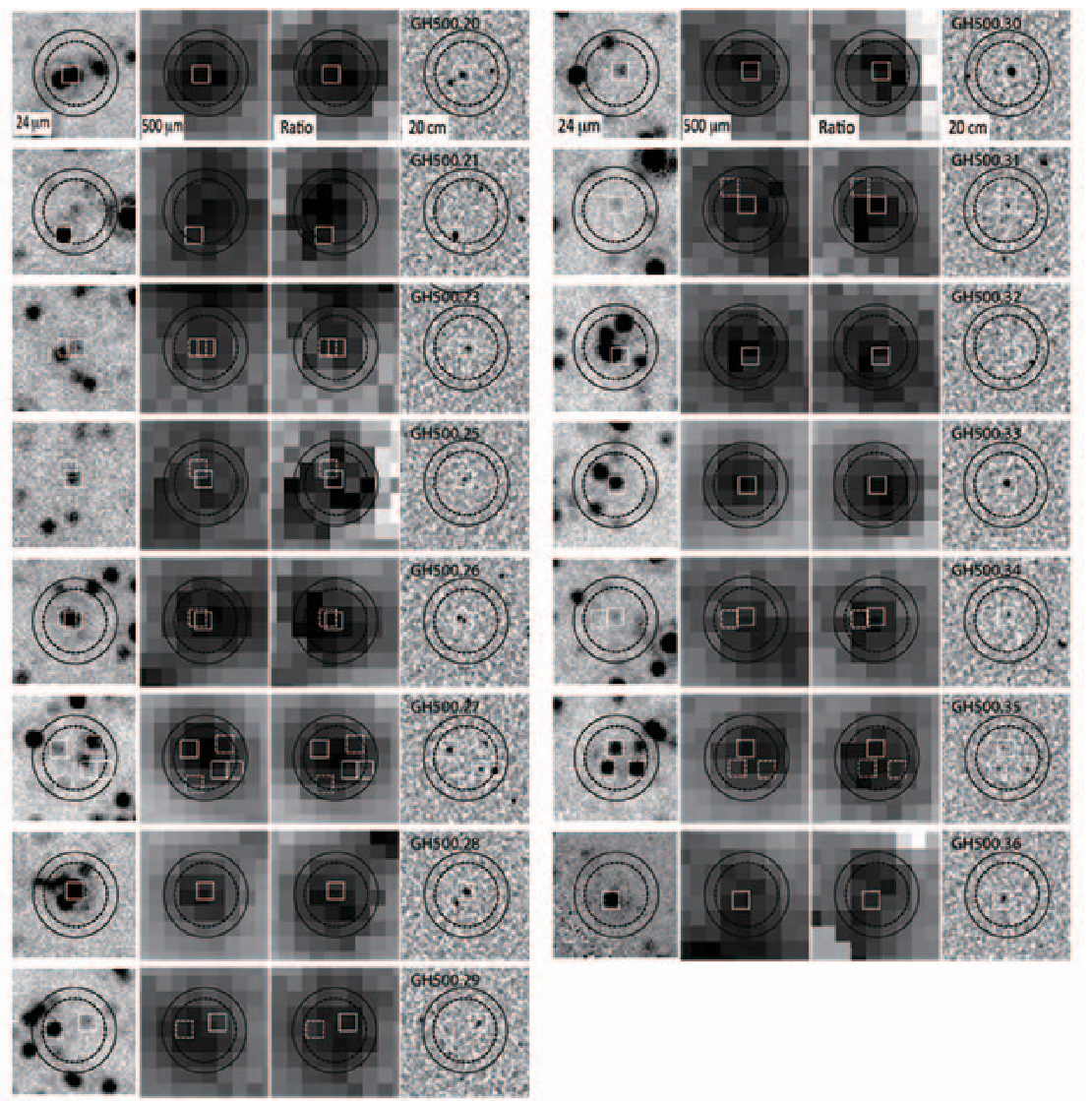}
\caption{
 {\it continued}
}
\end{figure*}
%
%\begin{figure}
%\centering{
%\includegraphics[scale=0.4, angle=0]{figure/ratioplot/contour/GHN_img_comb2.eps}
%\caption{
%--Continued
%Joint 99\% confidence contours of the \fekalfa emission line intensity vs. velocity width FWHM, obtained
%from Gaussian fits to the line observed with the \chandra HEG, as described in the text.
%The vertical dotted lines correspond to the FWHM of H$\beta$ line (Wang et al. 2007).}
%}}\label{fek}
%\end{figure}
\newpage
\clearpage
\section{Infrared-to-radio SEDs of high-redshift Herschel sources}
In Figure D1, we present the measured IR-to-radio photometry and best-fitting SED for each of the 500\um sources 
presented in Table 2. 
We find that the majority of our 500\um sources are well represented by the best-fit SED templates, 
and only seven ULIRGs display evidence of a radio excess above the best-fit SED.

\begin{figure*}[h]
\vspace{10pt}
%\centerline{\psfig{file=/Users/xshu/paper/goodsn/submit/1028/figure/sedfit/firsed/xshu/combine/fig_sed.eps,width=6.5in,height=5.5in,angle=0}}
%\centerline{\psfig{file=/Users/mac/paper/goodsn/submit/revise/0921/figure/sedfit/firsed/combine/fig_sed.eps,width=6.5in,height=5.5in,angle=0}}
%\centerline{\psfig{file=/Users/mac/paper/goodsn/submit/revise/0921/figure/sedfit0419/combine/fig_sed.eps,width=6.5in,height=5.5in,angle=0}}
\centerline{\psfig{file=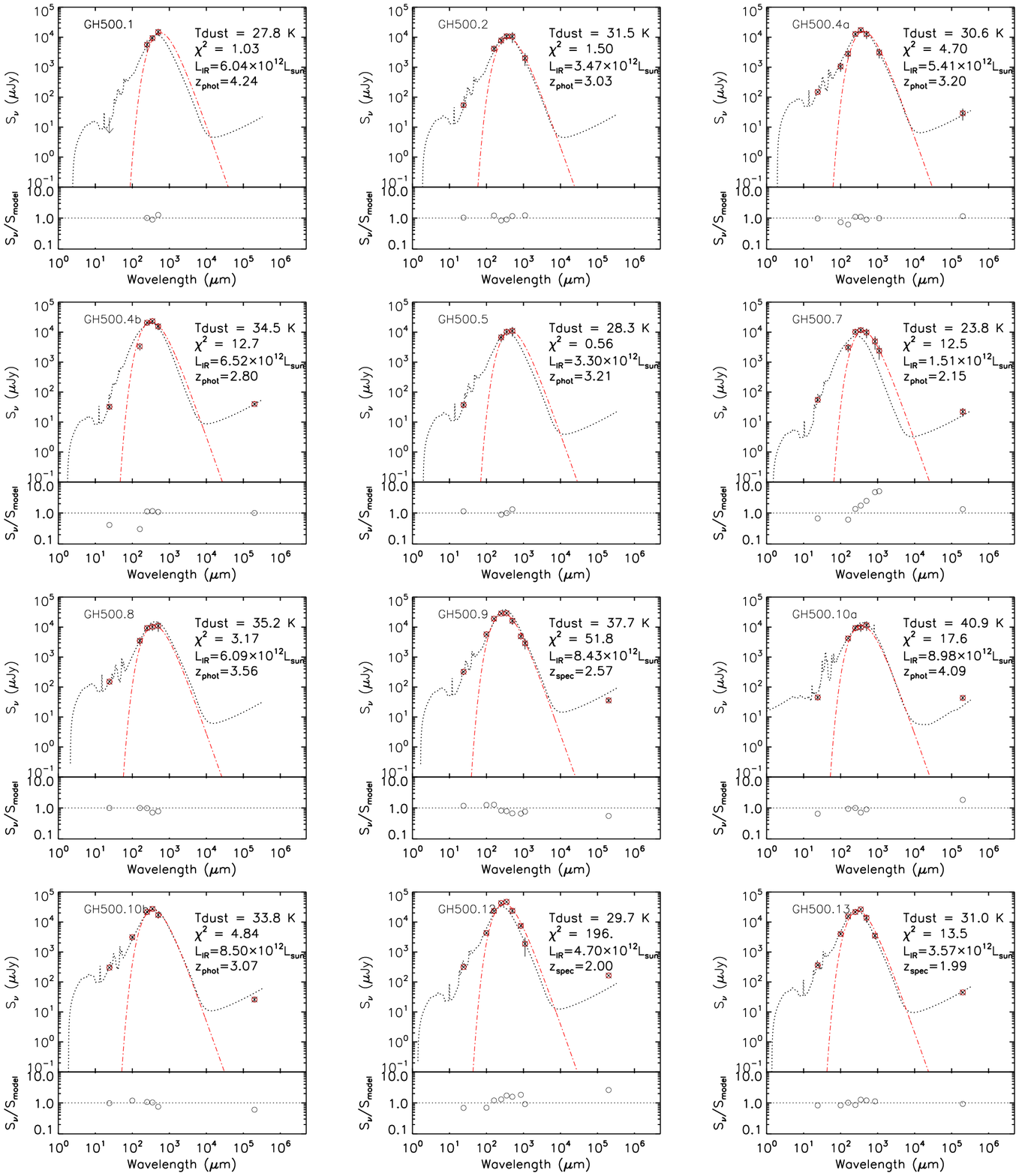,width=6.5in,height=5.5in,angle=0}}
%\centerline{\psfig{file=fig16_sed1.eps, width=6.5in,height=5.5in,angle=0}}
\caption{\footnotesize
Spectral energy distribution of our \ratio-identified $z\simgt2$ \herschel sources (open circles). 
Dotted lines represent the best-fit SED template to the photometry. 
The modified blackbody emission which is used to fit the far-IR-to-submm data
is shown by the red dashed line. 
}
\end{figure*}

\setcounter{figure}{0}
\begin{figure*}[tbh]
\vspace{10pt}
%\centerline{\psfig{file=/Users/xshu/paper/goodsn/submit/1028/figure/sedfit/firsed/xshu/combine/fig_sed2.eps,width=6.5in,height=5.5in,angle=0}}
%\centerline{\psfig{file=/Users/mac/paper/goodsn/submit/revise/0921/figure/sedfit/firsed/combine/fig_sed2.eps,width=6.5in,height=5.5in,angle=0}}
%\centerline{\psfig{file=/Users/mac/paper/goodsn/submit/revise/0921/figure/sedfit0419/combine/fig_sed2.eps,width=6.5in,height=5.5in,angle=0}}
\centerline{\psfig{file=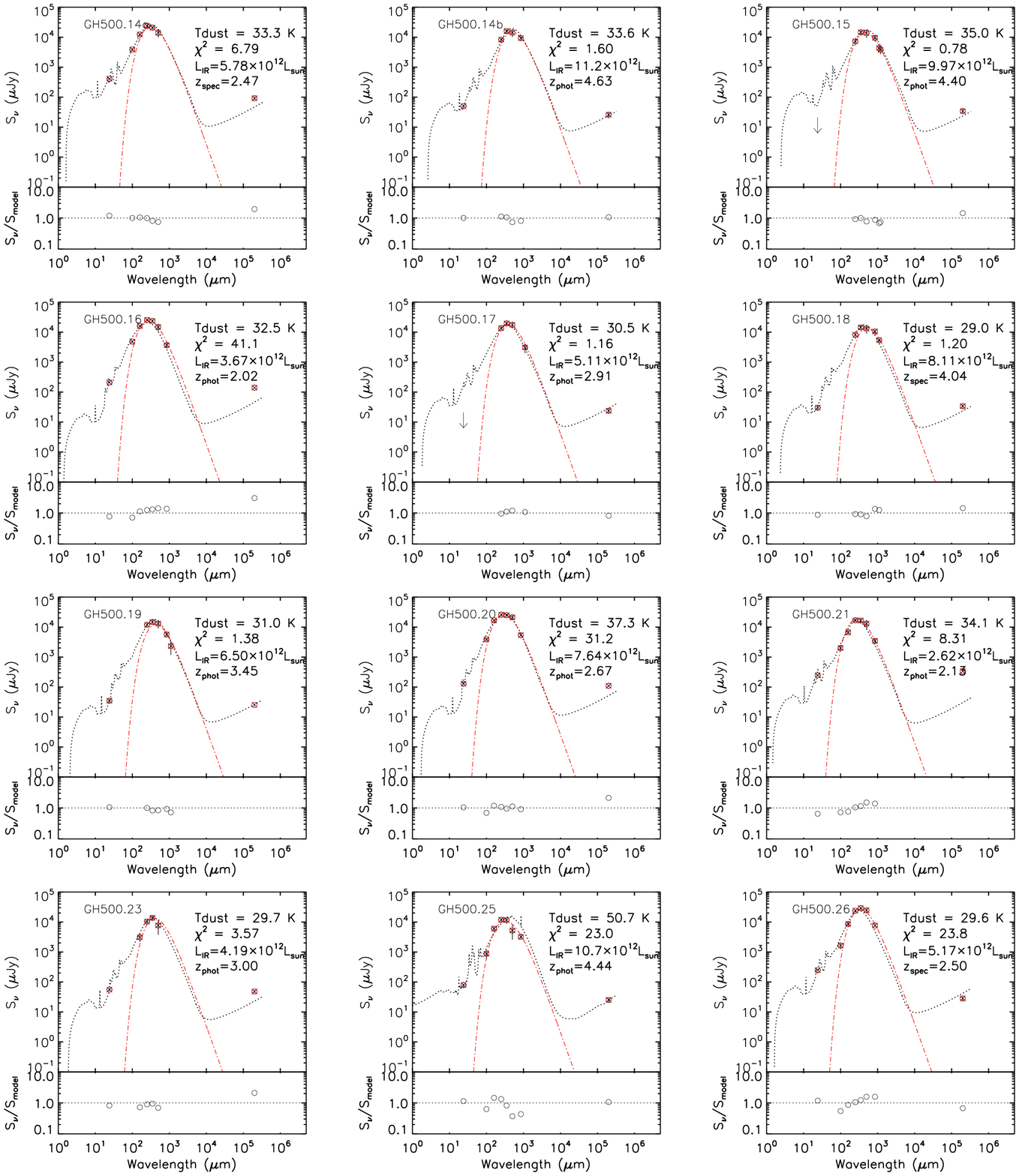,width=6.5in,height=5.5in,angle=0}}
\caption{
  {\it continued}
}
\end{figure*}

\setcounter{figure}{0}
\begin{figure*}[tbh]
\vspace{10pt}
%\centerline{\psfig{file=/Users/xshu/paper/goodsn/submit/1028/figure/sedfit/firsed/xshu/combine/fig_sed3.eps,width=6.5in,height=5.5in,angle=0}}
%\centerline{\psfig{file=/Users/mac/paper/goodsn/submit/revise/0921/figure/sedfit/firsed/combine/fig_sed3.eps,width=6.5in,height=5.5in,angle=0}}
\centerline{\psfig{file=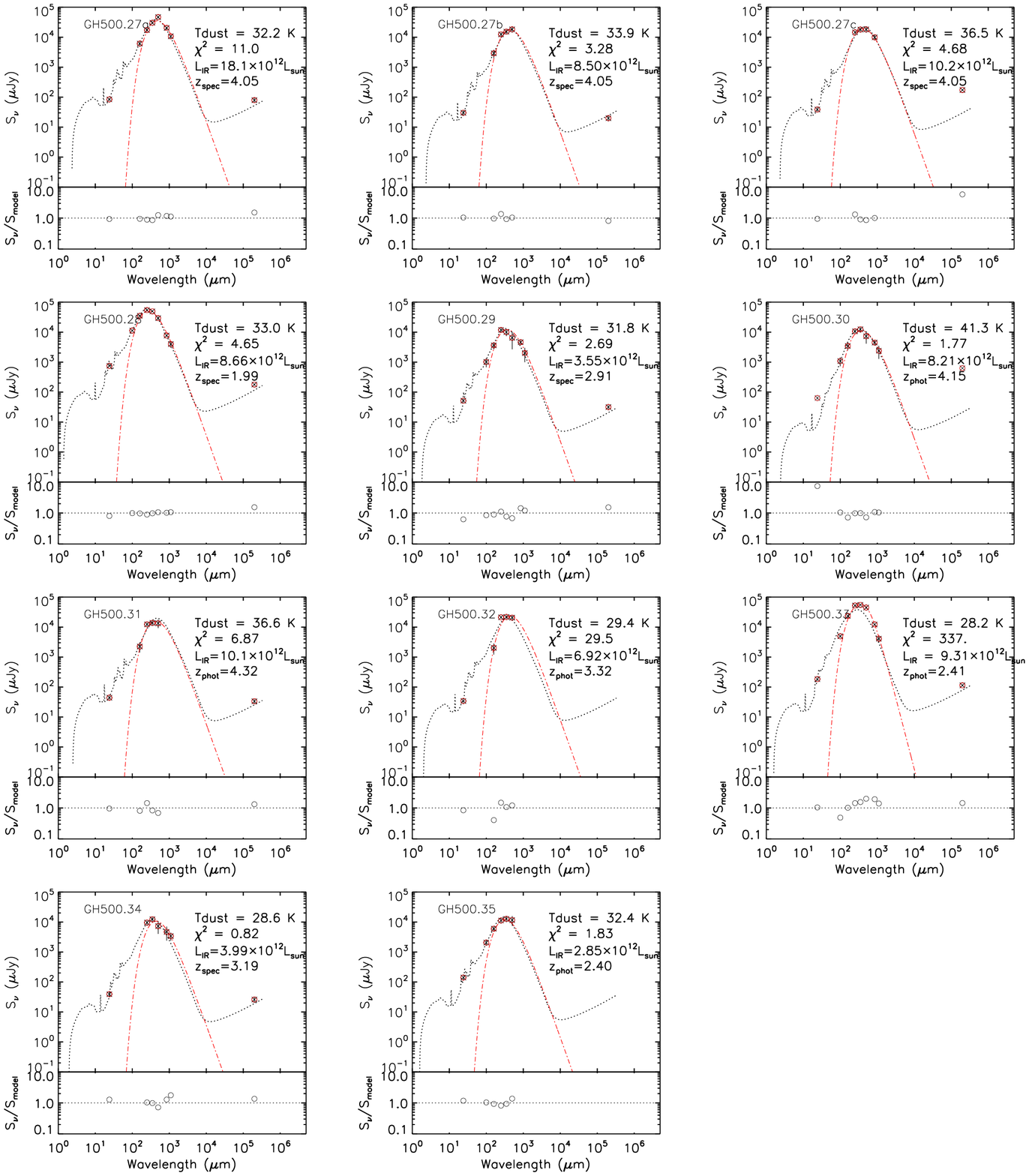,width=6.5in,height=5.5in,angle=0}}
%\centerline{\psfig{file=fig16_sed3.eps, width=6.5in,height=5.5in,angle=0}}
\caption{  {\it continued}}
\end{figure*}
\newpage
\clearpage

\section{Comparison between MIPS and IRAC identifications}

%\section{Validation of the method}

In Figure E1 (left), we plot the cumulative distribution of modified
$P$-values for all 24\um sources within the
search radius (solid line), and that for the IRAC 3.6\um counterparts
(dotted line). The later were calculated using the main-sequence
relation and a mass dependent dust extinction to predict the 500\um
flux from their stellar masses (see the text for details), in order to
taking into account alternative counterparts for a few sources with no
24\um counterpart satisfying $P\simlt$0.1. 
Using only the 24\um image for counterpart identifications, we 
find 27 out of 36 sources detected in the ratio map having
counterparts with $P\simlt$0.1. A $P$-value cut at 0.1 
corresponds to only 30 per cent of the cumulative distribution of
$P$-values for the 24\um catalog, and 7 per cent for the IRAC catalog.  
In comparison to the distribution of the traditional $P$-values for 
the radio catalogue (dashed line), it demonstrates that our approach
using 24\um data can
be useful for identifying secure 500\um counterparts in the absence of
the radio data, by effectively reducing the number of foreground sources falling into the 
search area by chance.  
Note that $P$-statistics for the IRAC catalog that has a high source surface
density (79/arcmin$^2$) is in agreement with the calculation starting
with the 24\um counterparts (Figure E1, right), albeit with a large
scatter. 
%the main-sequence
%relation can only provide an upper limit on the predicted 500\um fluxes and hence
%the inferred $P$-values (Figure E1, right) . 

Several studies have shown that high-redshift submm-selected galaxies
can be identified through their IR colors as measured by IRAC (e.g.,
Pope et al. 2006; Yun et al. 2008). 
By comparing with theoretical color tracks of dusty starbursts, Yun et al. (2008) 
proposed a dust-obscured young stellar population as the origin of the red IRAC color. 
%Candidate counterparts to high$-z$ starbursts can be identified 
In Figure E2, we show the 3.6- and 5.8-\um color-flux diagram for all IRAC sources 
within the search radius (grey dots), with secure counterparts identified with $P$-statistics 
highlighted in red circles (Table 2). It is apparent that candidate high$-z$ 500\um sources in the GOODS-North 
are typically redder than the field population. 
Within the region that was defined to select SMG counterparts (dotted line, Biggs et al. 2011), 
%rate from field galaxies, and 
we recover 91 per cent of the catalogued sources in Table 2, 
further supporting the reliability of our counterpart identifications.
%Candidate high-$z$ 500\um sources which
%are also detected at 450\um are marked with red circles.  
%Similar to Figure 5 (upper), a $P$-value cut at $\simlt$0.1 is
%sufficient to reveal most, if not all, $z\simgt$2
%candidates with major contributions to the 500\um emission. 
%galaxies detected in the ratio map are associated with the SCUBA-2 sources.  
%The photometric redshifts for most of the identified 500\um sources (16/19) are at 
%$z_{\rm phot}>2$ with a median of 
%These tests therefore suggest that our method can uncover
%potential high-$z$ \herschel 500\um sources with a high efficiency (identification rate of 18/22$\sim$82\%).    

%In Figure E3, we plot 
%We found that the photometric redshift distribution of the SCUBA-2 counterparts 
%We found that the photometric redshifts for most of
%the identified SCUBA-2 sources (16/19) are at $z_{\rm phot}>2$ with a median of $z=2.41\pm0.4$, 
%suggesting that our method is indeed able to uncover high-$z$ \herschel 500\um sources with 
%high efficiency (an identification rate of 18/22$\sim$82\%).    

\begin{figure}
\centering
\includegraphics[scale=0.81, angle=0]{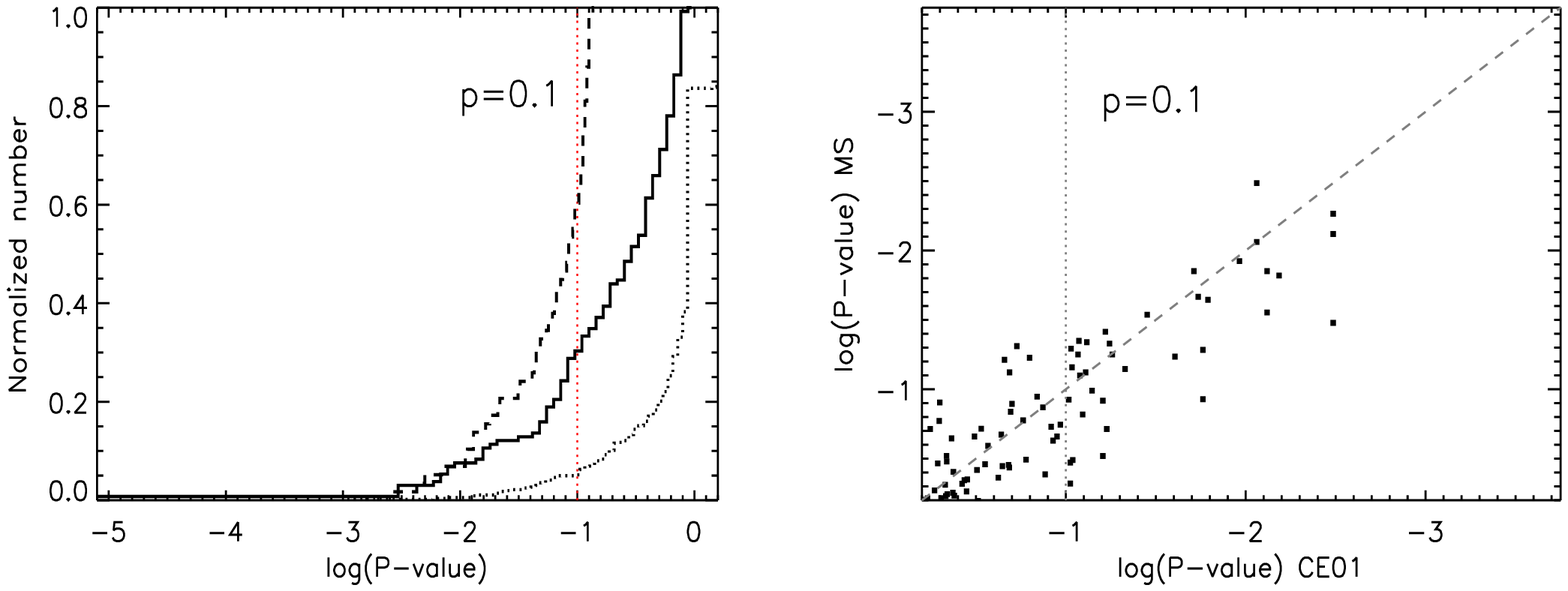}
\caption{
{\it Left:} Cumulative distribution of $P$-values for all counterparts
  within the search area for our cataloged sources. The solid line shows the distribution for
  24\um counterparts, while the dotted and dashed line is for the
  counterparts at the IRAC 3.6\um and radio, respectively.
{\it Right:} Comparison between the $P$-values calculated using the predicted $S_{\rm 500\mu m}$ 
starting from the 24\um flux and that starting from the stellar masses (Schreiber et al. 2015). Grey
dashed line is the one to one relation. 
}
\label{fig:fig_p1p2}
\end{figure}

\begin{figure}
\centering
\includegraphics[scale=0.61, angle=-90]{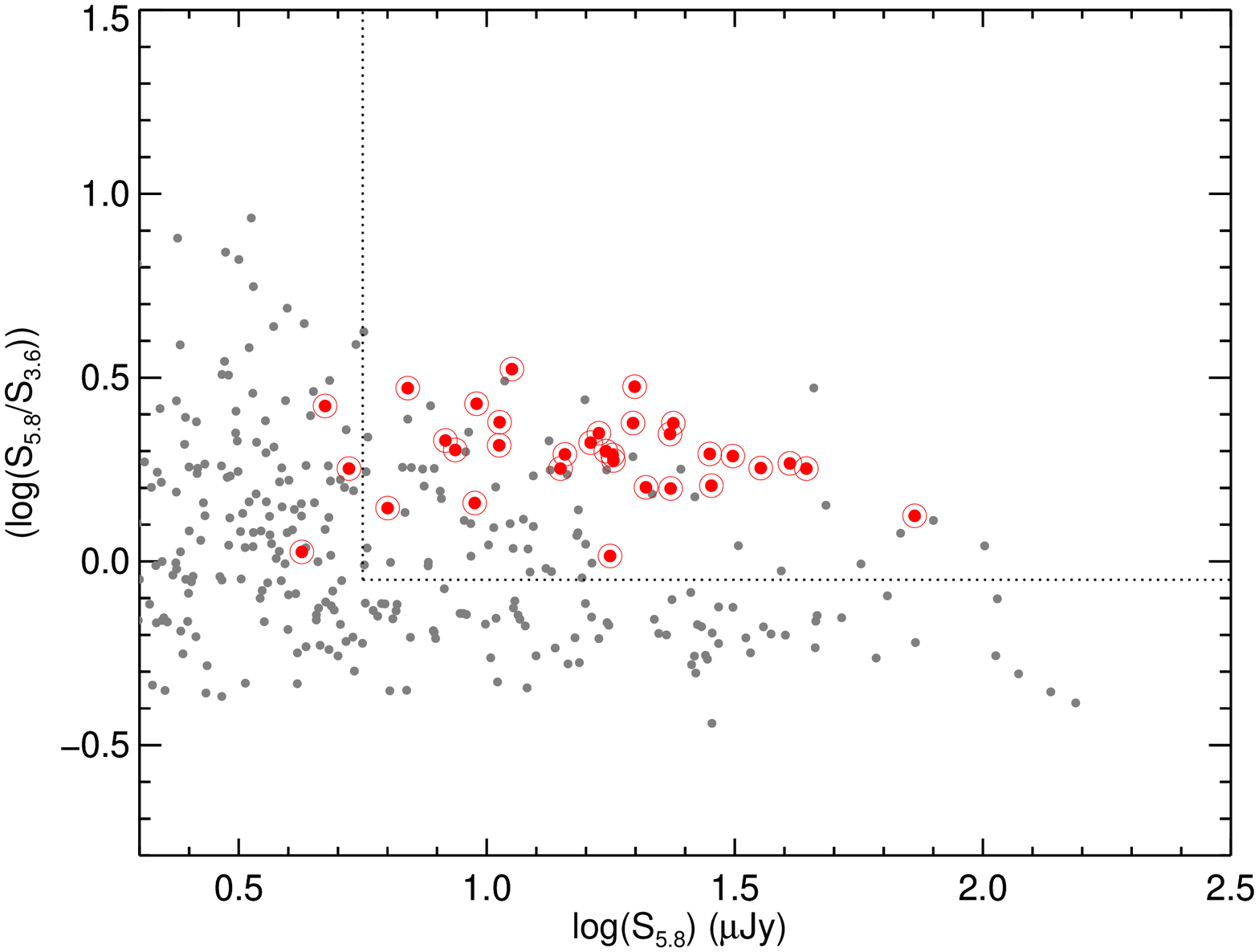}
\caption{
3.6 and 5.8\um colour-flux diagram for IRAC sources falling into the search area.
The secure radio and MIPS-identified counterparts are shown with red circles. The dotted line shows the region
to select SMGs defined by Biggs et al. (2011). 
}
\label{fig:fig_p1p2}
\end{figure}

\end{document}